%% file: ANA-HION-2018-22-PAPER.tex
\documentclass[PAPER, cernpreprint, atlasdraft=false, UKenglish, texmf, orcidlogo]{atlasdoc}
\usepackage{atlaspackage}
\usepackage{atlasbiblatex}

\usepackage{atlasphysics}
\graphicspath{{logos/}{figures/}}

\usepackage{ANA-HION-2018-22-PAPER-defs}

\AtlasTitle{Measurement of $J/\psi$--jet correlations in \(pp\) and \ce{Pb}+\ce{Pb}\xspace collisions at $\sqrt{s_{\text{NN}}}=5.02$ TeV with the ATLAS detector}

\AtlasAbstract{%
Yields of charmonia, bound states of $c\bar{c}$ quarks, are observed to be strongly suppressed in heavy-ion collisions relative to proton--proton collisions as a result of their interaction with the quark--gluon plasma produced in such collisions. Understanding the mechanisms responsible for this suppression requires a detailed understanding of charmonium production. To constrain the production mechanisms, it is essential to quantify how charmonium production correlates with jet activity.
Measurements of $J/\psi$ production in the dimuon decay channel are presented for inclusive $J/\psi$ mesons and for $J/\psi$ mesons that are either isolated or not isolated from jets, separately for prompt and non-prompt production.
Non-isolated $J/\psi$ mesons are defined by matching $J/\psi$ candidates to anti-$k_t$ jets with $R=0.2$ and $p_\mathrm{T}>20$~GeV, reconstructed without signal from the $J/\psi$ meson, allowing the quantification of jet activity accompanying $J/\psi$ production. The analysis uses Pb+Pb and $pp$ collisions at $\sqrt{s_{\text{NN}}}=5.02$~TeV recorded by the ATLAS experiment at the LHC, corresponding to integrated luminosities of $1.82$~nb$^{-1}$ and $0.26$~fb$^{-1}$, respectively.
The $J/\psi$ yields in Pb+Pb, cross sections in $pp$, and nuclear modification factors are presented for non-isolated, isolated, and inclusive $J/\psi$ up to $p_\mathrm{T}=60$~GeV. Isolated fractions and non-prompt fractions are reported in both $pp$ and Pb+Pb collisions. Strong $J/\psi$ suppression is observed to persist up to the highest measured $p_\mathrm{T}$ values, with differences observed between non-isolated, isolated, and inclusive $J/\psi$. These results provide new constraints on charmonium production mechanisms and their in-medium suppression.
}

\AtlasRefCode{HION-2018-22}

\PreprintIdNumber{CERN-EP-2026-167}

\HEPDataRecord{183386}

\AtlasJournalRef{\JHEP 11 (2026) 020}
\AtlasDOI{10.1007/JHEP11(2026)020}

\AtlasCoverEgroupAnalysisTeam{atlas-HION-2018-22-analysis-team@cern.ch}

\hypersetup{pdftitle={ATLAS document},pdfauthor={The ATLAS Collaboration}}

\begin{document}

\maketitle

\tableofcontents

\clearpage

%
%

%

% The next lines are included from the .//sections/01_Introduction.tex input file
\section{Introduction}

High-energy heavy-ion collisions at the Relativistic Heavy Ion Collider (RHIC)~\cite{Hahn:2003sc} and the Large Hadron Collider (LHC)~\cite{Evans:2008zzb} create conditions under which strongly interacting matter transitions into the deconfined phase of quantum chromodynamics (QCD), the quark--gluon plasma (QGP) \cite{Busza:2018rrf,Shuryak:2014zxa}. In this environment, hard probes --
particles or partonic systems produced in large-momentum-transfer scatterings that precede the formation of the medium --
serve as sensitive tools for characterising the medium. Energetic partons undergo substantial energy loss, giving rise to the well-established phenomenon of jet quenching \cite{Connors:2017ptx,Cunqueiro:2021wls,Apolinario:2024equ}, while heavy quark--antiquark bound states (quarkonia) are modified through colour screening, in-medium dissociation, and regeneration \cite{Brambilla:2010cs,ALICE:2019lga,ALICE:2019nrq}. The resulting suppression of quarkonium production is found to be pronounced and to persist up to transverse momenta of several tens of \GeV \cite{ ALICE:2023gco, ALICE:2022jeh, HION-2016-07, CMS-HIN-16-025, CMS-HIN-21-007, HION-2021-12, ALICE:2020wwx, CMS-HIN-16-023}. In the case of $c\bar{c}$ bound states (charmonia), besides the conventional suppression mechanisms associated with colour screening and in-medium dissociation, it has also been suggested that charmonia with high transverse momenta ($\pt$) may be affected by medium-induced gluon radiation, leading to an energy-loss mechanism analogous to that responsible for jet quenching \cite{Makris:2019ttx,Spousta:2016agr,Arleo:2017ntr}.

Understanding the origin of charmonia suppression requires a clear picture of how charmonia are produced in high-energy collisions. Charmonium states may be created directly in the hard scattering and subsequently evolve as either colour-singlet or colour-octet configurations \cite{Bodwin:1994jh,Cho:1995vh,Cho:1995ce,Brambilla:2010cs}, or they may be formed during a parton-shower evolution %
\cite{Baumgart:2014upa,Bain:2017wvk}. Although extensive theoretical work has been devoted to charmonia production, the dominant production mechanism and the possible interplay among different production mechanisms remain unresolved \cite{Lansberg:2019adr}. Additional complexity arises from the distinction between prompt charmonia, formed on the time scale of the strong interaction, and non-prompt charmonia, which appear much later through $b$-hadron decays. To constrain the production mechanisms, it is therefore important to experimentally quantify how charmonium production correlates with jet activity.

To quantify the correlation between charmonium production and jet activity, the CMS and LHCb collaborations measured the $\jpsi$--jet fragmentation function in 5.02~\TeV \pp collisions \cite{CMS-HIN-19-007} and 13~\TeV \pp collisions \cite{LHCb:2017llq}, respectively. In those measurements, the fragmentation function corresponds to the distribution of $z$, with $z$ defined as the ratio of the $\jpsi$ $\pt$ to the $\pt$ of the jet into which the $\jpsi$ is clustered. The prompt $\jpsi$ mesons were found to have more surrounding jet activity, i.e., to populate lower values of $z$ than predicted by \Pythia~8 \cite{Sjostrand:2007gs} simulations, suggesting that $\jpsi$ production late in the parton shower is generally underestimated in the model \cite{Bain:2017wvk}. The correlation between the $\jpsi$ production and jet activity was also studied at the level of yields in 8~\TeV \pp collisions by CMS \cite{CMS-BPH-15-003}. That measurement used events where at least one jet with $\pt>25$~\GeV was observed and evaluated the angular separation between the $\jpsi$ meson and the jet, finding that most prompt $\jpsi$ mesons with energy above 15~\GeV and rapidity $|y| < 1$ are contained in jets.

While the existing $\jpsi$--jet correlation measurements provide important constraints on theoretical models describing charmonium production, they still do not give complete information. In particular, the definition of whether the $\jpsi$ is contained in a jet depends on the jet definition. As the jet four-momentum is constructed from a sum containing the four-momentum of the $\jpsi$, an otherwise isolated $\jpsi$ with high enough $\pt$ is by definition a jet, thus in a jet, whereas the isolated $\jpsi$ with $\pt$ lower than the minimum value used to define a jet is not in a jet~\cite{Lansberg:2019adr}. In addition, studying the $\jpsi$ in an event sample that is triggered by the presence of a jet can introduce a possible event selection bias.

<<<<<<< HEAD
The first goal of the measurement presented in this \paper\ is to provide new information for quantifying the $\jpsi$--jet correlation by separating the kinematics of $\jpsi$ from the kinematics of jets and by analysing the $\jpsi$--jet correlations in all $\jpsi$ events rather than in events a priori containing jets. The second goal is to measure possible differences between the suppression of isolated $\jpsi$ and $\jpsi$ produced in association with jets in Pb+Pb collisions.

The measurement presented in this \paper\ uses \PbPb data collected by the ATLAS Experiment in 2015 and 2018 and \pp data collected in 2017, both at a centre-of-mass energy of 5.02~\TeV per colliding nucleon pair.  The Pb+Pb and \pp data samples correspond to integrated luminosities of $1.82$~nb$^{-1}$ and $0.26$~fb$^{-1}$, respectively. The measurement thus presents the quantification of $\jpsi$ suppression using the full LHC \RunTwo data sample of \PbPb collisions.
=======
The first goal of the measurement presented in this \paper\ is to provide new information for quantifying the $\jpsi$--jet correlation by separating the kinematics of $\jpsi$ from the kinematics of jets and by analysing the $\jpsi$--jet correlations in all $\jpsi$ events rather than in events a priori containing jets. The second goal is to measure possible differences between the suppression of isolated $\jpsi$ and $\jpsi$ produced in association with jets in \PbPb collisions.

The measurement presented in this \paper\ uses \PbPb data collected by the ATLAS Experiment in 2015 and 2018 and \pp data collected in 2017, both at a centre-of-mass energy of 5.02~\TeV per colliding nucleon pair.  The \PbPb and \pp data samples correspond to integrated luminosities of $1.82$~nb$^{-1}$ and $0.26$~fb$^{-1}$, respectively. The measurement thus presents the quantification of $\jpsi$ suppression using the full LHC \RunTwo data sample of \PbPb collisions.
>>>>>>> origin/master

% End of text imported from the .//sections/01_Introduction.tex input file

% The next lines are included from the .//sections/02_ATLASDetector.tex input file
%
\newcommand{\AtlasCoordFootnote}{
ATLAS uses a right-handed coordinate system with its origin at the nominal interaction point (IP)
in the centre of the detector and the \(z\)-axis along the beam pipe.
The \(x\)-axis points from the IP to the centre of the LHC ring,
and the \(y\)-axis points upwards.
Polar coordinates \((r,\phi)\) are used in the transverse plane,
\(\phi\) being the azimuthal angle around the \(z\)-axis.
The pseudorapidity is defined in terms of the polar angle \(\theta\) as \(\eta = -\ln \tan(\theta/2)\) and is equal to the rapidity
$ y = \frac{1}{2} \ln \left( \frac{E + p_z}{E - p_z} \right) $ in the relativistic limit.
Angular distance is measured in units of \(\Delta R \equiv \sqrt{(\Delta y)^{2} + (\Delta\phi)^{2}}\).}

\section{ATLAS detector}

The ATLAS detector~\cite{PERF-2007-01} at the LHC covers nearly the entire solid angle around the collision point.\footnote{\AtlasCoordFootnote}
It consists of an inner tracking detector surrounded by a thin superconducting solenoid, electromagnetic and hadronic calorimeters,
and a muon spectrometer incorporating three large superconducting air-core toroidal magnets.

The inner-detector system (ID) is immersed in a \qty{2}{\tesla} axial magnetic field
and provides charged-particle tracking in the range of \(|\eta| < 2.5\).
The high-granularity silicon pixel detector covers the vertex region and typically provides four measurements per track,
the first hit generally being in the insertable B-layer (IBL) installed before Run~2~\cite{ATLAS-TDR-19,PIX-2018-001}.
It is followed by the SemiConductor Tracker (SCT), which usually provides eight measurements per track.
These silicon detectors are complemented by the transition radiation tracker (TRT),
which enables radially extended track reconstruction up to \(|\eta| = 2.0\).
The TRT also provides electron identification information
based on the fraction of hits (typically 30 in total) above a higher energy-deposit threshold corresponding to transition radiation.

The calorimeter system covers the pseudorapidity range \(|\eta| < 4.9\).
Within the region \(|\eta|< 3.2\), electromagnetic calorimetry is provided by barrel and
endcap high-granularity lead/liquid-argon (LAr) calorimeters,
with an additional thin LAr presampler covering \(|\eta| < 1.8\)
to correct for energy loss in material upstream of the calorimeters.
Hadronic calorimetry is provided by the steel/scintillator-tile calorimeter,
segmented into three barrel structures within \(|\eta| < 1.7\), and two copper/LAr hadronic endcap calorimeters.
The solid angle coverage is completed with copper/LAr and tungsten/LAr calorimeter modules (FCal),
covering the forward regions of $3.1 < |\eta| < 4.9$.
The zero-degree calorimeters (ZDC) \cite{Avoni:2025lmu} measure neutral particles at pseudorapidities \(|\eta| \geq 8.3\) and consist of layers of alternating quartz rods and tungsten plates. ZDC modules are positioned symmetrically at \(z=\pm140~\mathrm{m} \).

The muon spectrometer (MS) comprises separate trigger and
high-precision tracking chambers measuring the deflection of muons in a magnetic field generated by the superconducting air-core toroidal magnets.
The field integral of the toroids ranges between \num{2.0} and \qty{6.0}{\tesla\metre}
across most of the detector.
Three layers of precision chambers, each consisting of layers of monitored drift tubes (MDTs), cover the region \(|\eta| < 2.7\),
complemented by cathode-strip chambers (CSCs) in the forward region, where the background is highest.
The muon trigger system covers the range \(|\eta| < 2.4\) with resistive-plate chambers (RPCs) in the barrel, and thin-gap chambers (TGCs) in the endcap regions.

The luminosity is measured mainly by the LUCID--2~\cite{LUCID2} detector that records Cherenkov light produced in the quartz windows of photomultipliers located close to the beampipe.

A two-level trigger system is used to select events of interest. The first-level (L1) trigger system is implemented in custom hardware, followed by selections made by software algorithms in the high-level trigger (HLT)~\cite{TRIG-2016-01}. The L1 trigger accepts events from the \qty{40}{\MHz} bunch crossings at a rate below \qty{100}{\kHz}, which the high-level trigger further reduces to record complete events to disk at about \qty{1}{\kHz}. The L1 muon trigger requires coincidences between hits on different RPC or TGC planes, which are used as a seed for the HLT algorithms. The HLT uses dedicated algorithms to incorporate information from both the MS and the ID, achieving position and momentum resolution close to that provided by the offline muon reconstruction~\cite{TRIG-2018-01}.

A software suite~\cite{SOFT-2022-02} is used in data simulation, in the reconstruction
and analysis of real and simulated data, in detector operations, and in the trigger and data acquisition
systems of the experiment.

% End of text imported from the .//sections/02_ATLASDetector.tex input file

% The next lines are included from the .//sections/03_DataSel.tex input file
\section{Data and simulated event samples}
\label{sec:data}

Events in \PbPb and \pp collisions are collected using a trigger requiring the presence of at least two muons, both with $\pTmu > 4$~\GeV. In the \pp sample, both muons must generate a spatially separated L1 muon trigger region of interest (RoI) and be confirmed by the HLT, while in the \PbPb sample, only one muon is required to be seen by the L1 muon trigger and confirmed by the HLT, the second muon is only required to satisfy the HLT requirements \cite{TRIG-2018-01}.
In addition to the muon trigger, two triggers are used in \PbPb collisions to select minimum-bias events. These are based on the presence of at least \qty{50}{\GeV} of transverse energy
in all sections of the calorimeter system ($|\eta| < 4.9$) or, for events that do not meet this condition, on the presence of energy deposition above a one-neutron threshold in both ZDC arms.

Both \PbPb and \pp data also contain contributions from multiple inelastic interactions per bunch crossing, also referred to as pile-up. The average number of inelastic interactions per bunch crossing reaches a maximum value of 0.004 in \PbPb collisions, and it varies between 1.4 and 4.4 in \pp collisions. Although only a small fraction of the \PbPb events ($<0.5$\%) contain pile-up, this contribution is suppressed by utilising the observed anti-correlation, expected from the nuclear geometry, between the total transverse energy deposited in both of the forward calorimeters, $\sumETF$, and the energy in the ZDC. Pile-up collisions are not rejected in \pp collisions.

Monte Carlo (MC) simulated event samples are used for performance studies and to correct for detector effects. Prompt $(pp \rightarrow J/\psi \rightarrow \mu\mu)$ and non-prompt $(pp \rightarrow b\bar{b} \rightarrow J/\psi \rightarrow \mu\mu)$ samples of \jpsi were produced using \Pythia~8.2 \cite{Sjostrand:2014zea} for event generation and \textsc{Photos} \cite{Barberio:1990ms} for electromagnetic radiation corrections. The A14 set of tuned parameters \cite{ATL-PHYS-PUB-2014-021} is used together with the CTEQ6L1 parton distribution function set \cite{Pumplin:2002vw}. The response of the ATLAS detector is simulated using \textsc{Geant4} \cite{SOFT-2010-01,Agostinelli:2002hh}. To simulate the \jpsi production in the high-multiplicity environment of \PbPb collisions, the simulated \pp events are overlaid with dedicated minimum-bias \PbPb data events \cite{Haas_2017}. The minimum-bias data events from \PbPb collisions are combined with the signal from the \Pythia~8 simulation of hard scattering events at the digitisation stage, and then reconstructed as a combined event.

In \PbPb collision data and MC samples, the transverse energy at the electromagnetic scale deposited in the forward calorimeter, $\sumETF$, is used to estimate the centrality of the collision. Centrality describes the degree of geometric overlap of two colliding nuclei in the plane perpendicular to the beam, with large overlap in central collisions and small overlap in peripheral collisions.
The measurement is performed in four exclusive centrality intervals, 0--10\%, 10--20\%, 20--40\%, 40--80\%, and an inclusive 0--80\% centrality interval. The centrality is defined such that smaller percentiles label more central collisions.
A weight is assigned to each MC event, ensuring that the event sample obtained from the minimum-bias trigger entering the MC overlay has the same centrality distribution as that measured in the data with $\jpsi$ meson candidates. The \textsc{TGlauberMC}~v3.2 package \cite{Loizides:2017ack}, which provides an improved implementation of the MC Glauber model \cite{Miller:2007ri}, is used to calculate the mean nuclear overlap function, $\avgTAA$ \cite{Miller:2007ri,HION-2020-06}, as well as its uncertainty, for each centrality interval as presented in Table~\ref{tab:centrality}. The quantity $\avgTAA$ is proportional to the mean number of binary nucleon--nucleon collisions divided by the inelastic nucleon--nucleon cross section, and accounts for the nuclear overlap geometry in a given centrality interval.

The number of minimum-bias events, $\Nevt$, is used to normalise the yield in the respective centrality interval. Minimum-bias events are selected by requiring that they pass at least one of the two minimum-bias triggers. The analysed data sample corresponds to $11.15\times 10^9$ \PbPb collisions after correction for the trigger prescale factor.

\begin{table}[htbp]
\centering
\caption{
The \avgTAA values and corresponding uncertainties for each centrality interval. The centrality-inclusive \avgTAA corresponds to the weighted sum of \avgTAA in individual centrality intervals.}
\begin{tabular}{c|c|c}
\toprule
Centrality & \avgTAA[$\mathrm{mb}^{-1}$] & $\sigma(\avgTAA)$ [$\mathrm{mb}^{-1}$] \\
\midrule
0--10\%  &  \qty{23.22}{}	& \qty{0.06}{} \\
10--20\% &  \qty{14.38}{}	& \qty{0.12}{} \\
20--40\% &  \qty{6.93}{}	& \qty{0.13}{} \\
40--80\% &  \qty{1.23}{}	& \qty{0.06}{} \\
0--80\%  &  \qty{7.05}{}	& \qty{0.08}{} \\
\bottomrule
\end{tabular}
\label{tab:centrality}
\end{table}

% End of text imported from the .//sections/03_DataSel.tex input file

% The next lines are included from the .//sections/04_Analysis.tex input file
\section{Data analysis}

The $\jpsi$ mesons are reconstructed in the dimuon ($\mu^+\mu^-$) decay channel.
The reconstructed muons are required to satisfy the \texttt{Medium} muon working point, ignoring the TRT requirements \cite{PERF-2015-10}, have $\pTmu > 4$~\GeV, $|\eta^{\mu}|$~$< 2.4$, and be matched to the muons reconstructed at the trigger level. Pairs of muons with opposite charges are used to construct dimuon objects. The transverse momentum of the dimuon, \pTmumu, is required to be larger than 9~\GeV, and the rapidity of the dimuon, $y_{\mu\mu}$, is required to have an absolute value of less than 2. In addition, the invariant mass of the dimuon, $m_{\mu\mu}$, is required to be between 2.5~\GeV and 3.5~\GeV. The selected dimuon is refitted to a common vertex.
The yields entering the evaluation of observables are extracted using two-dimensional unbinned maximum-likelihood fits \cite{Barlow:1993dm} in dimuon invariant mass $m_{\mu\mu}$ and pseudo-proper lifetime $\tau_{\mu\mu}$ defined as \cite{HION-2016-07}
\begin{equation}
\tau_{\mu\mu} = \frac{L_{xy} m_{\mu\mu}}{p_{\mathrm{T}}^{\mu\mu}},
\end{equation}
where $L_{xy}$ is the distance between the reconstructed primary vertex and the reconstructed dimuon vertex in the event on the transverse plane. It corresponds to the signed projection onto the \Jpsi momentum. The $\tau_{\mu\mu}$ is introduced to separate prompt and non-prompt \Jpsi in the fitting procedure.

The raw dimuon kinematic distributions entering the fitting procedure are corrected for detector-related effects, as described in Section~\ref{sec:raw}. The dimuon candidates are tagged for the presence of jets, which is described in Section~\ref{sec:jet}. The fit model and the extraction of corrected yields are described in Section~\ref{sec:fit}. Final observables extracted using corrected yields are defined in Section~\ref{sec:obs}.

\subsection{Corrections of raw \jpsi distributions}
\label{sec:raw}

Before the dimuon $m_{\mu\mu}$ and $\tau_{\mu\mu}$ distributions are fit to extract \jpsi yields, each candidate dimuon is corrected using a weight that accounts for the \jpsi dimuon acceptance, trigger efficiency, and reconstruction efficiency.
The dimuon acceptance $\mathcal{A}(\psi(nS))$(\absrapmumu,\xspace \pTmumu) is defined as the probability that the $\psi(nS)$ decay products fall within the fiducial volume (\pTmu ~$>$~\qty{4}{\GeV}, \xspace \absetamu~$<$~\qty{2.4}{}). The acceptance is obtained from a generator-level \enquote{accept-reject} simulation as described in Ref.~\cite{BPHY-2012-02}.
Finite detector and reconstruction resolution may lead to the \Jpsi acceptance correction being applied to the \psip candidates, and vice versa. Therefore, the upper mass boundary for the \Jpsi candidates of 3.5~\GeV is used, and the lower mass boundary for the \psip candidates is set to be 3.2~\GeV, resulting in the overlap range of 0.3~\GeV. The interpolation of the acceptances is then performed as
\begingroup
\normalsize
\begin{equation}
\mathcal{A}(\absrapmumu,\,\pTmumu) =
\begin{cases}
\mathcal{A}(J/\psi)
& m_{\mu\mu} \leq 3.2\,\text{\GeV}, \\[5pt]
\mathcal{A}(J/\psi)
\frac{3.5-m_{\mu\mu}\,[\text{\GeV}]}{0.3}
+ \mathcal{A}(\psip)
\frac{m_{\mu\mu}\,[\text{\GeV}]-3.2}{0.3}
& 3.2 < m_{\mu\mu} < 3.5\,\text{\GeV}.
\end{cases}
\label{eq:accinterpolation}
\end{equation}
\endgroup
To account for the final state radiation (FSR) of the \Jpsi decay products, the \PHOTOS 3.64 model \cite{Barberio:1990ms} is included in the \enquote{accept-reject} algorithm. In principle, the acceptance could also depend on the spin alignment of the $\jpsi$. In this analysis, $\jpsi$ mesons are assumed to be produced unpolarised, following Refs.~\cite{LHCb:2013izl,ALICE:2011gej,CMS-BPH-13-003}. The effects of variations to this assumption are considered and discussed in Section~\ref{sec:sys}.

The dimuon reconstruction efficiency, $\varepsilon_\mathrm{reco}(\mu_1,\mu_2)$, is determined as the product of two single-muon reconstruction efficiencies. Muon reconstruction in ATLAS is performed independently in the ID and MS. The total single-muon reconstruction efficiency can be factorised into the ID track reconstruction efficiency and MS reconstruction efficiency. %
The single muon reconstruction efficiency for the \pp data is obtained from the \Jmumu \Pythia~8 simulation. Additional data-to-MC scale factors are derived using \Jmumu tag and probe method \cite{MUON-2018-03} using 13~\TeV \pp data, as the previous study showed no dependence of the MS efficiency on the pile-up conditions \cite{HION-2019-58}. The combined ID+MS muon reconstruction efficiency for the \PbPb data is obtained using the tag and probe method in \Jmumu \Pythia~8 events overlaid with dedicated minimum-bias \PbPb data events as described in Section~\ref{sec:data}. Data-to-MC scale factors are determined in \PbPb data to account for the small differences between data and simulation. The scale factors are close to unity in the barrel region and reach values of 1.05 at low \pT in the endcap regions. The MS+ID efficiency at $\pTmu=4$~\GeV is about 55\% in the barrel region (\absetamu~$<1.05$) and 70\% in the endcap region ($1.05<\absetamu<2.4$). The efficiency saturates around 95\% at $\pTmu=9$~\GeV for both barrel and endcap regions. The performance of the muon reconstruction at low \pTmu in the barrel region is affected mainly by the muon energy loss in the calorimeter system and the presence of multiple scattering.

The dimuon trigger efficiency, $\varepsilon_\mathrm{trig}(\mu_1,\mu_2)$, is constructed from the corresponding single-muon trigger efficiencies in the input. The single-muon trigger employs two distinct logic schemes: a) an L1-seeded muon trigger, which requires the formation of an L1 RoI of the muon candidate that is subsequently confirmed in the HLT trigger sequence, and b) an HLT-only muon trigger, which is only performed at the HLT level without any L1 trigger requirement.
The single-muon trigger efficiency is determined in \Pythia MC simulation. Data-to-MC scale factors, defined as the ratio of the trigger efficiency evaluated in the \pp data to that evaluated in the MC, are used to correct for differences between the data and the simulation. Both scale factors and efficiencies are evaluated using a \Jmumu tag-and-probe method \cite{MUON-2018-03}.
The dimuon trigger efficiency in \pp collisions is factorised as the product of two L1-seeded muon trigger efficiencies. For the L1-seeded single-muon trigger efficiency, an additional centrality-dependent correction is determined as the ratio of the single-muon trigger efficiency measured in the \PbPb data as a function of centrality to the single-muon trigger efficiency measured in \pp data. The L1-seeded single-muon trigger efficiency varies from 60\% at $\pTmu=4$~\GeV to 70\% at higher \pTmu ($\pTmu=10$~\GeV) in the barrel region, and it varies from 50\% at $\pTmu=4$~\GeV to 90\% at higher \pTmu ($\pTmu=10$~\GeV) in the endcaps. The inefficiency originates primarily from the L1 trigger, where the RoI formation requires hit coincidences across multiple RPC and TGC stations. In 0--10\% central collisions, the centrality-dependent correction corresponds to 0.7 at $\pTmu=4$~\GeV, and it reaches unity at $\pTmu=10$~\GeV and $\pTmu=15$~\GeV in the barrel and endcap, respectively. In the 40--80\% centrality interval, the centrality-dependent correction is unity in both barrel and endcap regions. The centrality dependence arises from reduced tracking resolution in the trigger turn-on region, especially in central collisions, where the tracking resolution worsens due to an increase in random hit density from other tracks. The HLT-only single-muon trigger is only used in \PbPb collisions, and its efficiency is again determined using the tag and probe method. The HLT-only single-muon trigger efficiency plateau is found to be 90\% in both the barrel and endcap regions, regardless of energy threshold. The dimuon trigger efficiency in \PbPb collisions is factorised in terms of the L1-seeded and HLT-only muon trigger efficiencies, as previously used in Refs.~\cite{HION-2016-07, HION-2021-12, HION-2020-10}.

The total weight applied to raw dimuon distributions is then
\begin{equation}
\label{eq:w}
w = \frac{1}{\mathcal{A} \times \varepsilon_\mathrm{trig}(\mu_1,\mu_2) \times \varepsilon_\mathrm{reco}(\mu_1,\mu_2)}.
\end{equation}
The denominator of this weight is multiplied by jet-related corrections defined in the next section.

\subsection{Dimuon--jet association}
\label{sec:jet}

Jets entering this analysis are reconstructed using the \antikt\ algorithm~\cite{Cacciari:2008gp,Fastjet} with a radius parameter of $R=0.2$, which provides superior performance to $R=0.4$ jets in \PbPb collisions in terms of the jet energy resolution (JER) and jet reconstruction efficiency. The jet reconstruction procedure follows that used previously in \PbPb\ collisions~\cite{HION-2017-10,HION-2011-02}, and is summarised here. Calorimeter cells in all calorimeter layers, with their energy at the electromagnetic scale, are grouped into $\Delta\eta \times \Delta\phi = 0.1 \times \pi/32$ calorimetric towers. The towers are used as the input to the initial jet-finding. After the initial jet-finding, the contribution from the underlying event (UE) to the energy deposited in the towers is estimated on an event-by-event basis.
<<<<<<< HEAD
This is done by averaging over energies deposited in towers at a given $\eta$ position. During the averaging and subsequent subtraction, the $\phi$ dependence arising from the collective flow in Pb+Pb collisions is also taken into account.
=======
This is done by averaging over energies deposited in towers at a given $\eta$ position. During the averaging and subsequent subtraction, the $\phi$ dependence arising from the collective flow in \PbPb collisions is also taken into account.
>>>>>>> origin/master
Information from towers within $\Delta R = 0.4$ of jet candidates is excluded from the estimate of UE energy. The energies in towers inside jets are updated to subtract the estimated UE contribution, and the UE procedure is iterated using a better-defined set of jets to define the exclusion regions. The resulting jet kinematics are corrected using $\pt$- and $\eta$-dependent factors, determined from simulation, to account for the response of the calorimeter to jets %
\cite{JETM-2018-05}.
An additional correction for the absolute response in data is based on \textit{in situ} studies of \pp\ collisions where jets recoil against photons, $Z$ bosons, or jets in other regions of the calorimeter~\cite{JETM-2018-05}. This calibration is followed by a cross-calibration that relates the jet energy scale (JES) in high-luminosity 13~\TeV \pp\ collisions~\cite{ATLAS-CONF-2015-016}, used for \textit{in situ} studies, to that of jets reconstructed in the 5.02~\TeV data.     %
The same jet reconstruction procedure, without the azimuthal modulation of the UE, is also applied to \pp\ collisions. The UE subtraction in \pp\ collisions removes the pile-up contribution to the jet.
\enquote{Truth}-level jets are defined in the MC samples before detector simulation by applying the \antikt\ algorithm with $R=0.2$ to stable particles with a proper lifetime greater than 30~ps, but excluding muons and neutrinos, which do not leave significant energy deposits in the calorimeter. After the detector simulation, the truth jets are matched to the nearest reconstructed jet in $\Delta R <0.15$.

Jets are used to tag dimuon candidates before the $\jpsi$ signal is extracted using the fitting procedure described in Section~\ref{sec:fit}. For each dimuon candidate passing the selection criteria, a search is performed for a reconstructed jet with transverse momentum $\pT>\qty{20}{\GeV}$ matching the angular distance $\Delta R(\mathrm{dimuon},\mathrm{jet})<0.4$.
The $\Delta R$ requirement is chosen to associate $J/\psi$ candidates with nearby jets using a cone size comparable to a jet radius commonly used in LHC jet measurements \cite{STDM-2016-03,CMS-SMP-15-007}.

Presence of a matching jet then allows the dimuon candidates to be separated into three classes: \textit{non-isolated}, \textit{isolated}, and \textit{inclusive}.
Non-isolated dimuon candidates refer to those with a matched jet, while the inclusive class refers to all dimuon candidates. The yield in the isolated class is calculated as the difference between the inclusive and non-isolated yields as determined in the fit procedure described in Section~\ref{sec:fit}.

To correct the dimuon tagging for the finite jet momentum resolution and jet reconstruction efficiency, a two-dimensional bin-by-bin correction is derived using the non-prompt \Jpsi MC samples defined in Section~\ref{sec:data}. The correction factors are obtained, in bins of reconstructed dimuon \pt and reconstructed jet \pt, as the ratio of the yield of reconstructed dimuons associated with truth jets in a given truth-jet \pt bin to the yield associated with reconstructed jets in the corresponding reconstructed-jet \pt bin. The correction therefore accounts for migrations between jet-\pt bins, including migrations across the $\pt>20$~\GeV selection threshold, as well as for jet reconstruction inefficiency. The factors are derived separately for \pp\ collisions and for each \PbPb\ centrality interval, thus accounting for the different jet momentum resolutions in the respective data samples, and are applied as an additional weight $w_\mathrm{jet}$ to non-isolated dimuon candidates entering the fitting procedure described in Section~\ref{sec:fit}. Since no differential jet-\pt distribution is reported, a full unfolding of the jet-\pt spectrum is not performed. To reduce the dependence of the bin-by-bin correction on the simulated jet-\pt spectrum, the reconstructed jet-\pt spectra in simulation are reweighted to those observed in data, and the effect of this reweighting is included as a systematic uncertainty. The correction factors also account for muon energy deposits in the calorimeter. The procedure is validated by an MC closure test, with the closure consistent with unity.

\subsection{\jpsi signal extraction}
\label{sec:fit}

<<<<<<< HEAD
The yields of prompt and non-prompt \jpsi mesons are determined using unbinned maximum-likelihood fits to the weighted (as defined in Equation~\ref{eq:w}) dimuon $m_{\mu\mu}$ and $\tau_{\mu\mu}$ distributions, following the same procedure for both \pp and Pb+Pb data. The fits are performed separately for each \pTmumu and centrality interval and for the non-isolated and inclusive classes. The yield of the isolated $\jpsi$, $N_\mathrm{iso}$, is extracted from the yield of the non-isolated $\jpsi$, $N_\mathrm{non-iso}$, and inclusive yield, $N_\mathrm{incl}$, using the difference
=======
The yields of prompt and non-prompt \jpsi mesons are determined using unbinned maximum-likelihood fits to the weighted (as defined in Equation~\ref{eq:w}) dimuon $m_{\mu\mu}$ and $\tau_{\mu\mu}$ distributions, following the same procedure for both \pp and \PbPb data. The fits are performed separately for each \pTmumu and centrality interval and for the non-isolated and inclusive classes. The yield of the isolated $\jpsi$, $N_\mathrm{iso}$, is extracted from the yield of the non-isolated $\jpsi$, $N_\mathrm{non-iso}$, and inclusive yield, $N_\mathrm{incl}$, using the difference
>>>>>>> origin/master
\begin{equation}
N_\mathrm{iso} = N_\mathrm{incl} - N_\mathrm{non-iso}.
\end{equation}
The probability distribution function (PDF) for the fit, as previously validated in Refs.~\cite{BPHY-2012-02,HION-2016-07}, is defined as a normalised sum of five terms corresponding to one prompt signal source, one non-prompt signal source, one prompt background source and two non-prompt background sources. Each term is factorised into dimuon $m_{\mu\mu}$-dependent and $\tau_{\mu\mu}$-dependent functions. The compact form of the PDF can be written as
\begin{equation}
\text{PDF}(\twomass{\mu}{\mu},\taumumu) = \sum_{i=1}^{5}\mathcal{N}_{i}h_{i}(\twomass{\mu}{\mu})\cdot t_{i}(\taumumu) \ast g_{i}(\taumumu),
\end{equation}
where $\mathcal{N}_{i}$ is the normalisation factor, $h_{i}(\twomass{\mu}{\mu})$ is a function describing the dimuon \twomass{\mu}{\mu} signal or background distribution, $t_{i}(\taumumu)$ is a function describing the dimuon $\taumumu$ signal or background distribution, and $g_{i}(\taumumu)$ is a double Gaussian function describing the detector resolution. The symbol \enquote{$\ast$} denotes the convolution. The distribution functions $h_i$ and $t_i$ are defined using the following distribution functions: the Crystal Ball (CB) function \cite{Oreglia:1980cs}, the Gaussian (G) distribution function, and the exponential (E) distribution function. All individual components are fully defined in Table~\ref{tab:pdf}. The fit is performed using the RooFit framework \cite{RooFit}. To stabilise the fit model and reduce parameter correlation, a number of component terms listed in Table~\ref{tab:pdf} share common parameters. Some terms are scaled to each other by a multiplicative parameter, while others are fixed to the values observed in the MC simulation.

The signal mass shape is modelled as the sum of the Crystal-Ball distribution and a single Gaussian distribution for the prompt ($i=1$ in Table~\ref{tab:pdf}) and non-prompt ($i=2$) signal. These CB and Gaussian distributions share a common mean. The $\alpha$ parameter represents the fractional contribution of the CB and Gaussian signal distributions. The width term in the CB distribution corresponds to the width of the Gaussian distribution multiplied by a scaling factor. The tail parameters of the CB distribution, as well as the width-scaling factor, are fixed based on the MC simulations. Variations of these parameters are considered as part of the fit model systematic uncertainties. The background mass shape is divided into one prompt ($i=3$) and two non-prompt parts ($i=4$, $i=5$), modelled by three exponentials.
No separate $\psi(2S)$ signal component is included in the fit, since the fitted mass range ends well below the $\psi(2S)$ mass. Any residual contribution from its low-mass tail is absorbed by the smooth background parameterisation.

The $\taumumu$ distribution is composed of the prompt and non-prompt components. The prompt component is modelled as a delta function ($i=1$ in Table~\ref{tab:pdf}) convolved with the double-Gaussian resolution function. The resolution function has a fixed mean at $\tau = 0$. The width parameters are free but are related by the fixed scaling factor between two Gaussian distributions. The variation of this parameter is considered part of the fit model systematic uncertainty. The non-prompt pseudo-proper decay time is described as a single-sided exponential distribution ($i=2$) convolved with a double-Gaussian resolution function. The background $\taumumu$ distribution is also composed of the prompt and non-prompt components. The prompt component is described by a delta function convolved with the double-Gaussian resolution function ($i=3$). The non-prompt component is modelled by a single-sided exponential distribution for positive values of $\tau$ ($i=4$), and by a double-sided exponential ($i=5$) that accounts for mis-reconstructed dimuons or Drell--Yan muons and combinatorial background. The single-sided and double-sided exponential distributions are convolved with the same double-Gaussian resolution distributions that are used for the signal part.

\begin{table}[htbp]
\centering
\caption{PDFs of the individual components corresponding to the prompt (P) and non-prompt (NP) sources of the fit model used in this analysis. The $\alpha$ represents the fractional contribution of the Crystal-Ball (CB) and Gaussian (G) signal distributions, the $\delta$(\taumumu) is the Dirac delta function, and the E denotes the exponential function.}
\begin{tabular}{c|c|c|c|c}
\toprule
$i$ & Type & Source & $h_{i}(\twomass{\mu}{\mu})$ & $t_{i}(\taumumu)$ \\
\midrule
1 & Signal & P & $\alpha\text{CB}_{1}(\twomass{\mu}{\mu})+(1-\alpha)\text{G}_{1}(\twomass{\mu}{\mu})$ & $\delta$(\taumumu) \\
2 & Signal & NP & $\alpha\text{CB}_{1}(\twomass{\mu}{\mu})+(1-\alpha)\text{G}_{1}(\twomass{\mu}{\mu})$ & $\text{E}_{1}$(\taumumu) \\
\midrule
3 & Background    & P & $\text{E}_{2}$(\twomass{\mu}{\mu}) & $\delta$(\taumumu) \\
4 & Background    & NP & $\text{E}_{3}$(\twomass{\mu}{\mu}) & $\text{E}_{4}$(\taumumu) \\
5 & Background    & NP & $\text{E}_{5}$(\twomass{\mu}{\mu}) & $\text{E}_{6}$(|\taumumu|) \\
\bottomrule
\end{tabular}
\label{tab:pdf}
\end{table}

\begin{figure}[ht!]
\centering
\subfloat[]{\includegraphics[width=0.49\textwidth]{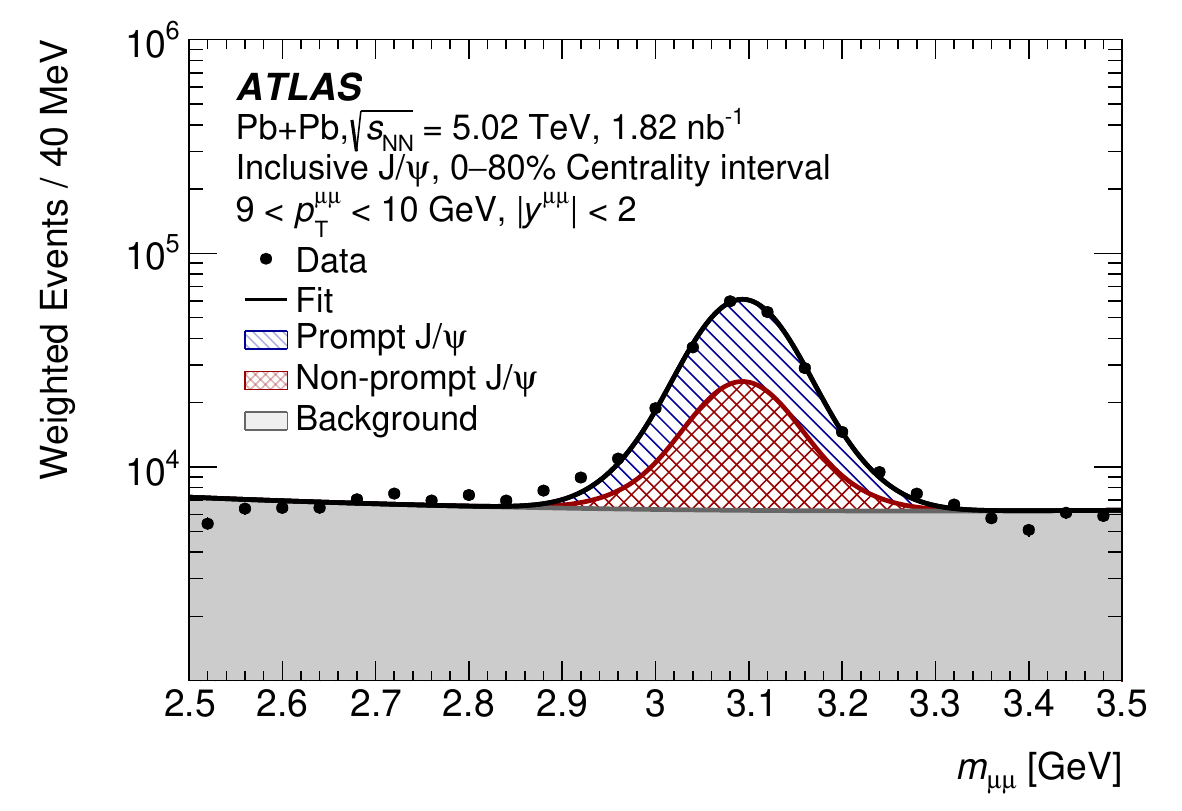}}
\subfloat[]{\includegraphics[width=0.49\textwidth]{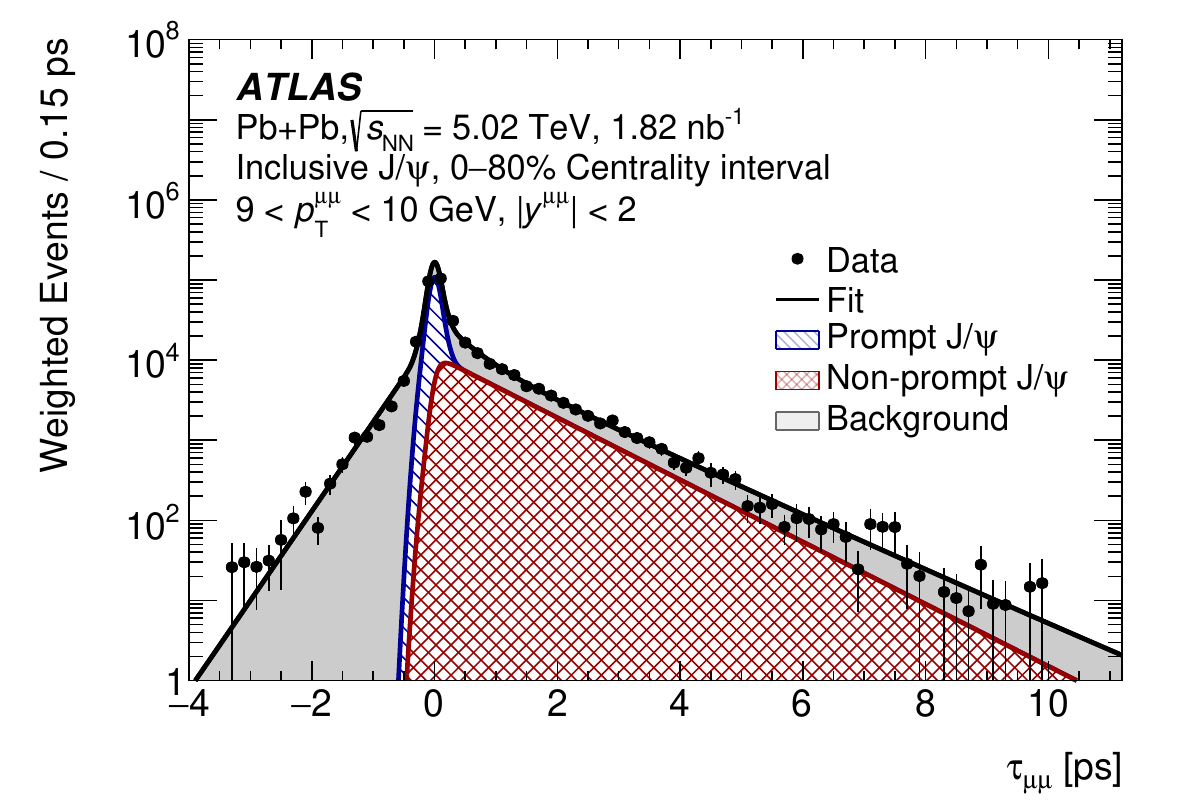}}
\vspace{0.7em}
\subfloat[]{\includegraphics[width=0.49\textwidth]{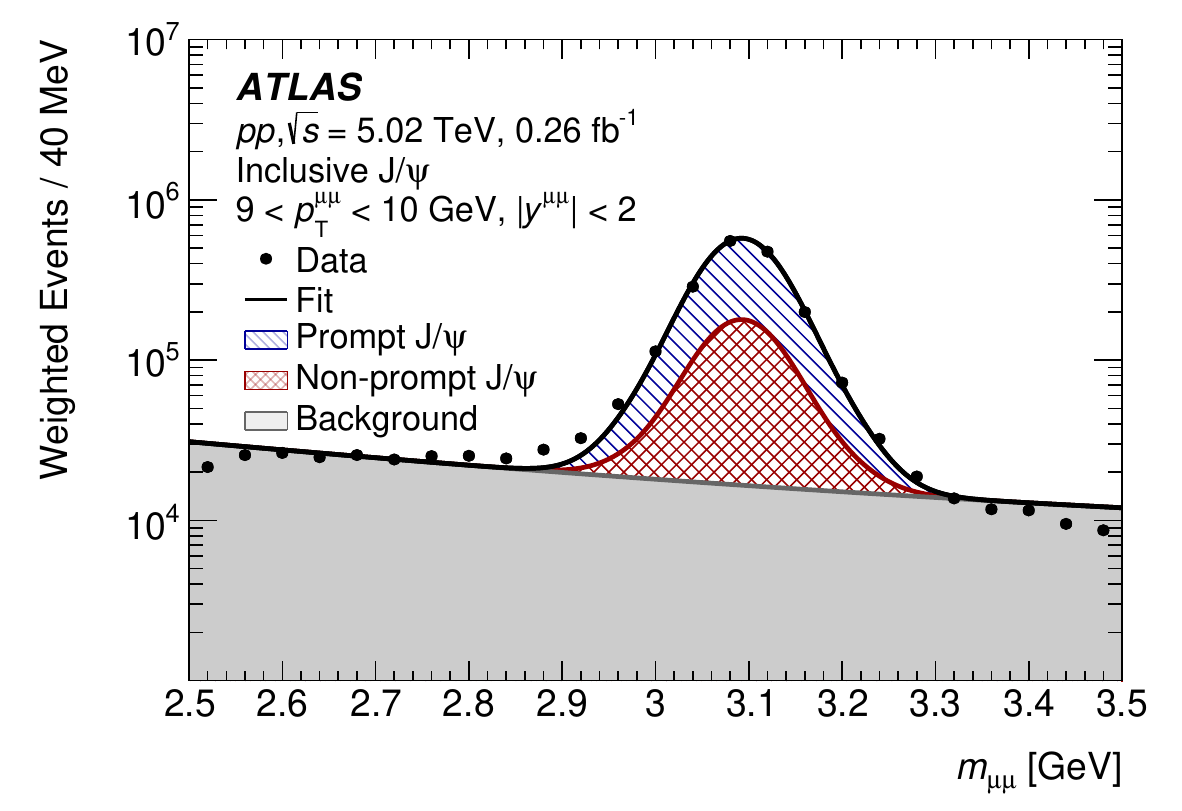}}
\subfloat[]{\includegraphics[width=0.49\textwidth]{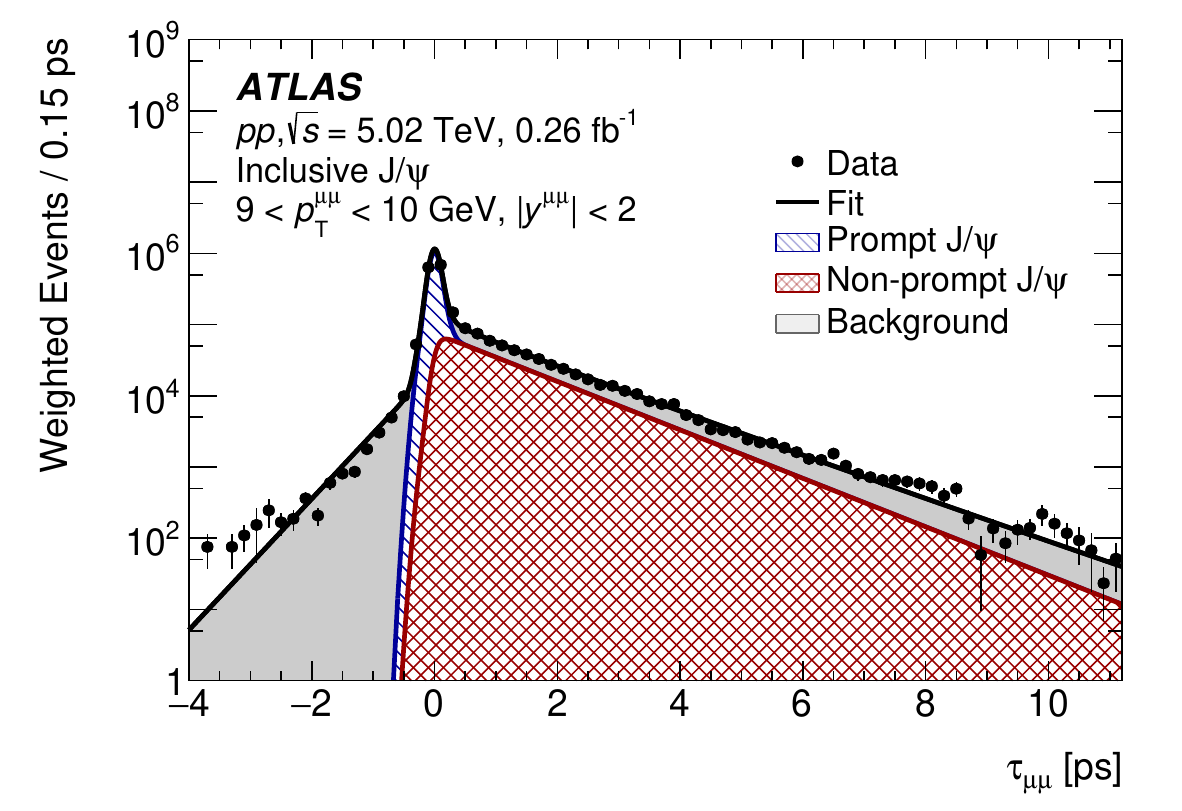}}

\caption{Binned representation of (a,c) the weighted dimuon invariant-mass \twomass{\mu}{\mu} distribution and (b,d) the dimuon pseudo-proper decay-time \taumumu distribution in the fitted range for candidates in the transverse-momentum interval 9~\GeV~$<\pTmumu<$~10~\GeV and $\absrapmumu<2$ in (a,b) 0--80\% Pb$+$Pb collisions and in (c,d) $pp$ collisions. The fit results are shown as solid lines, while the coloured filled areas indicate the extracted signal and background components. The error bars correspond to the statistical uncertainties of the data points. }
\label{fig:FitPbPb}
\end{figure}

An example of the fit for the 0--80\% centrality interval of \PbPb collisions, projected onto the binned data distributions, is shown in panels (a,b) in Figure~\ref{fig:FitPbPb}, a corresponding example for \pp data is shown in panels (c,d) in Figure~\ref{fig:FitPbPb}.

The UE in \PbPb collisions, pile-up in \pp collisions, and multi-parton interactions (MPI) in both \PbPb and \pp collisions may lead to the presence of jets produced in $\Delta R<0.4$ from the $\jpsi$, but not originating from the same hard process as the $\jpsi$. The yield of these background jets is estimated using the orthogonal-cone method~\cite{HION-2012-08}. The background jet yield is estimated as a yield of jets in the event with $\jpsi$ in a cone of radius $R=0.4$ centred at coordinates $(\phi_{\jpsi},-\eta_{\jpsi})$ in the azimuth--pseudorapidity space. In the region near $\eta=0$ where the background cone would overlap with the signal cone, the $\eta$ coordinate is shifted by a constant value and the yield is re-weighted by a factor compensating for the difference in the average yields taken from simulation. %
The estimated background-jet yield is expressed as a background-jet rate, $\rho$, quantifying the probability for a genuinely isolated \Jpsi to be associated with a background jet and therefore classified as non-isolated. The observed and corrected yields are then related by
\begin{equation}
o_i = r_i(1-\rho), \qquad o_n = r_n + r_i\rho ,
\end{equation}
where $o_i$ and $o_n$ denote the observed yields in the isolated and non-isolated categories, respectively, and $r_i$ and $r_n$ denote the corresponding corrected yields.
The yield of background jets is estimated to be less than 1\% of the isolated $\jpsi$ yield in \pp collisions and less than 10\% in 0--10\% central \PbPb collisions.
The procedure is tested on jet-tagged dimuon candidates in MC, resulting in full recovery of the jet-tagged signal distributions. The difference in the background yield magnitude between data and MC is less than 2\%.

\begin{figure}[ht!]
\centering
\includegraphics[width=0.49\linewidth]{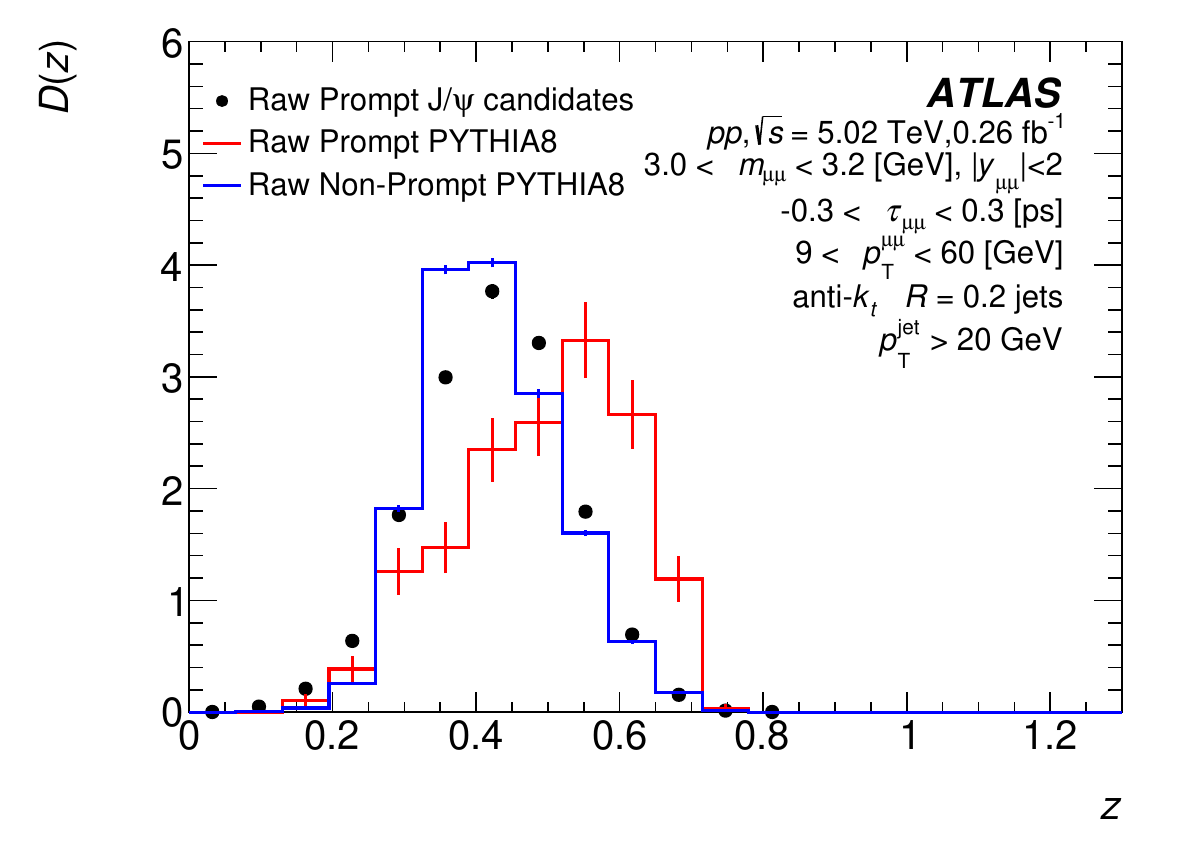}
\caption{Raw fragmentation function evaluated for non-isolated prompt $J/\psi$ candidates in $pp$ collisions. The $J/\psi$ candidates were selected requiring $3.0<\twomass{\mu}{\mu}<3.2 $~\GeV, $-0.3<\taumumu<0.3$~ps and $9<\pTmumu<60$~\GeV. Jets with $p_\mathrm{T}^\mathrm{jet}>20$~\GeV are used. Results are compared with the LO QCD \Pythia prediction of raw distributions for prompt and non-prompt dimuon candidates with the same kinematic requirements as those applied to the data. The vertical lines represent statistical uncertainties. The statistical uncertainties of the black markers are negligible and thus not visible.}
\label{fig:fragmentfunction}
\end{figure}

To facilitate comparison with previous publications \cite{CMS-HIN-19-007, LHCb:2017llq, CMS-BPH-15-003} and provide a qualitative test of the jet association procedure, the raw fragmentation function of \Jpsi, i.e. the fragmentation function not corrected for the detector response, efficiency, and acceptance, is evaluated. In this case, the fraction of jet transverse momentum carried by a $\jpsi$ candidate is estimated using $z$ defined as
\begin{equation}
\label{eq:z}
z = \frac{\pTmumu}{p_\mathrm{T}^\mathrm{jet} + \pTmumu},
\end{equation}
where $p_\mathrm{T}^\mathrm{jet}$ represents the momentum of the jet reconstructed without the signal from the $\jpsi$ meson, and the denominator of Equation~\ref{eq:z} thus follows the definition used by the previous measurements \cite{CMS-HIN-19-007, LHCb:2017llq, CMS-BPH-15-003}.
The raw fragmentation function is then defined as
\begin{equation}
D(z) = \frac{1}{N_\mathrm{jet}}\frac{\text{d} N(z)}{\text{d} z},
\end{equation}
where $N(z)$ is the yield of $\jpsi$ mesons carrying a momentum fraction $z$ normalised by the number of jets ($N_\mathrm{jet}$) containing a $\jpsi$. The raw fragmentation functions are evaluated for non-isolated prompt $\jpsi$ candidates selected based on conditions $3.0<m_{\mu\mu} < 3.2$~\GeV, $-0.3 < \taumumu < 0.3$ ps and $9<\pTmumu<60$~\GeV. These distributions are not unfolded for the detector response and are not corrected for dimuon background, non-prompt contribution, and jet background. The resulting raw fragmentation functions evaluated in \pp\ collisions are shown in Figure~\ref{fig:fragmentfunction}.
The raw fragmentation function measured in \pp\ collisions is directly compared with raw fragmentation functions evaluated from leading-order (LO) QCD \Pythia~8 samples as defined in Section~\ref{sec:data} using the same kinematic selections. The measured raw fragmentation functions for prompt dimuon candidates are qualitatively consistent with the fragmentation function simulated in \Pythia~8 for non-prompt $\jpsi$ production and differ significantly from those simulated for prompt $\jpsi$ production. This confirms the earlier findings of the CMS Collaboration reported in Ref.~\cite{CMS-HIN-19-007}.

\subsection{Definition of observables}
\label{sec:obs}

The measurement presented in this paper quantifies \Jpsi cross section and yields in \pp and \PbPb collisions, respectively. These cross sections and yields are used to evaluate the nuclear modification factor of the \jpsi meson. These observables are evaluated for prompt and non-prompt \jpsi, which also allows the determination of the fraction of non-prompt \jpsi mesons. These observables are also evaluated for non-isolated, isolated, and inclusive \jpsi mesons, which allows the determination of the fraction of isolated \jpsi mesons. The definition of all observables is given in this section.

\newcommand{\XY}{ }
\newcommand{\rapmumu}{y_{\mu\mu}}

The differential cross section for \Jmumu production in \pp collisions is defined as
\begin{equation}
\mathcal{B} \cdot
\frac{\mathrm{d^{2}}\sigma_{\Jpsi}^{\XY}}{\mathrm{d}\pTmumu\mathrm{d}\rapmumu} = \frac{N_{\XY\Jpsi}^{\mathrm{corr}}}{\mathrm{\Delta}\pTmumu \times \mathrm{\Delta}\rapmumu \times L},
\end{equation}
where $\mathcal{B}(\Jmumu)=(5.93 \pm 0.06) \%$ \cite{PDG2024} is the \Jpsi dimuon decay branching fraction, $N_{\XY\Jpsi}^{\mathrm{corr}}$ is the corrected \Jpsi yield estimated by using the procedure discussed in Section~\ref{sec:fit}, $\pTmumu$ and $\rapmumu$ are the dimuon transverse momentum and rapidity, respectively, and $L$ is the integrated luminosity of the \pp data sample.

The yield characterising the \Jmumu production in \PbPb collisions is defined as
\begin{equation}
\mathcal{B} \cdot \left. \frac{\mathrm{d}^{2}N_{\Jpsi}^{\XY}}{\mathrm{d}\pTmumu\mathrm{d}\rapmumu} \right|_{\mathrm{cent}} = \left.\frac{N_{\XY\Jpsi}^{\mathrm{corr}}}{\mathrm{\Delta}\pTmumu \times \mathrm{\Delta}\rapmumu } \right|_{\mathrm{cent}},
\end{equation}
where \enquote{cent} refers to a specific centrality interval.

In nucleus--nucleus collisions, the suppression of \jpsi production is quantified using the nuclear modification factor, \RAA, which is defined for each centrality interval as
\begin{equation}
\left. \RAA^{\XY} \right|_{\mathrm{cent}} = \frac{\left. \frac{1}{N_{\mathrm{evt}}}\frac{\mathrm{d}^{2}N_{\Jpsi}^{\XY}}{\mathrm{d}\pTmumu\mathrm{d}\rapmumu} \right|_{\mathrm{cent}}}{\avgTAA|_{\mathrm{cent}} \times \frac{\mathrm{d^{2}}\sigma_{\Jpsi}^{\XY}}{\mathrm{d}\pTmumu\mathrm{d}\rapmumu}},
\end{equation}
where $N_{\mathrm{evt}}$ is the total number of minimum-bias events for a given centrality interval and the values of the mean nuclear overlap function \avgTAA are provided in Section~\ref{sec:data}.

The non-prompt fraction, $f^{\mathrm{\Jpsi}}_{\mathrm{NP}}$, is defined as the ratio of the corrected yield of non-prompt (NP) $\jpsi$ mesons to the sum of the corrected prompt (P) and non-prompt $\jpsi$ yields,
\begin{equation}
f^{\mathrm{\Jpsi}}_{\mathrm{NP}} = \frac{N_{\Jpsi}^{\mathrm{NP,corr}}}{N_{\Jpsi}^{\mathrm{NP,corr}} + N_{\Jpsi}^{\mathrm{P,corr}}}.
\end{equation}

The isolated fraction, $f^{\Jpsi}_\mathrm{iso}$, is defined as the ratio of the corrected yield of isolated (iso) $\jpsi$ mesons the corrected inclusive $\jpsi$ yields (inc),

\begin{equation}
f^{\mathrm{\Jpsi}}_{\mathrm{iso}} = \frac{N_{\Jpsi}^{\mathrm{iso,corr}}}{N_{\Jpsi}^{\mathrm{inc,corr}}}.
\end{equation}
Binomial error propagation is used to obtain the statistical uncertainties of $f^{\mathrm{\Jpsi}}_{\mathrm{NP}}$ and $f^{\Jpsi}_\mathrm{iso}$.
%

% End of text imported from the .//sections/04_Analysis.tex input file

% The next lines are included from the .//sections/05_Systematics.tex input file
\section{Systematic uncertainties}
\label{sec:sys}

The following sources of systematic uncertainties are considered in this measurement:
\Jpsi kinematic acceptance uncertainty, muon reconstruction efficiency uncertainty, muon trigger efficiency uncertainty, fit model uncertainty, jet-related uncertainties, UE/MPI subtraction uncertainty, \pp luminosity uncertainty, and \avgTAA uncertainty. Systematic uncertainties in the measured distributions can be categorised into two classes: bin-wise correlated uncertainties and uncertainties that affect the overall normalisation of distributions. Uncertainties due to the determination of \avgTAA and \pp luminosity belong to the second class, all other uncertainties belong to the first. The systematic uncertainties are propagated, taking the correlations among sources into account, leading to an appropriate reduction in systematic uncertainties in ratios.

\subsection{Acceptance uncertainty}
The uncertainty in acceptance has two components. The first component evaluates the impact of spin alignment on the acceptance. To evaluate this component, six alternative scenarios are evaluated corresponding to extreme cases of spin-alignment as described in Ref.~\cite{BPHY-2012-02}. The uncertainty is calculated as the maximum deviation from the assumption of unpolarised production. Since the polarisation of charmonia in \pp collisions is measured to be small \cite{CMS-BPH-13-003,ALICE:2011gej,LHCb:2013izl}, its modification due to the nuclear environment is neglected, and the spin-alignment uncertainty is assumed to cancel out in ratios involving \pp and \PbPb yields. This uncertainty is presented on the cross sections and yields separately from other uncertainties. The second source is the acceptance interpolation in Equation~\ref{eq:accinterpolation}, which may introduce an artificial slope to the background when the signal-to-background ratio is low. To address this potential bias, an alternative set of weights is derived without the interpolation. The acceptance uncertainties cancel fully in all ratios; they affect only the per-event yields in \PbPb and the cross sections in \pp collisions. The dominant systematic uncertainty corresponds to the unknown \Jpsi polarisation.

\subsection{Muon reconstruction efficiency uncertainty}
The systematic uncertainty associated with the muon reconstruction efficiency in \pp collisions accounts for possible biases in the tag-and-probe method used to evaluate it. The uncertainty is below 1\%, and further details of its estimation are given in Ref.~\cite{MUON-2018-03}.

Several sources of uncertainty are considered in evaluating the total muon reconstruction efficiency uncertainty in \PbPb collisions. This uncertainty is evaluated following Ref.~\cite{MUON-2018-03} by changing the tag-muon selection criteria and modifying the line shapes in the extraction fit procedure. The statistical uncertainties of the scale factors are also included as a source of systematic uncertainty. Due to the small probe yield in the \PbPb data, the statistical uncertainties of the scale factors dominate the uncertainty in the muon reconstruction efficiency correction.
The final systematic uncertainty is evaluated as a quadrature sum of individual sources and propagated through the data-to-MC scale factors.

\subsection{Muon trigger efficiency uncertainty}

The systematic uncertainty in the trigger efficiency correction is estimated in a similar manner to the uncertainty in the muon reconstruction efficiency. The fit model and tag-muon selection criteria are modified in the tag-and-probe method. For the muon trigger efficiency in \PbPb collisions, an additional uncertainty associated with the centrality-dependent correction is added to the total systematic uncertainty. Statistical uncertainties in the muon trigger efficiency scale factors are also propagated to the final result. All sources of systematic uncertainty in the muon trigger are combined in quadrature and propagated through the data-to-MC scale factors.

\subsection{Fit model uncertainty}
The uncertainty associated with the fit model is obtained by relaxing its parameter constraints and introducing alternative PDFs. In each modification of the fit model, all measured quantities are recalculated and compared with the nominal fit. In total, ten variations of the fit model are considered. The weighted sum of the CB distribution and the Gaussian distribution is changed to the weighted sum of two Gaussian distributions, and the parameters of the CB distribution, which were originally fixed, are left free. The scaling factor relating the width of the Gaussian distribution to the width parameter of the CB distribution is varied. For the background $m_{\mu\mu}$ PDF, the single exponential distribution is replaced by the first-order Chebyshev polynomial. For the time-resolution PDF, the double Gaussian distribution is replaced with a single Gaussian distribution, and the width parameter is varied. The dominant source of uncertainty is the variation in the signal mass shape. The root-mean-square (RMS) of all variations is then assigned as the systematic uncertainty associated with the fit model.

\subsection{Jet correction uncertainty}
Three sources of jet-related uncertainty are considered: JES, JER, and the shape of the jet \pt spectrum.
The JES uncertainty has three components. The first component is a centrality-independent baseline determined from calorimetric studies~\cite{JETM-2018-05} and from relative energy-scale differences between heavy-ion and 13~\TeV \pp\ jet reconstruction~\cite{ATLAS-CONF-2015-016}.
The second component accounts for the dependence of the calorimeter response on the parton-shower composition~\cite{HION-2017-10}. Since the flavour composition of the jet population selected in this analysis is not known a priori, the conservative \textit{unknown flavour composition} prescription is used~\cite{PERF-2014-02}.
The third, centrality-dependent component accounts for differences in the calorimetric response to jets with parton showers modified by quenching in \PbPb\ collisions~\cite{ATLAS-CONF-2015-016,HION-2017-10}.

The JER  uncertainty is derived from dijet energy-balance studies in 13~\TeV \pp\ data~\cite{JETM-2018-05} and includes additional contributions to account for the different heavy-ion jet reconstruction procedure.
The JES and JER uncertainties are propagated through the analysis by re-evaluating the bin-by-bin correction factors after the corresponding variations of the reconstructed jet momentum.
Since the bin-by-bin correction is also sensitive to the shape of the reconstructed jet \pt spectrum, the spectrum in MC is reweighted to match that measured in data, as described in Section~\ref{sec:jet}. The correction is additionally evaluated without this reweighting. The resulting variation provides a conservative estimate of the sensitivity to residual differences between the reconstructed jet \pt spectra in data and simulation, including effects not fully captured by the individual JES and JER uncertainty components. The dominant jet-related uncertainty is JER.

To account for the finite size of the MC samples used to evaluate the bin-by-bin correction, the uncertainty in this correction was propagated to the final result using the bootstrap method \cite{ATL-PHYS-PUB-2021-011}. To preserve the correlation between the numerator and the denominator of the correction, a random Poisson-distributed weight with a mean of unity is assigned to each event. A total of 500 varied correction maps are generated, and the full analysis procedure is repeated using various maps. The RMS of the results obtained with varied maps is assigned as the systematic uncertainty associated with the finite size of the MC sample.

The quadrature sum of the individual source uncertainties is taken as the final jet correction uncertainty.

\subsection{Background subtraction uncertainty}
The uncertainty in the jet background subtraction, described in Section~\ref{sec:fit}, is evaluated by varying the background-jet rate within its overall statistical uncertainty. For \pp collisions, this uncertainty is also studied and found to be negligible.

\subsection{\pp luminosity and \avgTAA uncertainties}
The uncertainty in \avgTAA arises from geometric modelling uncertainties (e.g. nucleon–nucleon inelastic cross section, Woods–Saxon parameterisation of the nucleon distribution) and the uncertainty in the fraction of selected inelastic \PbPb collisions.
The values of these uncertainties are presented in Table~\ref{tab:centrality}. The integrated luminosity of the \pp data is calibrated using data from dedicated beam separation scans and is determined using the procedures described in Ref.~\cite{DAPR-2021-01}. The relative systematic uncertainty is 1.0\%.

\begin{figure}[!ht]
\centerline{
\subfloat[]{\includegraphics[width=0.49\textwidth]{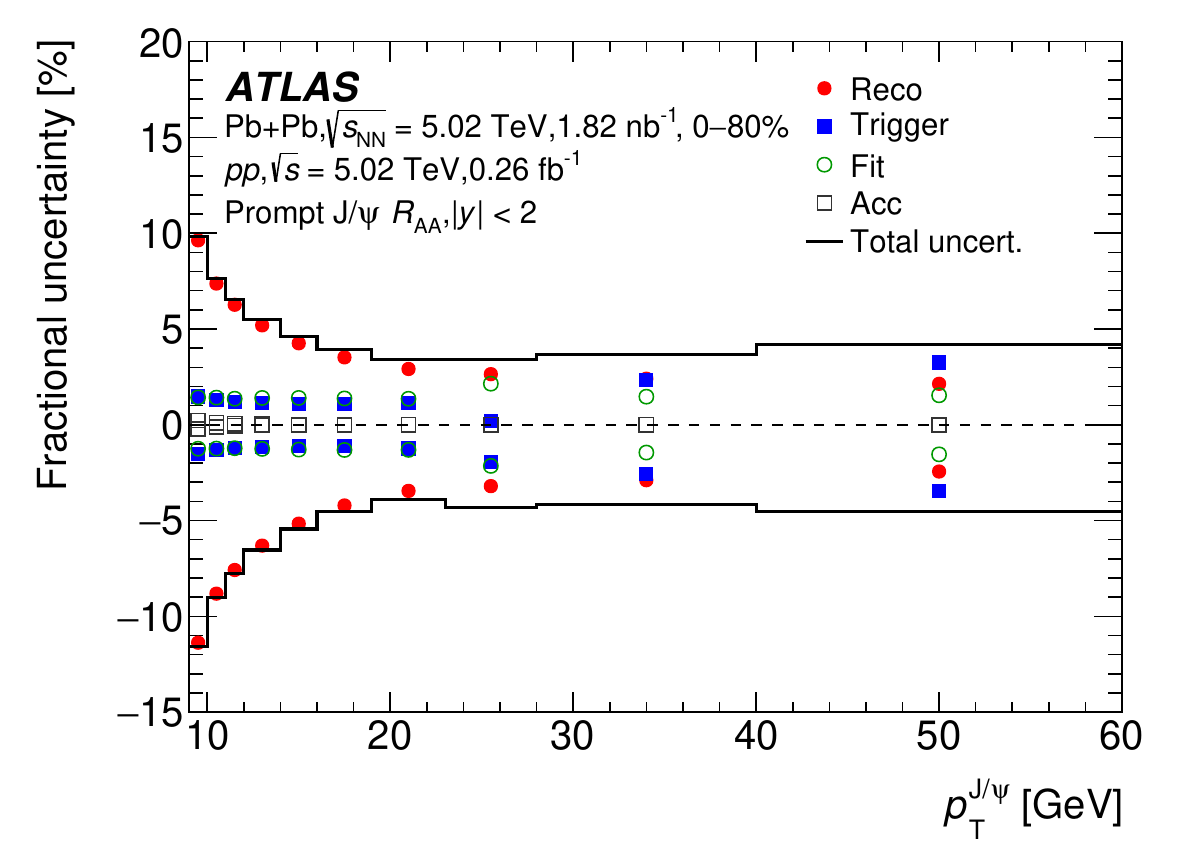}}
\subfloat[]{\includegraphics[width=0.49\textwidth]{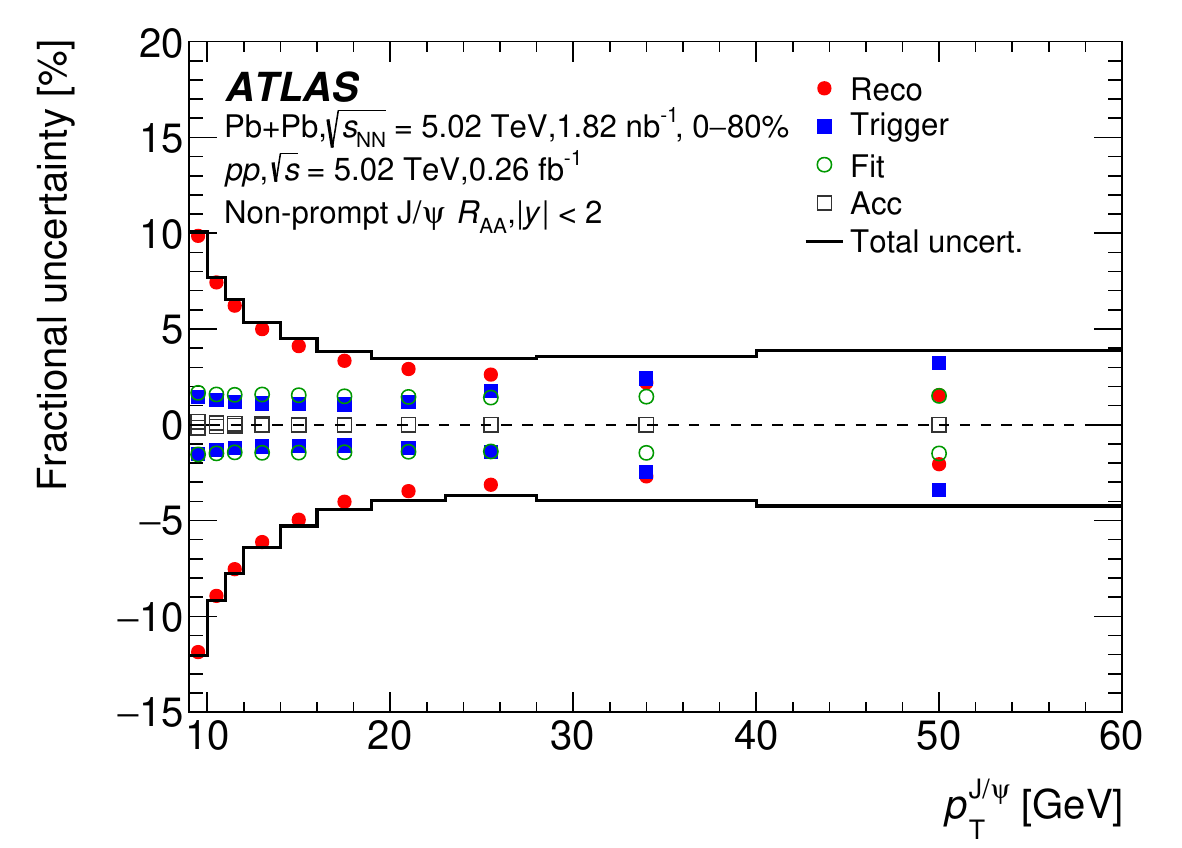}}
}
\caption{Summary of the sizes of different sources of bin-wise systematic uncertainty for (a) the inclusive prompt $J/\psi$ $R_{\mathrm{AA}}$  and for (b) the inclusive non-prompt $J/\psi$ $R_{\mathrm{AA}}$ evaluated in the 0--80\% centrality interval. The solid line denotes total uncertainty.}
\label{fig:IncluRaaSyst}
\end{figure}

\begin{figure}[!ht]
\centerline{
\subfloat[]{\includegraphics[width=0.49\textwidth]{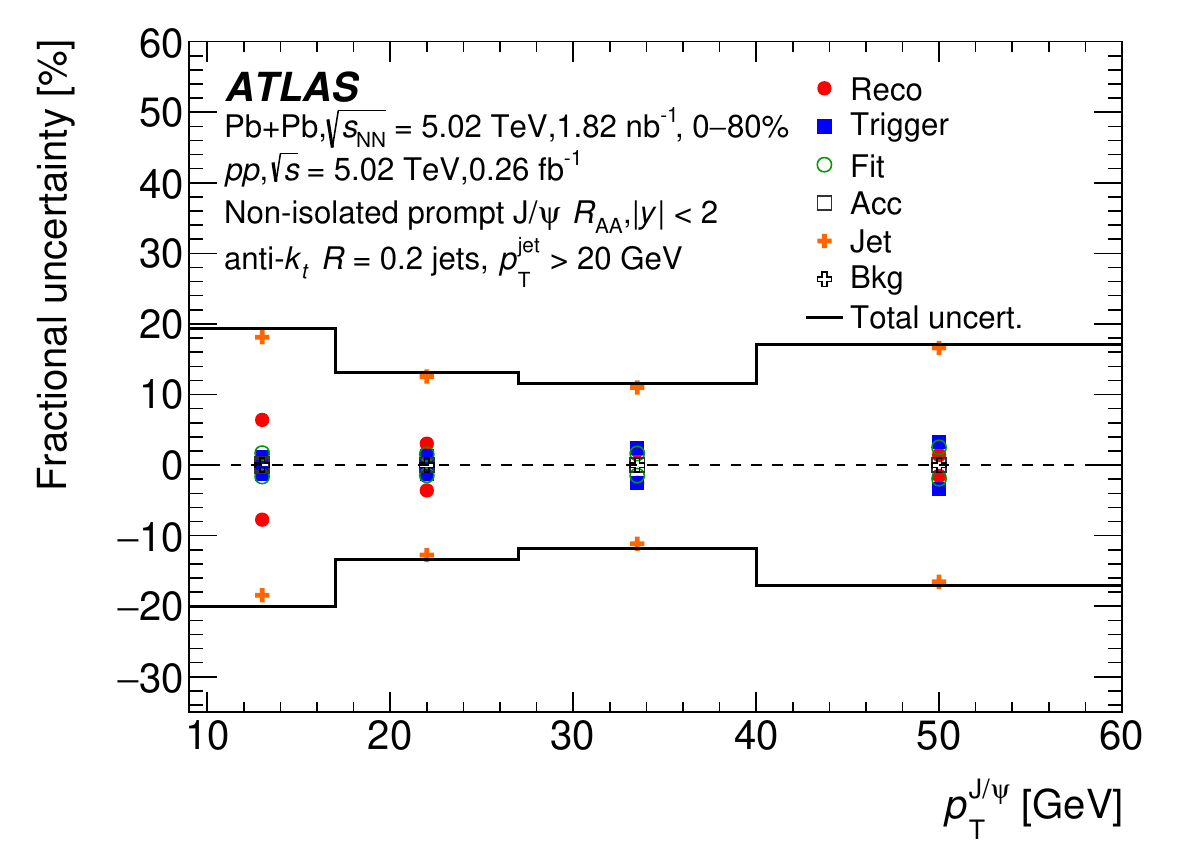}}
\subfloat[]{\includegraphics[width=0.49\textwidth]{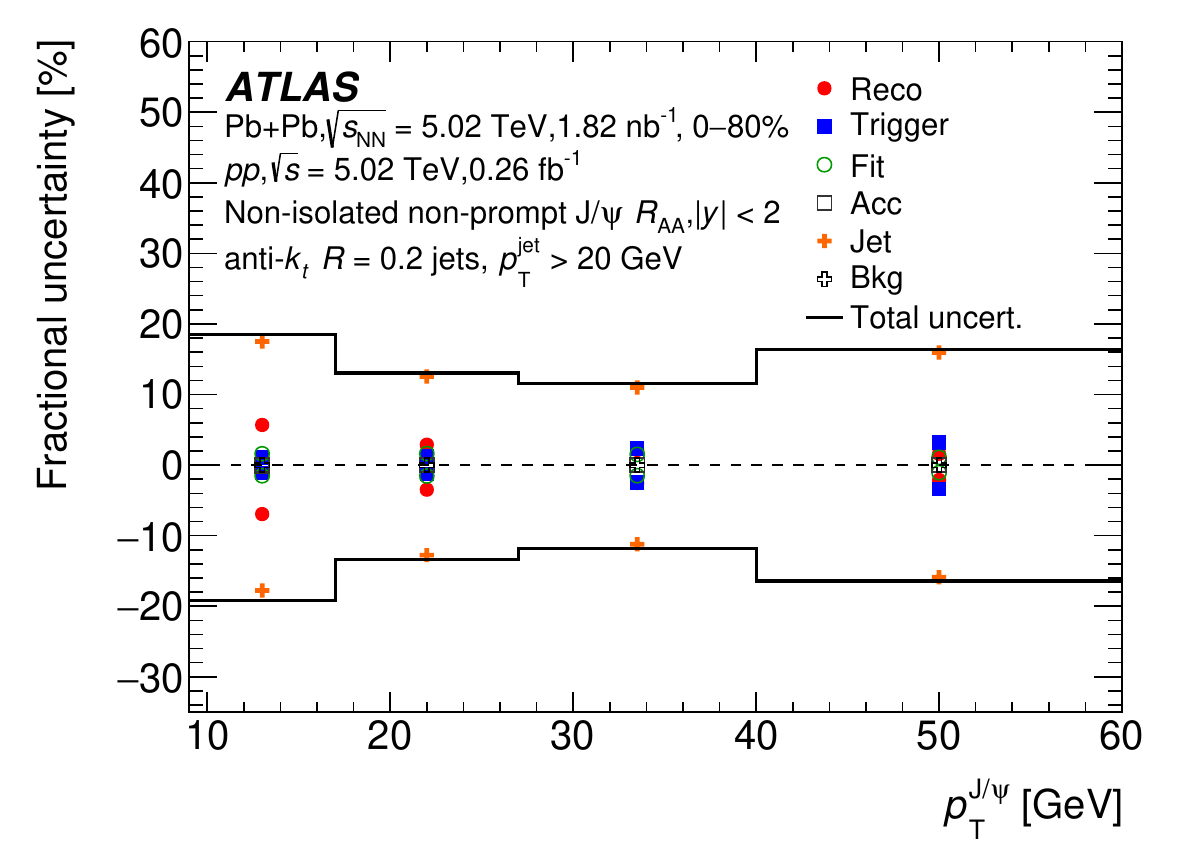}}
}
\caption{Summary of the sizes of different sources of bin-wise systematic uncertainty for (a) the non-isolated prompt $J/\psi$ $R_{\mathrm{AA}}$ and for (b) the non-isolated non-prompt $J/\psi$ $R_{\mathrm{AA}}$ evaluated in the 0--80\% centrality interval. The solid line denotes total uncertainty.}
\label{fig:NonIsoRaaSyst}
\end{figure}

Table~\ref{tab:systematics} summarises the magnitude of bin-wise correlated systematic uncertainties in measured prompt \Jpsi yields in \pp and \PbPb collisions. The corresponding uncertainties for non-prompt \Jpsi production are of similar magnitude.
Figure~\ref{fig:IncluRaaSyst} summarises the magnitudes of bin-wise systematic uncertainties for the inclusive prompt and non-prompt \Jpsi \RAA, whereas Figure~\ref{fig:NonIsoRaaSyst} summarises the magnitudes of bin-wise systematic uncertainties for the non-isolated prompt and non-prompt \Jpsi \RAA.
Because some components of the systematic uncertainties partially cancel in the construction of \RAA, the uncertainties shown in Figures~\ref{fig:IncluRaaSyst} and \ref{fig:NonIsoRaaSyst} are smaller than the corresponding yield uncertainties summarised in Table~\ref{tab:systematics}.

\begin{table}[h!]
\centering
\caption{Summary of the sources of bin-wise correlated systematic uncertainty for the prompt $J/\psi$ yields.}
\begin{tabular}{l|l|c|c|c}
\toprule
Collision type & Source & Inclusive \Jpsi[\%]  & Isolated \Jpsi[\%] & Non-isolated \Jpsi[\%]  \\ \midrule
\multirow{6}{*}{\pp collisions} & Acceptance & $<0.1$ & $<0.1$ & $<0.1$  \\
& Trigger eff. & 1.1--3.4 & 1.1--3.4 & 1.0--3.4  \\
& Reconstruction eff. & 0.1--0.2 & 0.1--0.2 & 0.1--0.2  \\
& Signal extraction & 1.5--2.2 & 1.5--2.4 & 1.6--2.5  \\
& Jet correction & -- & 0.0--5.4 & 11--37  \\
& Background & -- & -- & --  \\
\midrule
\multirow{6}{*}{\PbPb collisions} & Acceptance & $<0.5$ & $<0.5$ & $<0.5$  \\
& Trigger eff. & 0.2--1.3 & 0.2--1.3 & 0.2--2.0  \\
& Reconstruction eff. & 2.1--12 & 2.5--12 & 1.2--12  \\
& Signal extraction & 1.2--2.8 & 1.1--2.9 & 0.3--3.4  \\
& Jet correction & -- & 0.1--6.5 & 11--50  \\
& Background & -- & 0.1--0.3 & 0.1--30  \\
\toprule
\end{tabular}
\label{tab:systematics}
\end{table}

% End of text imported from the .//sections/05_Systematics.tex input file

% The next lines are included from the .//sections/06_Results.tex input file
\section{Results}
\begin{figure}[htbp]
\centering

\subfloat[]{
\includegraphics[width=0.48\textwidth]{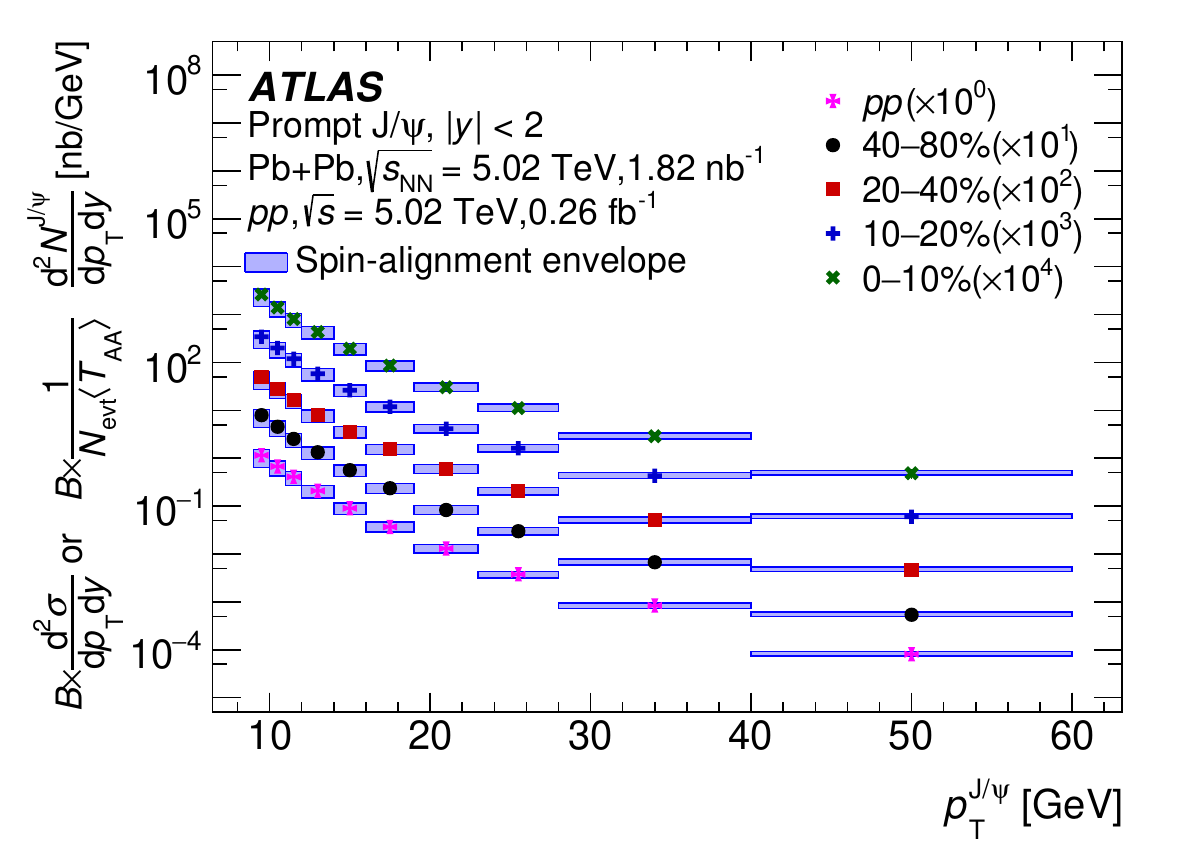}
}
\hfill
\subfloat[]{
\includegraphics[width=0.48\textwidth]{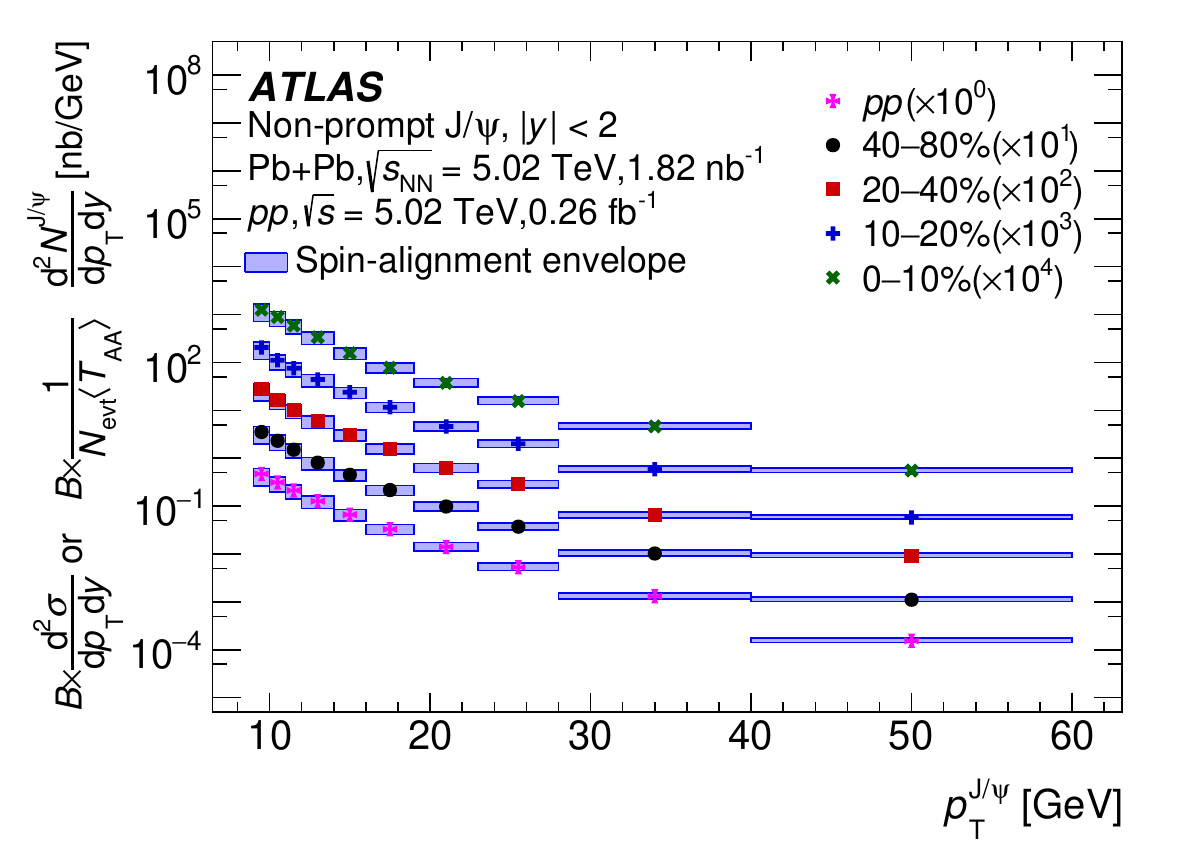}
}

\vspace{0.7em}

\subfloat[]{
\includegraphics[width=0.48\textwidth]{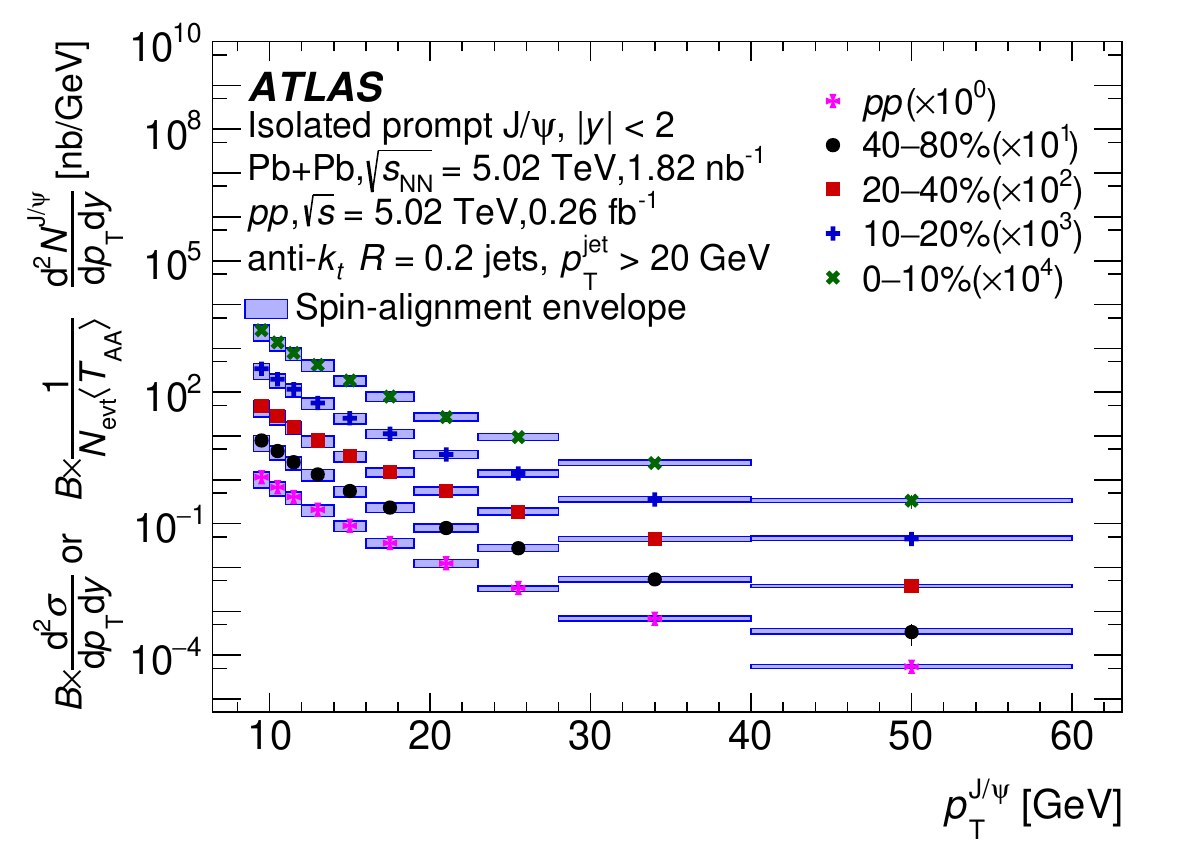}
}
\hfill
\subfloat[]{
\includegraphics[width=0.48\textwidth]{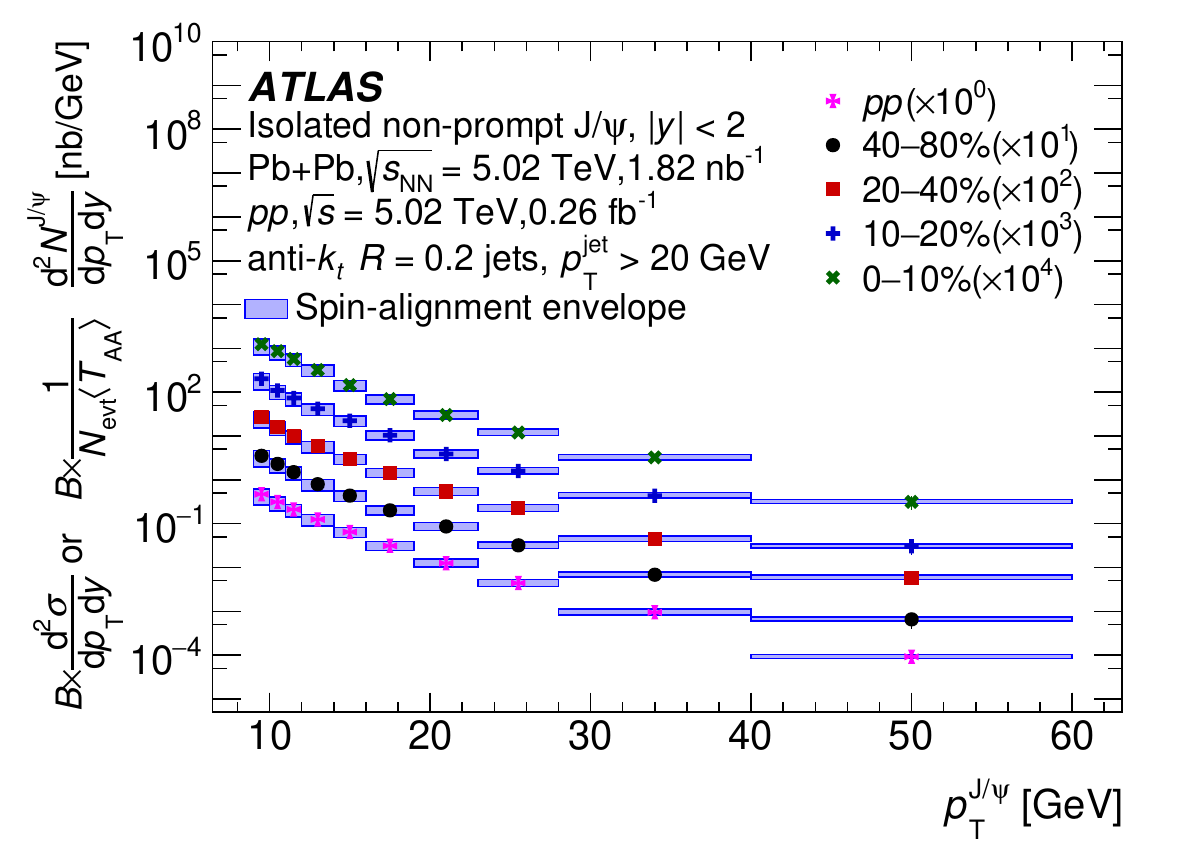}
}

\vspace{0.7em}

\subfloat[]{
\includegraphics[width=0.48\textwidth]{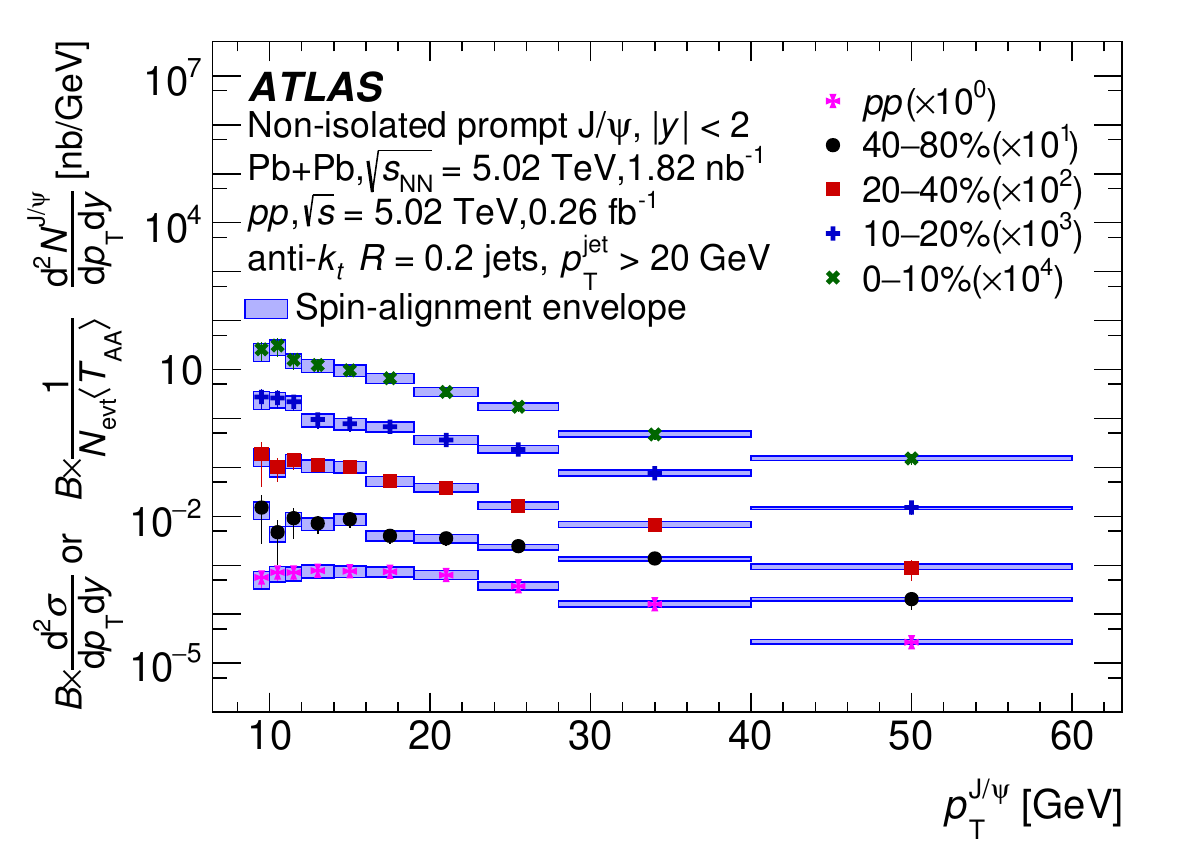}
}
\hfill
\subfloat[]{
\includegraphics[width=0.48\textwidth]{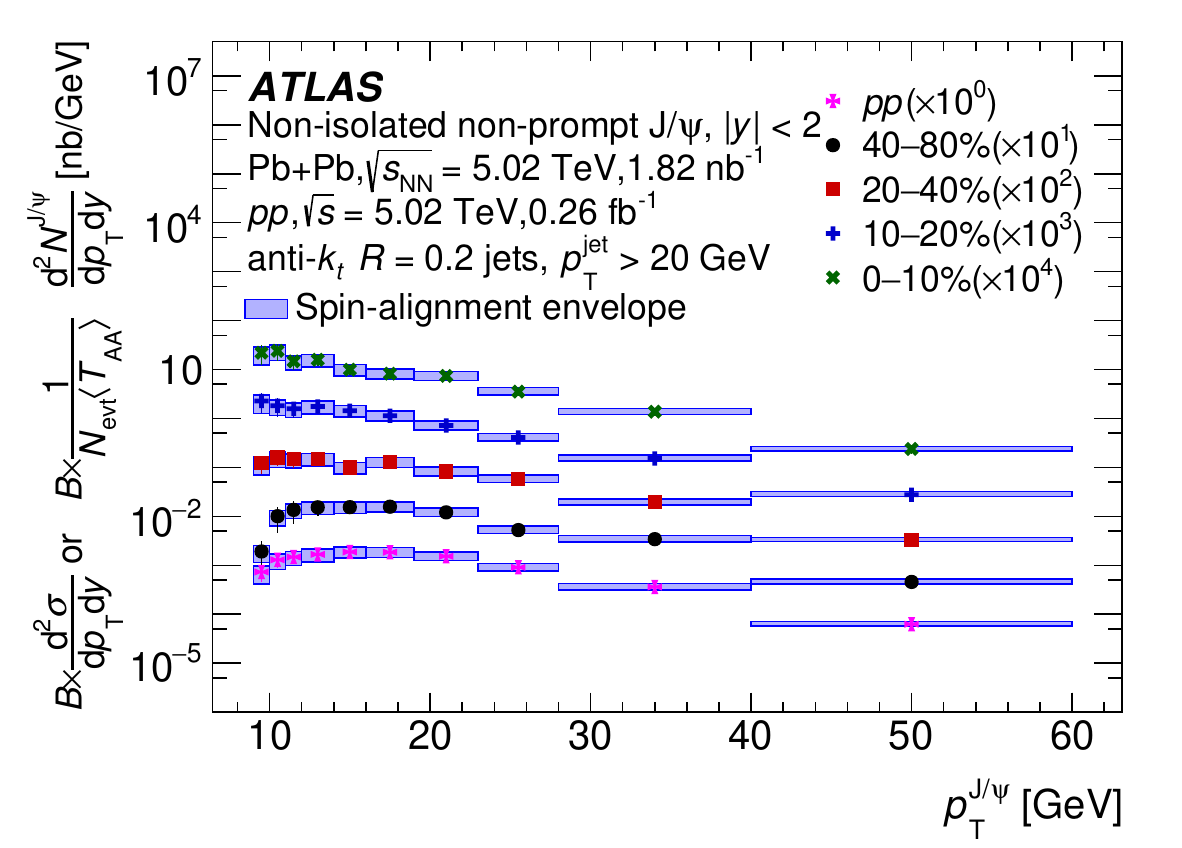}
}

\caption{$N_\mathrm{evt}$- and $\avgTAA$-normalised yields in different centrality intervals of Pb$+$Pb collisions and cross section in $pp$ collisions for (a,b) Inclusive, (c,d) Isolated, (e,f) Non-isolated, (a,c,e) Prompt, and (b,d,f) Non-prompt $J/\psi$ production measured as a function of $J/\psi$ $\pT$.
Each distribution is scaled by an additional factor $10^{n}$, indicated in the legend, to improve visibility.
The boxes represent the variation in the results for various $J/\psi$ spin polarisations.
The vertical error bars denote combined statistical and systematic uncertainties excluding the uncertainty due to $J/\psi$ spin polarisations. }
\label{fig:Yields}
\end{figure}

To study the $\jpsi$ production mechanism, yields normalised by $N_\mathrm{evt}$ and $\avgTAA$, measured in four centrality intervals of Pb+Pb collisions, and cross sections measured in \pp collisions are shown in Figure~\ref{fig:Yields}.
Yields and cross sections evaluated as a function of \Jpsi \pt are reported for prompt and non-prompt $\jpsi$ and for three classes characterising the correlation of $\jpsi$ production with jet production: non-isolated, isolated, and inclusive. A similar power-law-like decrease in measured distributions can be seen for inclusive and isolated $\jpsi$ mesons, while an increasing trend in the distributions with increasing $\pt$ in the region of $\pt<20$~\GeV is seen for non-isolated $\jpsi$ mesons. This increasing trend may be interpreted as a consequence of the increasing phase-space accessibility of the selected $\jpsi$+jet topology, which is restricted by the requirement of an accompanying jet with $\pt > 20$~\GeV. This requirement becomes less restrictive with increasing $\jpsi$ $\pt$.

\begin{figure}[htbp]
\centering

\subfloat[]{
\includegraphics[width=0.48\textwidth]{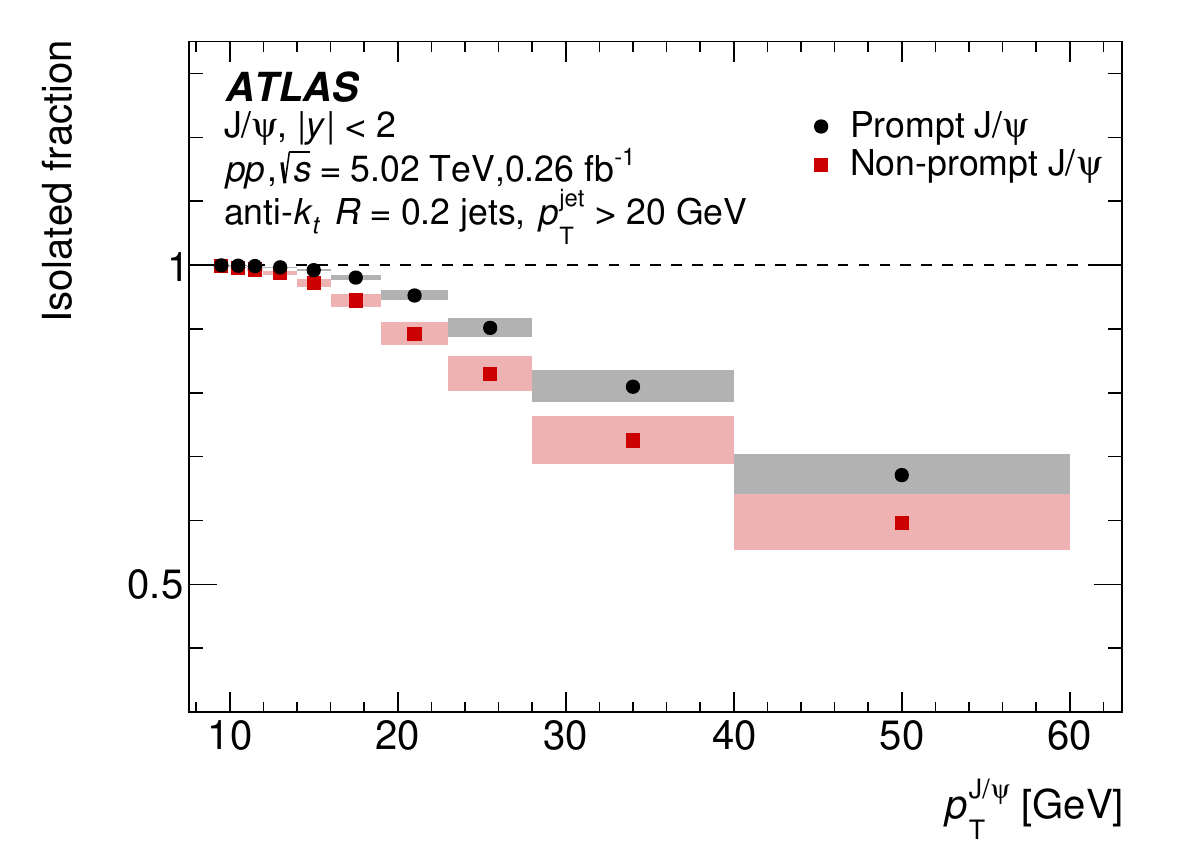}
}
\hfill
\subfloat[]{
\includegraphics[width=0.48\textwidth]{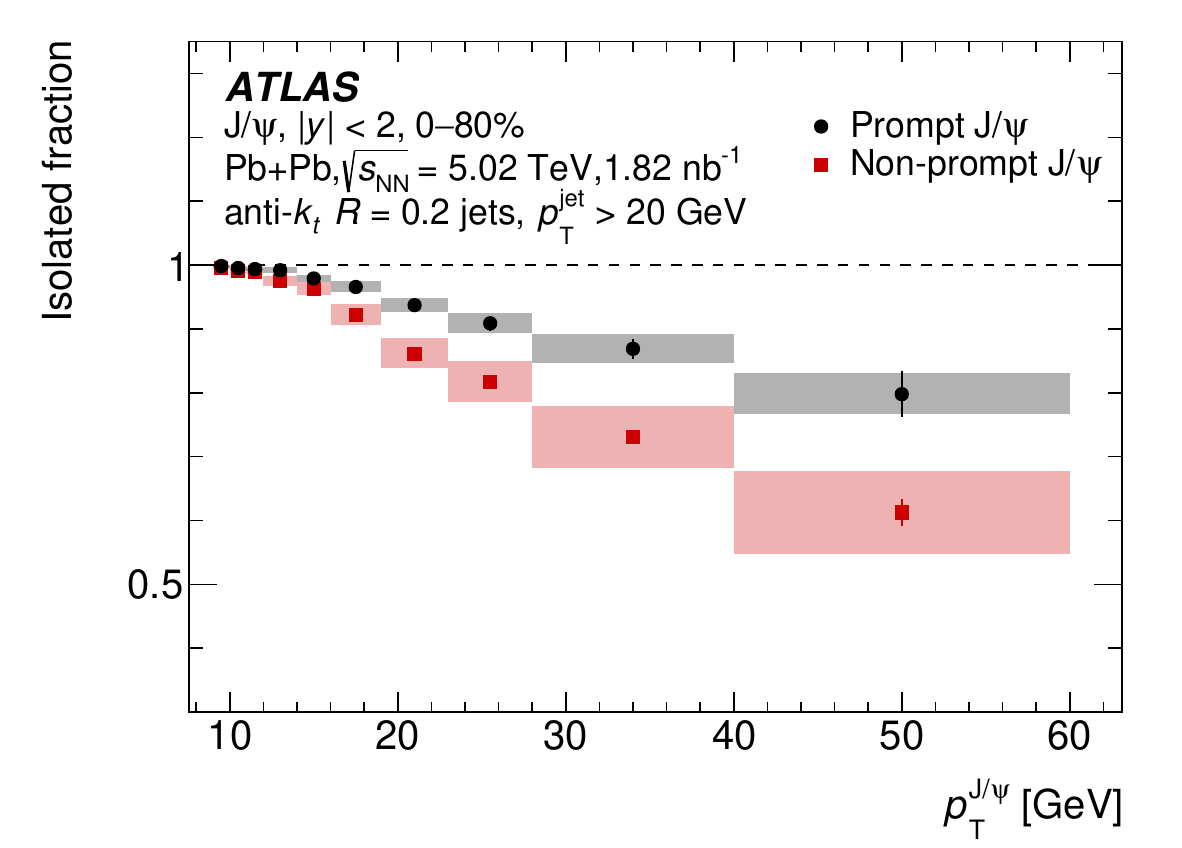}
}

\vspace{0.7em}

\subfloat[]{
\includegraphics[width=0.48\textwidth]{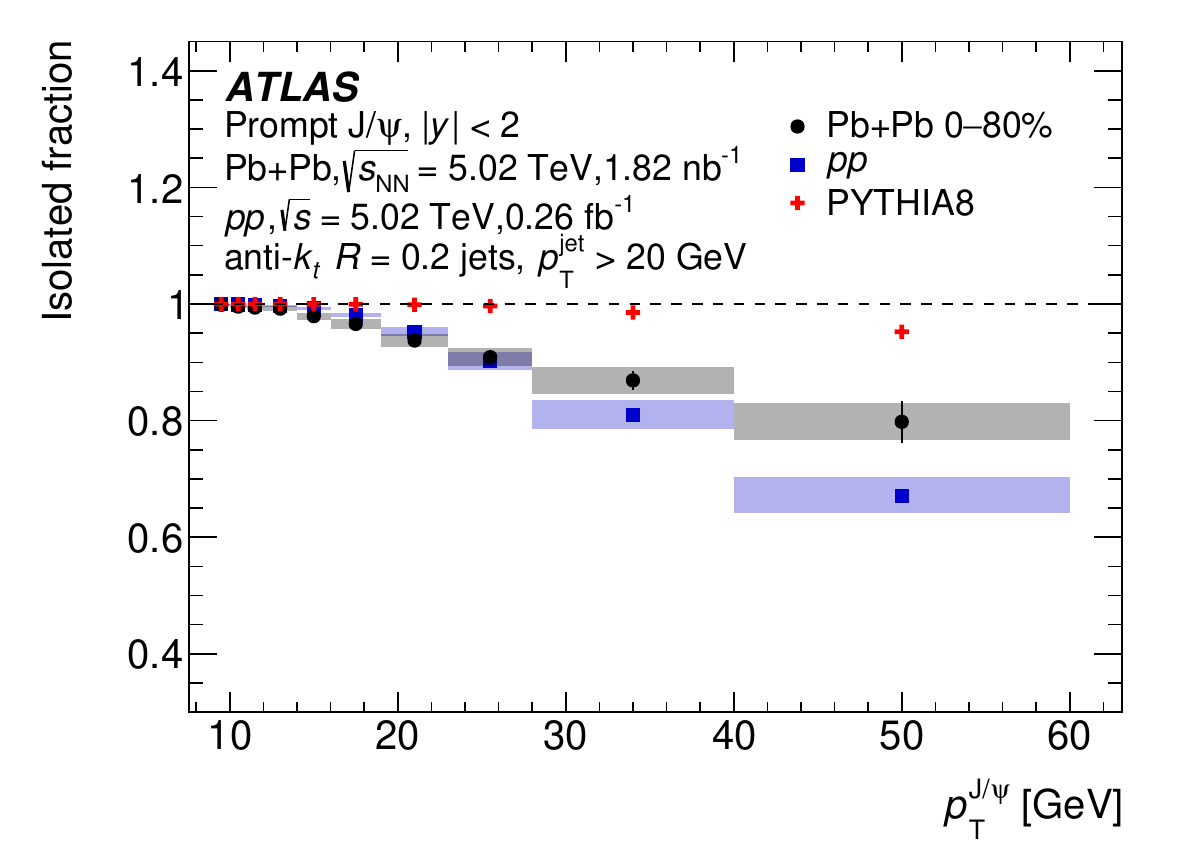}
}
\hfill
\subfloat[]{
\includegraphics[width=0.48\textwidth]{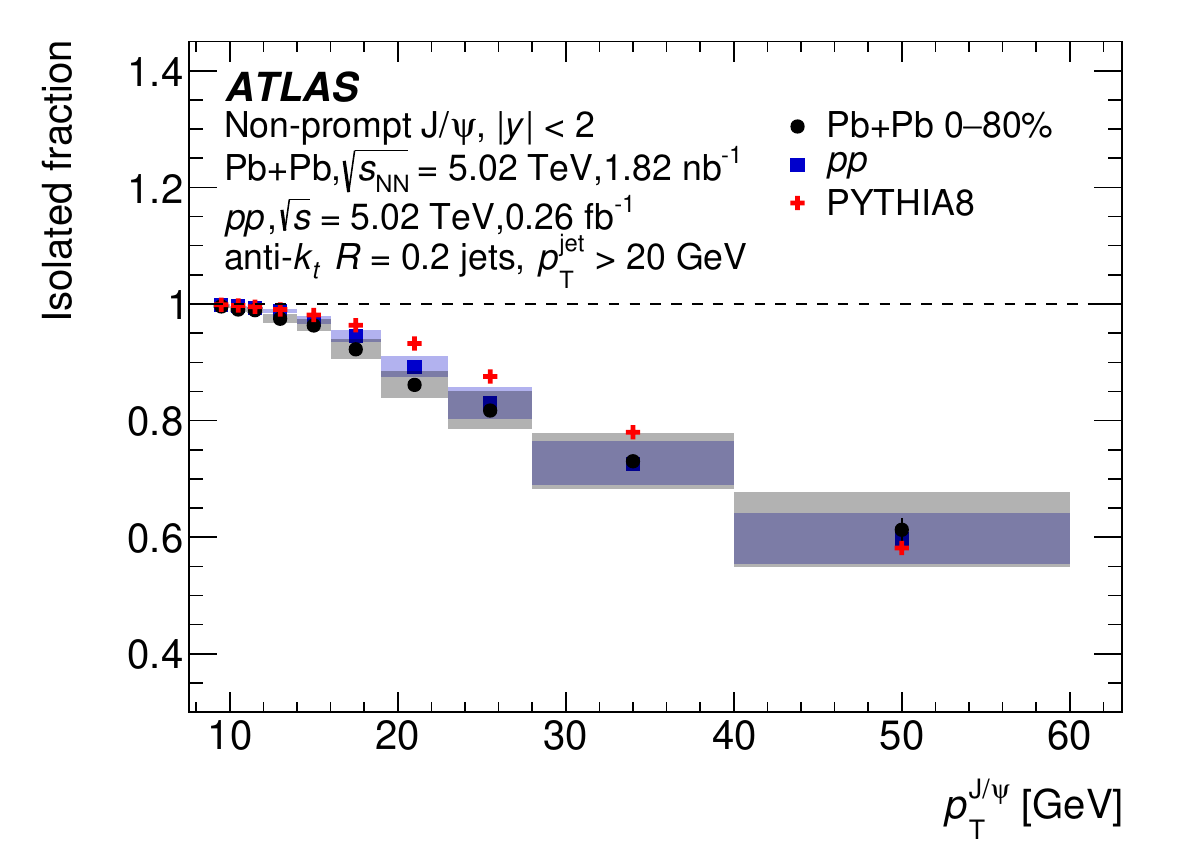}
}

\vspace{0.7em}

\subfloat[]{
\includegraphics[width=0.48\textwidth]{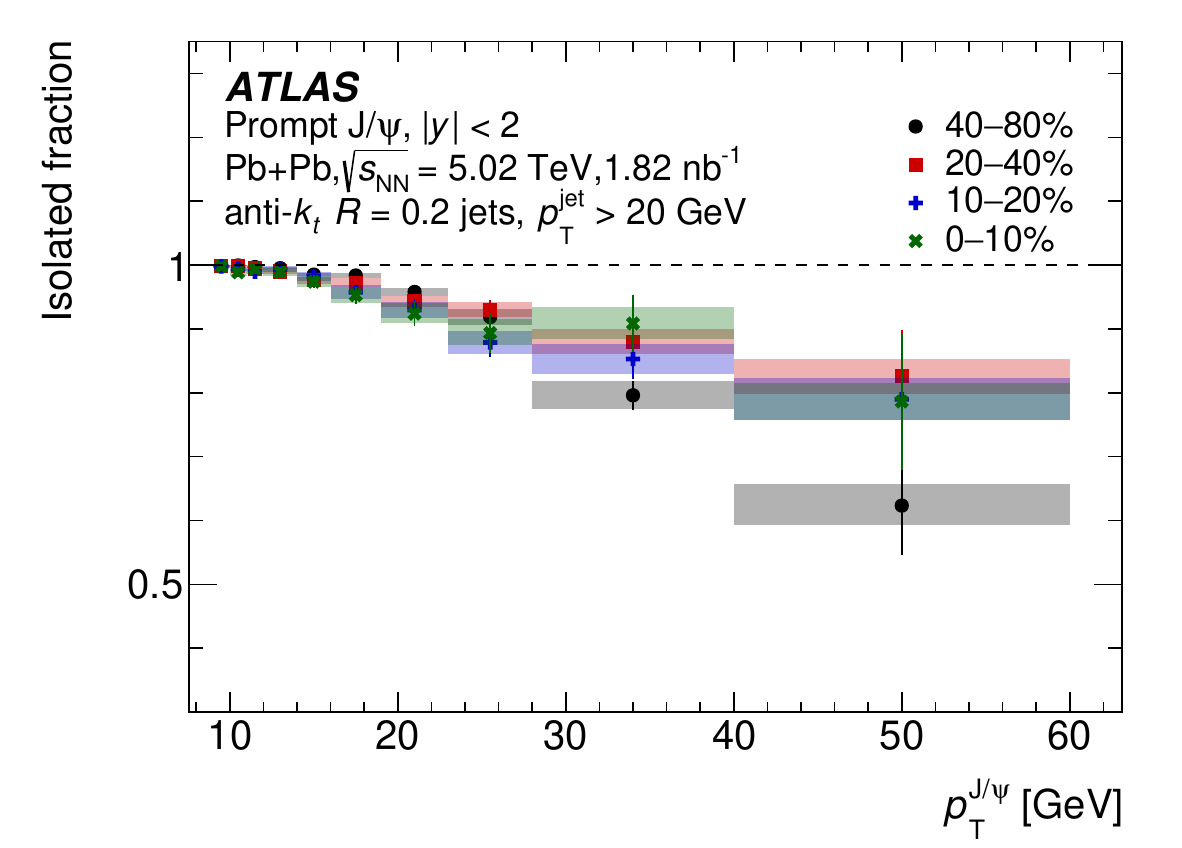}
}
\hfill
\subfloat[]{
\includegraphics[width=0.48\textwidth]{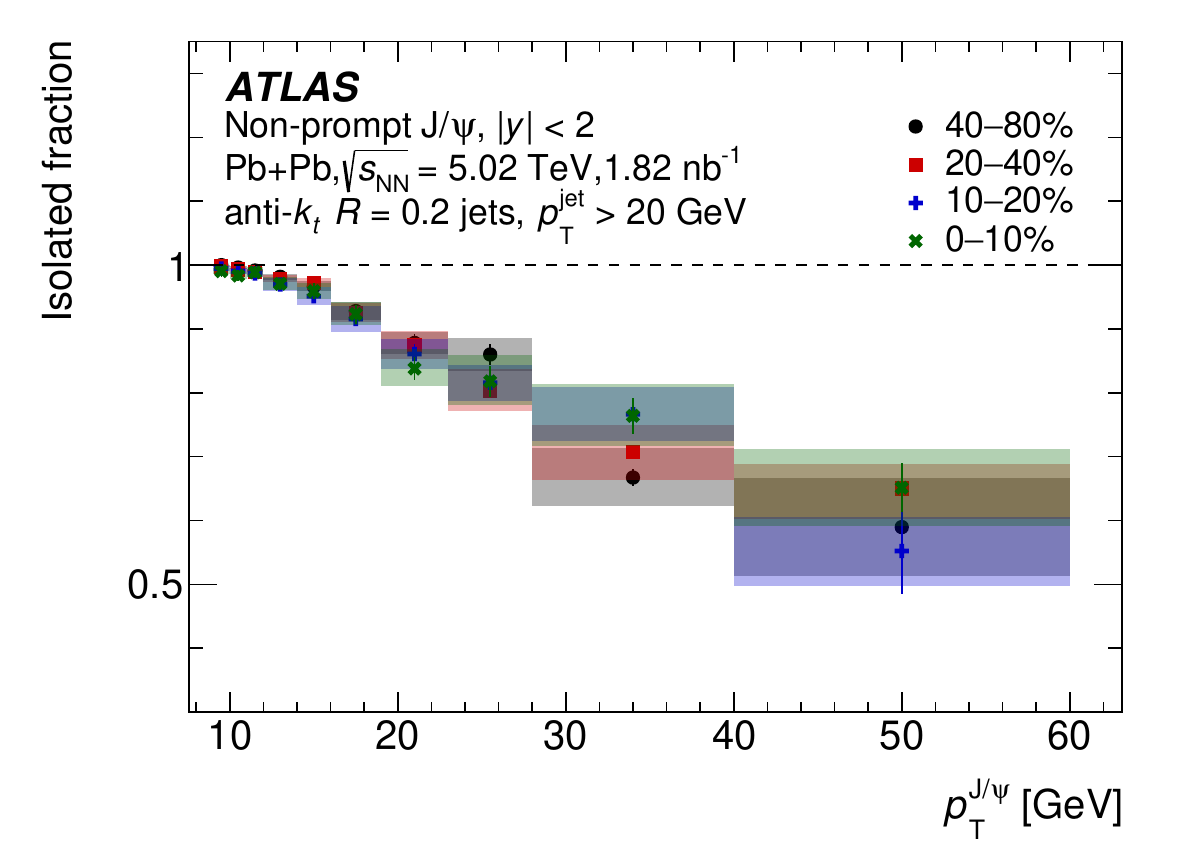}
}

\caption{Isolated fraction evaluated as a function of $J/\psi$ \pT for Prompt and Non-prompt $J/\psi$ in (a) $pp$ and (b) Pb$+$Pb collisions in 0--80\% centrality interval.
Comparison of the Isolated fraction evaluated as a function of $J/\psi$ \pT for (c) Prompt and (d) Non-prompt $J/\psi$ in $pp$ and Pb$+$Pb collisions in 0--80\% centrality interval compared to the LO prediction from the \Pythia~8.
Isolated fraction evaluated as a function of $J/\psi$ \pT for (e) Prompt and (f) Non-prompt $J/\psi$ in Pb$+$Pb collisions for various centrality intervals.
The error bars denote the statistical uncertainties, and the boxes represent the systematic uncertainties.}
\label{fig:IsoFrac}
\end{figure}

Yields and cross sections reported in Figure~\ref{fig:Yields} are used to evaluate the fraction of isolated $\jpsi$ (isolated fraction) defined in Section~\ref{sec:obs}, which is reported in Figure~\ref{fig:IsoFrac}. Overall, the isolated fraction starts at a value of 1 for $\jpsi$ with $\pt \approx 10$~\GeV and decreases with increasing $\pt$, reaching a value between 0.6 and 0.8 at $\pt \approx 50$~\GeV.
The increasing trend in distributions in Figure~\ref{fig:Yields}, attributed to the opening of the phase-space accessibility, suggests that full isolation seen for $\jpsi$ with $\pt < 20$~\GeV is largely connected with the minimum jet-$\pt$ requirement of 20~\GeV.

The upper panels of Figure~\ref{fig:IsoFrac} compare the isolated fraction for prompt and non-prompt $\jpsi$, showing that it is systematically 10--20\% larger for prompt $\jpsi$ mesons. No differences in the isolated fraction are found between \PbPb and \pp collisions or among different \PbPb centrality intervals, as shown in the middle panels and lower panels of Figure~\ref{fig:IsoFrac}, respectively. The middle panels also show a comparison with the LO QCD prediction from \Pythia~8. This generator-level study is performed using the same kinematic selection criteria as those applied to the data. \Pythia~8 is found to overestimate the isolated fraction for prompt \Jpsi, but provides a good prediction of the non-prompt \Jpsi isolated fraction.

\begin{figure}[!ht]
\centering
\subfloat[]{\includegraphics[width=0.49\textwidth]{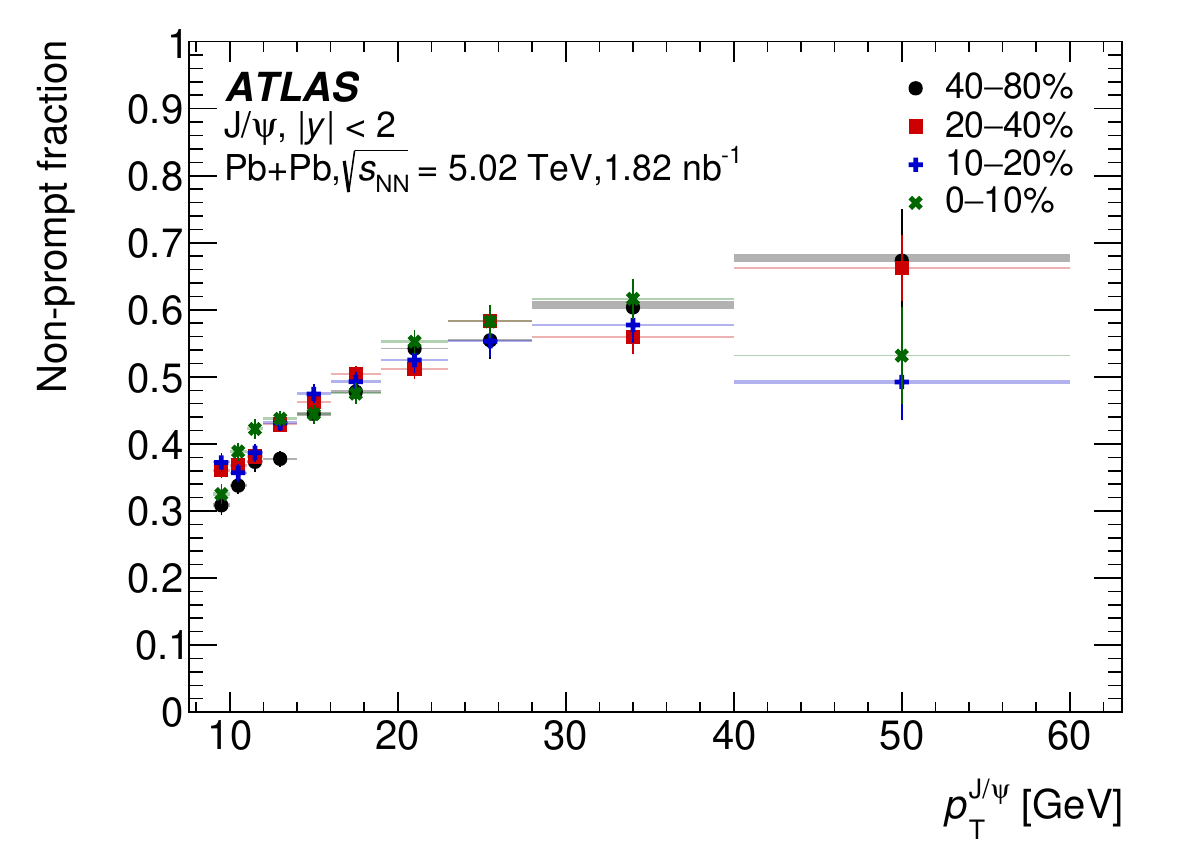}}
\subfloat[]{\includegraphics[width=0.49\textwidth]{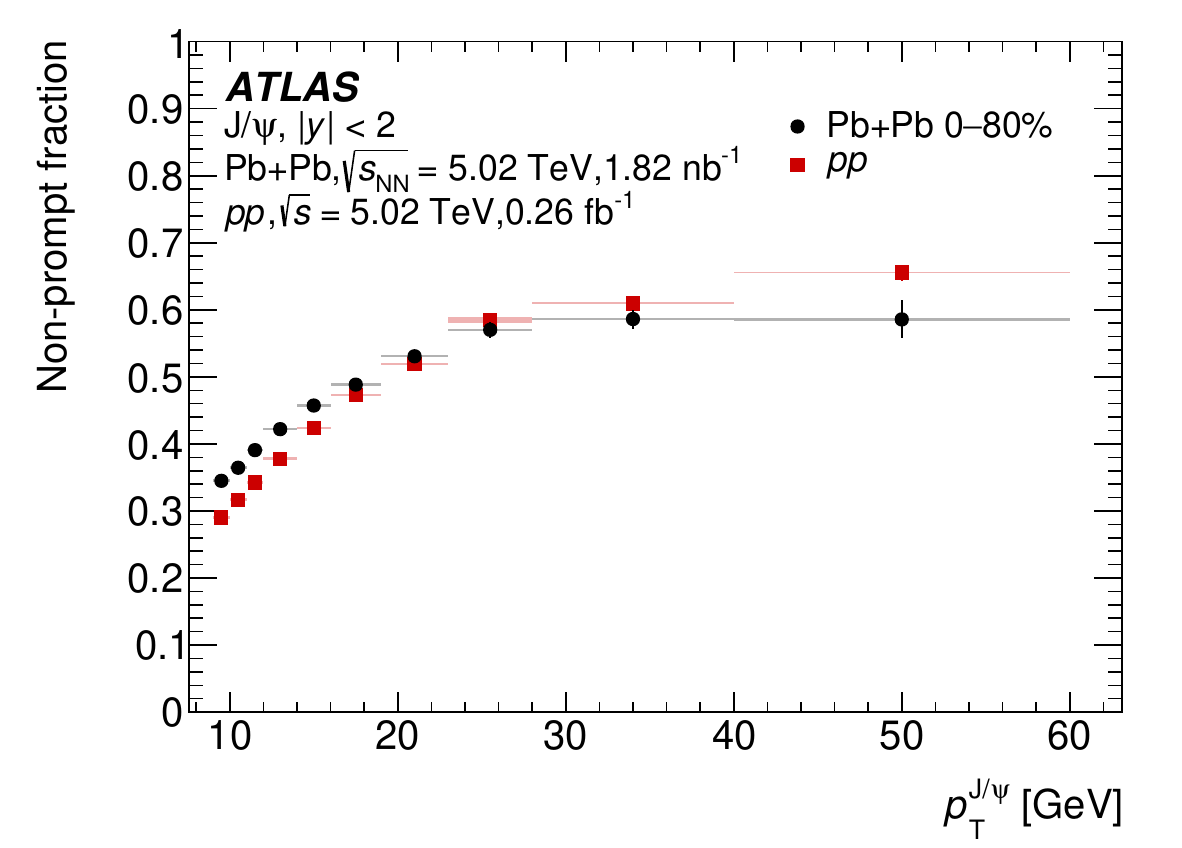}}
\caption{(a) Non-prompt fraction of Inclusive $J/\psi$ production as a function of $J/\psi$ \pT in Pb$+$Pb collisions for various centrality intervals. (b) Non-prompt fraction of Inclusive $J/\psi$ production as a function of $J/\psi$ \pT in $pp$ and Pb$+$Pb in the 0--80\% centrality interval. The error bars denote the statistical uncertainties, and the boxes represent the systematic uncertainties.}
\label{fig:NPFracCent}
\end{figure}

\begin{figure}[!ht]
\centering
\subfloat[]{\includegraphics[width=0.49\textwidth]{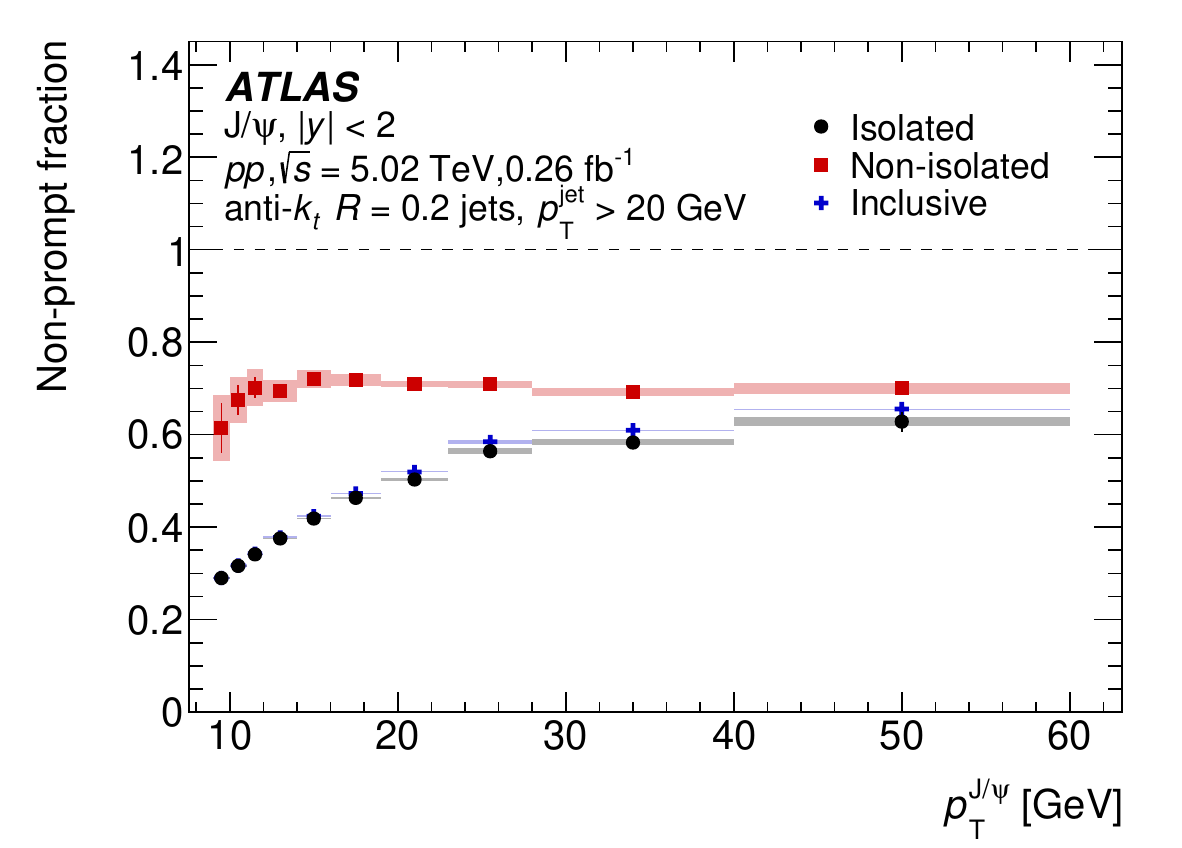}}
\subfloat[]{\includegraphics[width=0.49\textwidth]{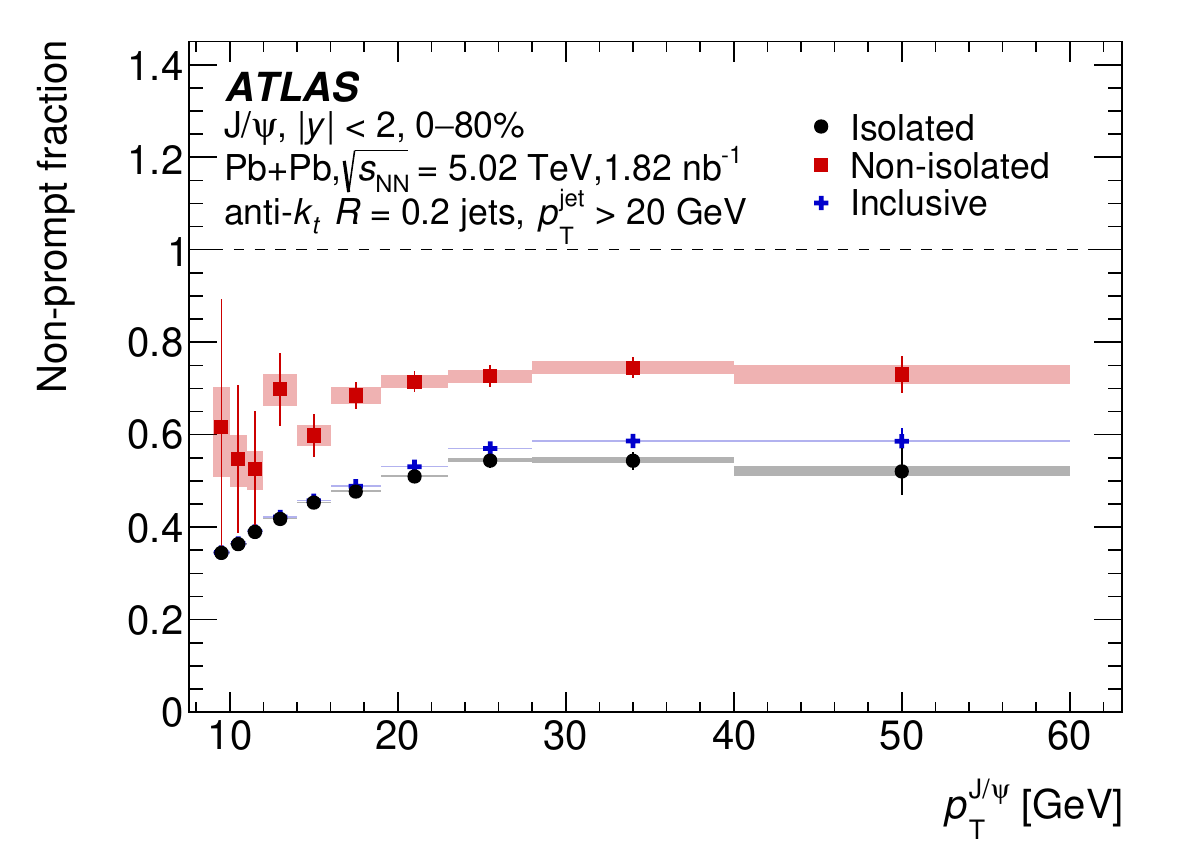}}
\caption{Non-prompt fraction of Inclusive, Isolated and Non-isolated $J/\psi$ production as a function of $J/\psi$ \pT in (a) $pp$ and (b) Pb$+$Pb collisions in the 0--80\% centrality interval. The error bars denote the statistical uncertainties, and the boxes represent the systematic uncertainties.}
\label{fig:NPFracAll}
\end{figure}

The non-prompt fraction for inclusive \Jpsi measured in four centrality intervals of \PbPb collisions is shown in the left panel of Figure~\ref{fig:NPFracCent}. The non-prompt fraction starts at a value of approximately $0.3$ at $\pt^{\Jpsi}=10$~\GeV, and it increases, reaching a value of approximately 0.6 at \Jpsi $\pt \approx 50$~\GeV. No significant centrality dependence is observed, consistent with the previous measurement in Ref.~\cite{HION-2016-07}. The non-prompt fraction for inclusive \Jpsi measured in \pp collisions and 0--80\% \PbPb collisions is shown in the right panel of Figure~\ref{fig:NPFracCent}. The non-prompt fraction measured in \PbPb collisions has a slightly different slope from that measured in \pp collisions, as previously observed in Ref.~\cite{HION-2016-07}. The left and right panels of Figure~\ref{fig:NPFracAll} show the non-prompt fraction in \pp and 0--80\% \PbPb collisions, respectively, evaluated for non-isolated, isolated, and inclusive $\jpsi$ mesons. While the non-prompt fraction is consistent between the isolated and inclusive classes, it differs significantly for non-isolated $\jpsi$. For $\pt>15$~\GeV, the non-prompt fraction of non-isolated $\jpsi$ is consistent with no $\pt$ dependence, and has a value of approximately 0.7.
These results provide complementary information to the evaluation of the isolated fraction presented in Figure~\ref{fig:IsoFrac}.

\begin{figure}[!ht]
\centering
\subfloat[]{\includegraphics[width=0.49\textwidth]{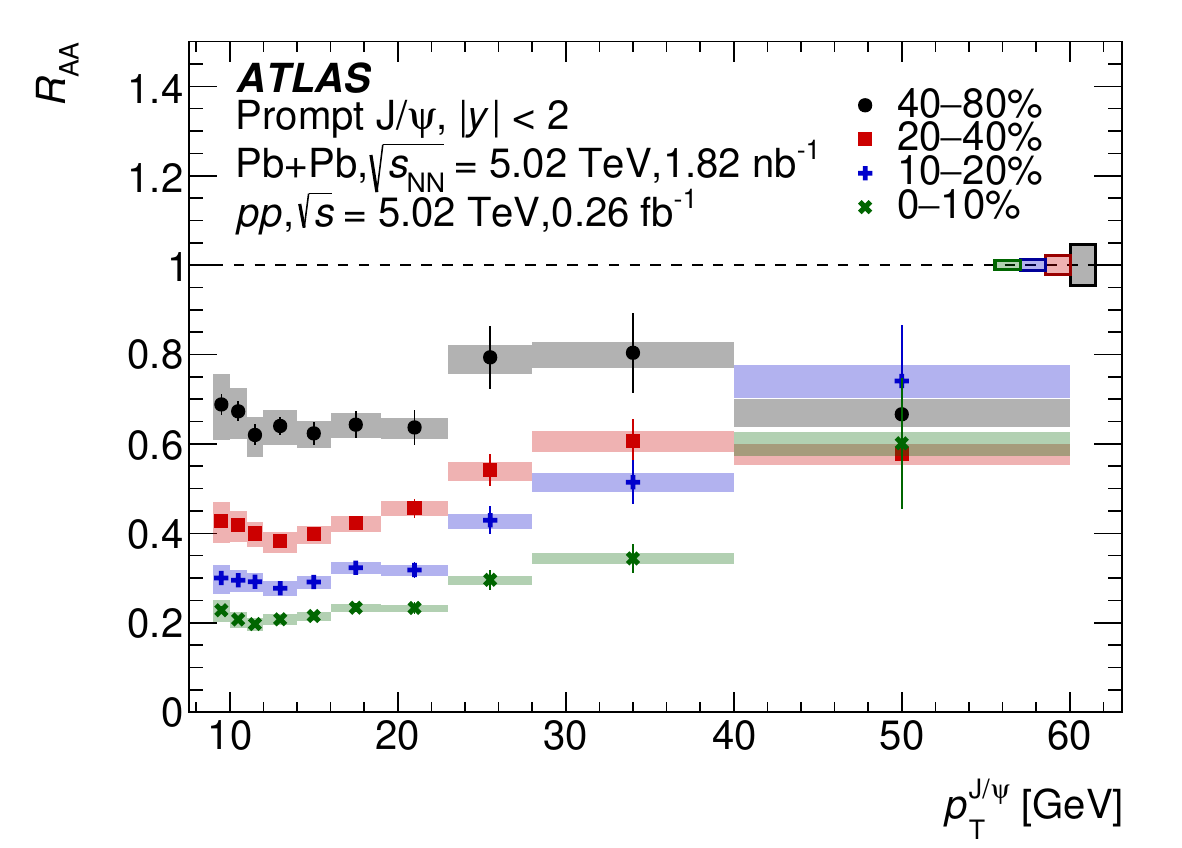}}
\subfloat[]{\includegraphics[width=0.49\textwidth]{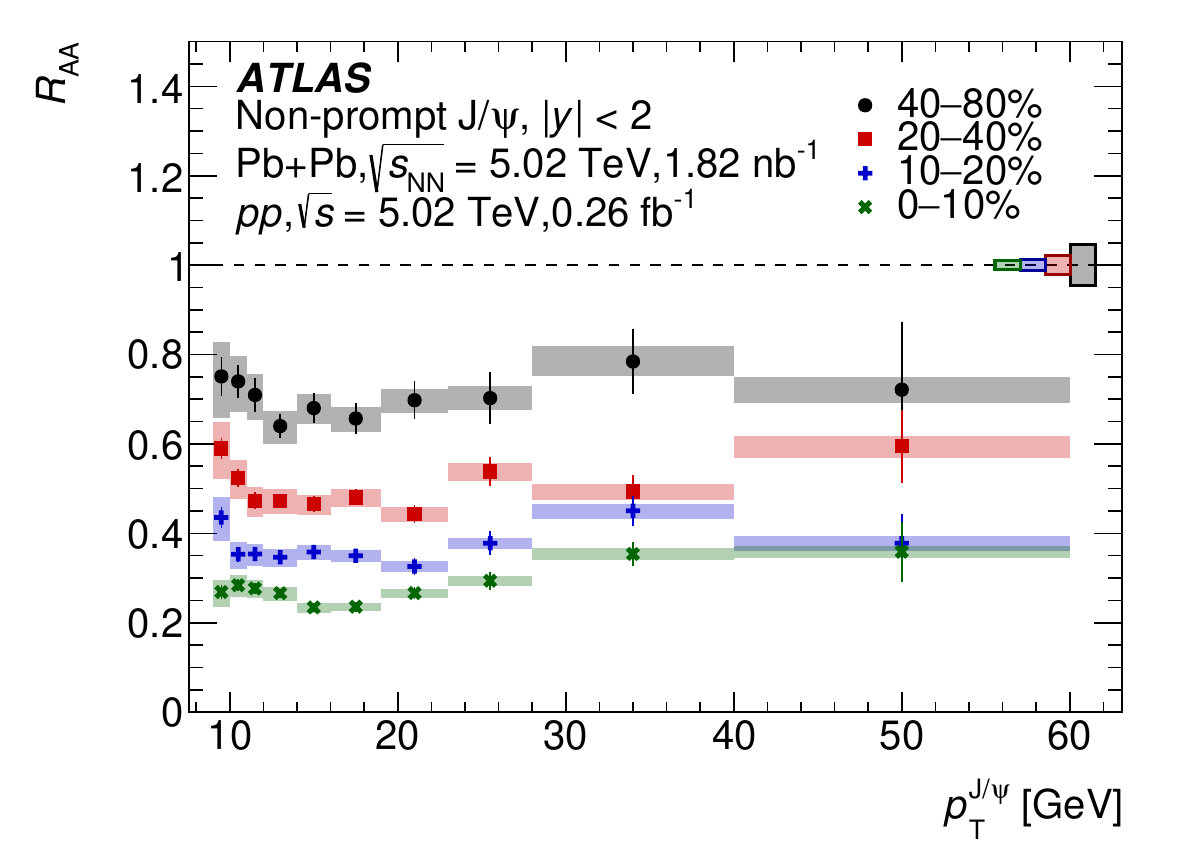}}

\caption{The nuclear modification factor as a function of $J/\psi$ \pT for (a) Prompt and (b) Non-prompt $J/\psi$ evaluated for different centrality intervals. The error bars denote the statistical uncertainties, and the boxes represent the systematic uncertainties. Boxes at $R_{\mathrm{AA}} = 1$ represent global scale uncertainties, namely the $pp$ luminosity uncertainty and \avgTAA uncertainty.}
\label{fig:RaaPandNP}
\end{figure}

Figure~\ref{fig:RaaPandNP} shows the $\RAA$, constructed from \PbPb yields and \pp cross sections in \pp, as a function of $\jpsi$ $\pt$  for inclusive prompt and non-prompt $\jpsi$. These new results improve the quantification of $\jpsi$ suppression for $\pt>30$~\GeV with the $\RAA$ values reaching approximately 0.6 and 0.4 at $\pt \approx 50$~\GeV for prompt and non-prompt $\jpsi$, respectively, in the 0--10\% centrality interval.

The $\RAA$ evaluated as a function of \Jpsi $\pt$ for non-isolated, isolated and inclusive $\jpsi$ in the 0--80\% centrality interval is shown in Figure~\ref{fig:RaaPandNP080All}, and figures of the \RAA for other centrality intervals are presented in the Appendix of this paper. The two panels of Figure~\ref{fig:RaaPandNP080All} compare prompt and non-prompt $\jpsi$. In both cases, the $\RAA$ values for the isolated and the inclusive classes are consistent over the full $\pt$ range. For $\pt>30$~\GeV, the central values of $\RAA$ of non-isolated prompt $\jpsi$ appear to be systematically lower than those for isolated prompt $\jpsi$. To quantify this difference, the $\pt$ dependence of $\RAA$ is fitted separately for the isolated and non-isolated samples using polynomial functions. The quadratic and higher-order terms are consistent with zero within uncertainties, indicating that linear functions provide a good description of the data in the kinematic range considered. The slopes extracted for the isolated and non-isolated samples differ by 3.2 standard deviations for the 0--80\% centrality interval as well as for the individual centrality intervals excluding the most peripheral one. As a cross-check, Stouffer's test~\cite{stouffer1949,Bityukov:2008zz} is applied directly to the data points, yielding a significance of 3.5 standard deviations.
Such a difference might be interpreted as a larger suppression of \Jpsi produced in association with
a more developed parton shower or fragmentation process.
In the case of non-prompt $\jpsi$, no significant differences are observed among the three isolation classes of $\jpsi$ mesons. A difference in $\RAA$ between non-isolated and isolated $\jpsi$ is also observed for $\pt<20$~\GeV in both the prompt and non-prompt samples, although with a lower significance than in the high-$\pt$ prompt sample. The $\RAA$ in that region tends to grow, which may be related to a complex interplay between phase-space effects and quenching of jets used to tag the $\jpsi$.



\begin{figure}[!ht]
\centering
\subfloat[]{\includegraphics[width=0.49\textwidth]{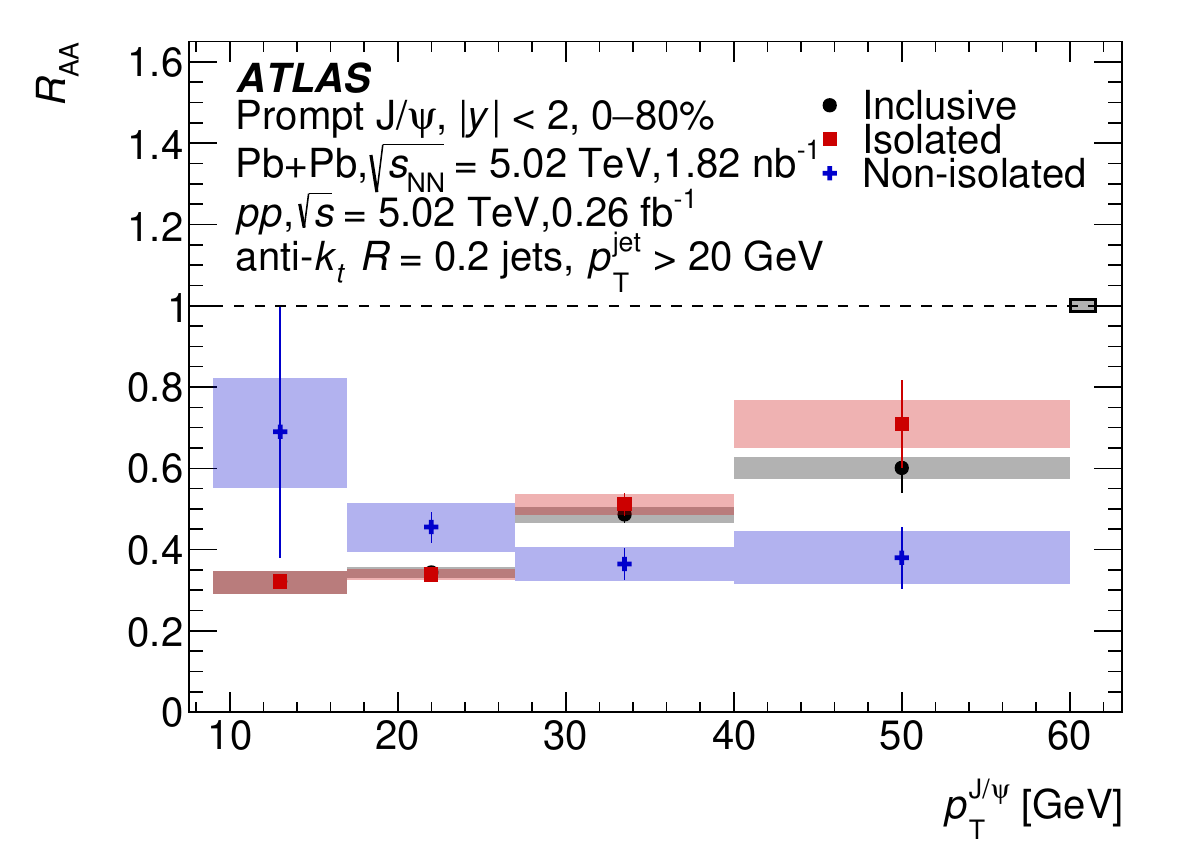}}
\subfloat[]{\includegraphics[width=0.49\textwidth]{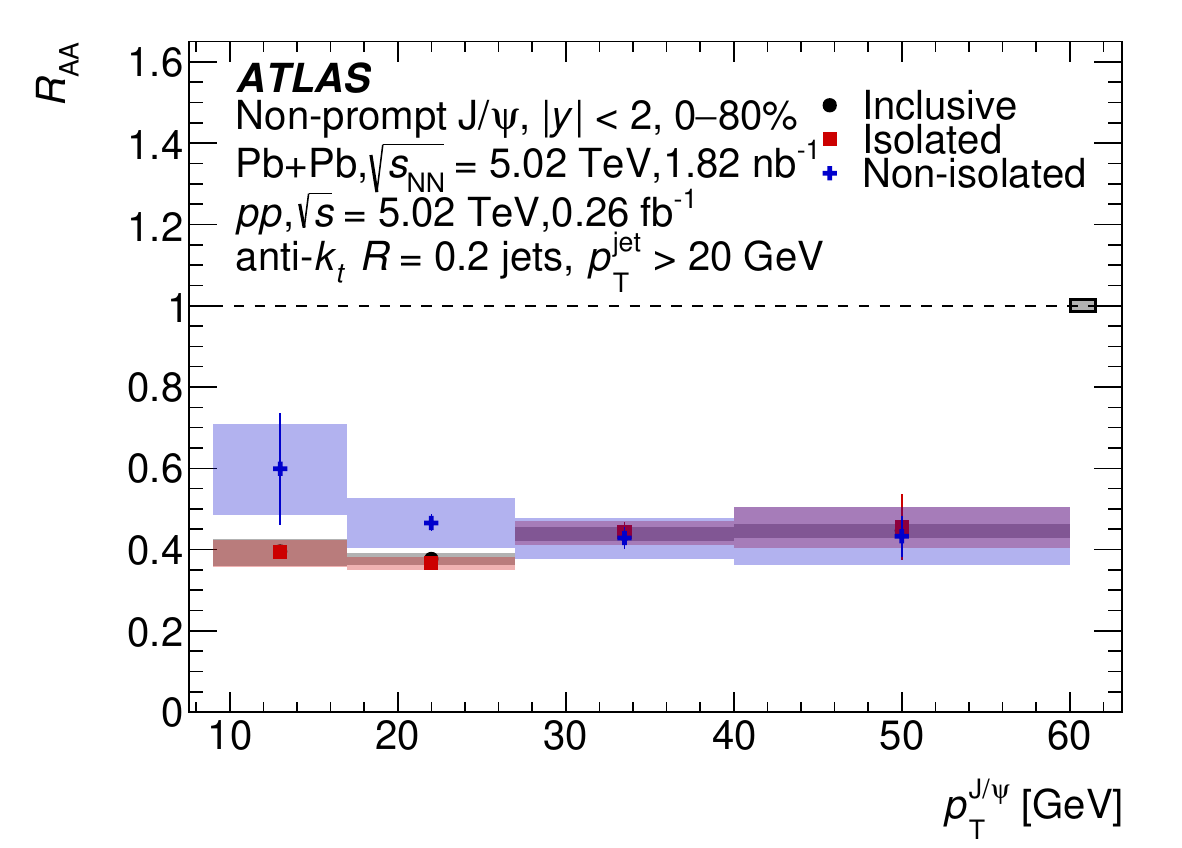}}

\caption{The nuclear modification factor as a function of \pT for Inclusive, Isolated and Non-isolated (a) Prompt and (b) Non-prompt $J/\psi$ evaluated for 0--80\% centrality interval. The error bars denote the statistical uncertainties, and the boxes represent the systematic uncertainties. Boxes at $R_{\mathrm{AA}} = 1$ represent global scale uncertainties, namely the $pp$ luminosity uncertainty and \avgTAA uncertainty.}
\label{fig:RaaPandNP080All}
\end{figure}

% End of text imported from the .//sections/06_Results.tex input file

% The next lines are included from the .//sections/07_Conclusions.tex input file
\FloatBarrier
\section{Conclusion}

This paper presents ATLAS measurements of $\jpsi$ meson production in the dimuon decay channel in 5.02~\TeV \pp\ and \PbPb\ collisions, for $\jpsi$ mesons with $9<\pt<60$~\GeV and $|y|<2$. The measurements are performed for three classes of $\jpsi$ mesons, defined according to their association with an $R=0.2$ jet: non-isolated, isolated, and inclusive. Each of these classes is further divided into prompt and non-prompt $\jpsi$. The \Jpsi non-isolation is defined by the presence of a jet with $\pt>20$~\GeV within $\Delta R < 0.4$ from $\jpsi$.
The jet association is corrected for jet reconstruction efficiency and migrations across the $\pt>20$~\GeV threshold, so that it corresponds to the truth-level jet selection, and the jet \pt does not include the \pt of the \Jpsi.
The presence of jets from multi-parton interactions, or from the underlying event, which may correlate the $\jpsi$ with a jet not originating from the same hard process, is corrected for using the orthogonal-cone method. In this set-up, more than 60\% of non-prompt $\jpsi$ are found to be isolated from jets at the highest measured $\pt$ of $\jpsi$. The isolated fraction increases systematically with decreasing $\pt$ of $\jpsi$ and reaches unity for $\jpsi$ with $\pt=9$~\GeV. For \Jpsi $\pt \geq 20$~\GeV, the isolated fraction is found to be approximately $10-20$\% larger for prompt compared to non-prompt $\jpsi$ with no differences found between \pp\ and \PbPb\ collisions. The $\RAA$ of $\jpsi$ is measured to be consistent between the isolated and the inclusive case. The observed difference in prompt \Jpsi $\RAA$ between the non-isolated $\jpsi$ and isolated \Jpsi class has a significance of $3.5\sigma$.

These new results complement previous measurements of quarkonia--jet correlations that required events to contain a jet and the $\pt$ of jets included the $\pt$ of the $\jpsi$ meson. Results presented here provide new input to constrain theoretical models describing \Jpsi production in \pp and \PbPb collisions, and provide new information for modelling \Jpsi suppression in \PbPb\ collisions.

% End of text imported from the .//sections/07_Conclusions.tex input file

%

%
%
%
%
%
\section*{Acknowledgements}
%

%

% The next lines are included from the .//acknowledgements/Acknowledgements.tex input file
%
%

%
%

We thank CERN for the very successful operation of the LHC and its injectors, as well as the support staff at
CERN and at our institutions worldwide without whom ATLAS could not be operated efficiently.

The crucial computing support from all WLCG partners is acknowledged gratefully, in particular from CERN, the ATLAS Tier-1 facilities at TRIUMF/SFU (Canada), NDGF (Denmark, Norway, Sweden), CC-IN2P3 (France), KIT/GridKA (Germany), INFN-CNAF (Italy), NL-T1 (Netherlands), PIC (Spain), RAL (UK) and BNL (USA), the Tier-2 facilities worldwide and large non-WLCG resource providers. Major contributors of computing resources are listed in Ref.~\cite{ATL-SOFT-PUB-2026-001}.

We gratefully acknowledge the support of ANPCyT, Argentina; YerPhI, Armenia; ARC, Australia; BMWFW and FWF, Austria; ANAS, Azerbaijan; CNPq and FAPESP, Brazil; NSERC, NRC and CFI, Canada; CERN; ANID, Chile; CAS, MOST, NSFC and FRFCU, China; Minciencias, Colombia; MEYS CR, Czech Republic; DNRF and DNSRC, Denmark; IN2P3-CNRS and CEA-DRF/IRFU, France; SRNSFG, Georgia; BMFTR, HGF and MPG, Germany; GSRI, Greece; RGC and Hong Kong SAR, China; ICHEP and Academy of Sciences and Humanities, Israel; INFN, Italy; MEXT and JSPS, Japan; CNRST, Morocco; NWO, Netherlands; RCN, Norway; MNiSW, Poland; FCT, Portugal; MNE/IFA, Romania; MSTDI, Serbia; MSSR, Slovakia; ARIS and MVZI, Slovenia; DSI/NRF, South Africa; MICIU/AEI, Spain; SRC and Wallenberg Foundation, Sweden; SERI, SNSF and Cantons of Bern and Geneva, Switzerland; NSTC, Taipei; TENMAK, T\"urkiye; STFC/UKRI, United Kingdom; DOE and NSF, United States of America.

Individual groups and members have received support from BCKDF, CANARIE, CRC and DRAC, Canada; CERN-CZ, FORTE and PRIMUS, Czech Republic; COST, ERC, ERDF, Horizon 2020 and Marie Sk{\l}odowska-Curie Actions, European Union; Investissements d'Avenir Labex, Investissements d'Avenir Idex and ANR, France; DFG and AvH Foundation, Germany; Herakleitos, Thales and Aristeia programmes co-financed by EU-ESF and the Greek NSRF, Greece; BSF-NSF and MINERVA, Israel; NCN and NAWA, Poland; La Caixa Banking Foundation, CERCA and AGAUR programs from Generalitat de Catalunya and PROMETEO and GenT Programmes Generalitat Valenciana, Spain; G\"{o}ran Gustafssons Stiftelse, Sweden; The Royal Society and Leverhulme Trust, United Kingdom; Eric and Wendy Schmidt Fund for Strategic Innovation, United States of America.

In addition, individual members wish to acknowledge support from Chile: Agencia Nacional de Investigaci\'on y Desarrollo (ANID FONDECYT reg. 1230987, FONDECYT 1230812, FONDECYT 1240864, Fondecyt 3240661, Fondecyt Regular 1240721); China: Fundamental Research Funds for the Central Universities (010-63263105), Chinese Ministry of Science and Technology (MOST-2023YFA1605700, MOST-2023YFA1609300), National Natural Science Foundation of China (NSFC 12275265, NSFC-W2543005); Czech Republic: Czech Science Foundation (GACR - 24-11373S), Ministry of Education Youth and Sports (ERC-CZ-LL2327, FORTE CZ.02.01.01/00/22\_008/0004632); EU: H2020 European Research Council (ERC - 101002463); European Union: European Research Council (BARD No. 101116429, ERC - 101219398, ERC 101089007), European Regional Development Fund (HE COFUND GA No.101081355, ERDF), Marie Sklodowska-Curie Actions (GAP-101168829); France: Agence Nationale de la Recherche (ANR-21-CE31-0013, ANR-22-EDIR-0002, ANR-24-CE31-0504-01); Germany: Deutsche Forschungsgemeinschaft (DFG - 469666862); China: Research Grants Council (GRF); Italy: Istituto Nazionale di Fisica Nucleare (LHC-MIUR - 28003/2025), Ministero dell'Università e della Ricerca (NextGenEU I53D23000820006 M4C2.1.1, SOE2024\_0000023); Japan: Japan Society for the Promotion of Science (JSPS KAKENHI  JP25H0063, JSPS KAKENHI JP22H04944, JSPS KAKENHI JP24K23939, JSPS KAKENHI JP24KK0251, JSPS KAKENHI JP25H00650, JSPS KAKENHI JP25H01291, JSPS KAKENHI JP25K01011, JSPS KAKENHI JP25K01023, JSPS KAKENHI JP25KK0047); Poland: Polish National Science Centre (NCN OPUS 2023/51/B/ST2/02507, NCN OPUS nr 2022/47/B/ST2/03059, UMO-2019/34/E/ST2/00393, UMO-2021/42/E/ST2/00350, UMO-2022/47/O/ST2/00148, UMO-2023/49/B/ST2/04085, UMO-2023/51/B/ST2/00920, UMO-2024/53/N/ST2/00869); Spain: Agència de Gestió d'Ajuts Universitaris i de Recerca. (AGAUR - 2023 BP 00141), Ministry of Science and Innovation (RYC2019-028510-I, RYC2020-030254-I, RYC2021-031273-I, RYC2022-038164-I), Ministerio de Ciencia, Innovación y Universidades/Agencia Estatal de Investigaci\'on (EU NextGenerationEU (PRTR-C17.I1), PID2022-142604OB-C22); Sweden: Carl Trygger Foundation (Carl Trygger Foundation CTS 22:2312), Swedish Research Council (Swedish Research Council 2023-04654, VR 2021-03651, VR 2022-03845, VR 2022-04683, VR 2023-03403, VR 2024-05451, VR 2025-05940), Knut and Alice Wallenberg Foundation (KAW 2023.0366); United Kingdom: The Binks Trust, Royal Society (NIF-R1-231091); United States of America: U.S. Department of Energy (ECA DE-AC02-76SF00515), John Templeton Foundation (John Templeton Foundation 63206), Neubauer Family Foundation.

%
%

% End of text imported from the .//acknowledgements/Acknowledgements.tex input file

%
\clearpage
\section*{Appendix}
\addcontentsline{toc}{section}{Appendix}

\begin{figure}[htbp]
\centering

\subfloat[]{
\includegraphics[width=0.48\textwidth]{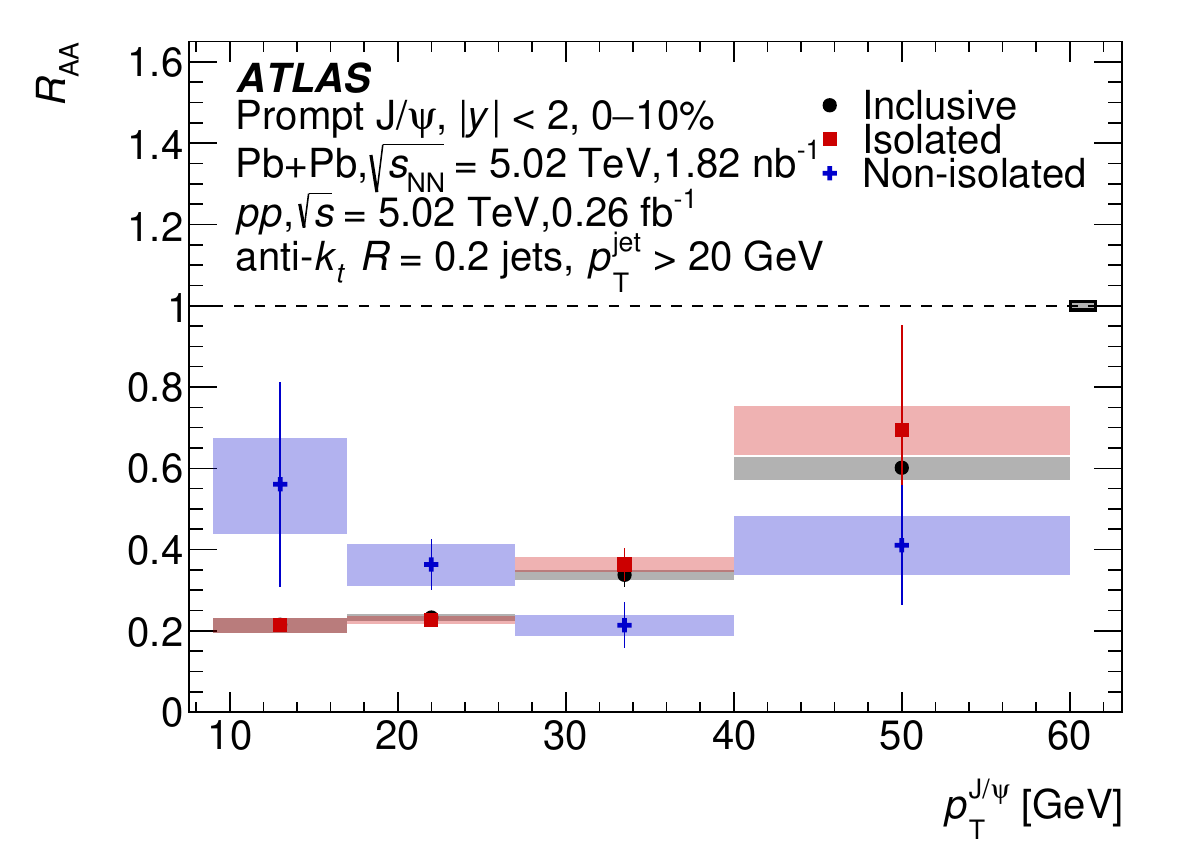}
}
\hfill
\subfloat[]{
\includegraphics[width=0.48\textwidth]{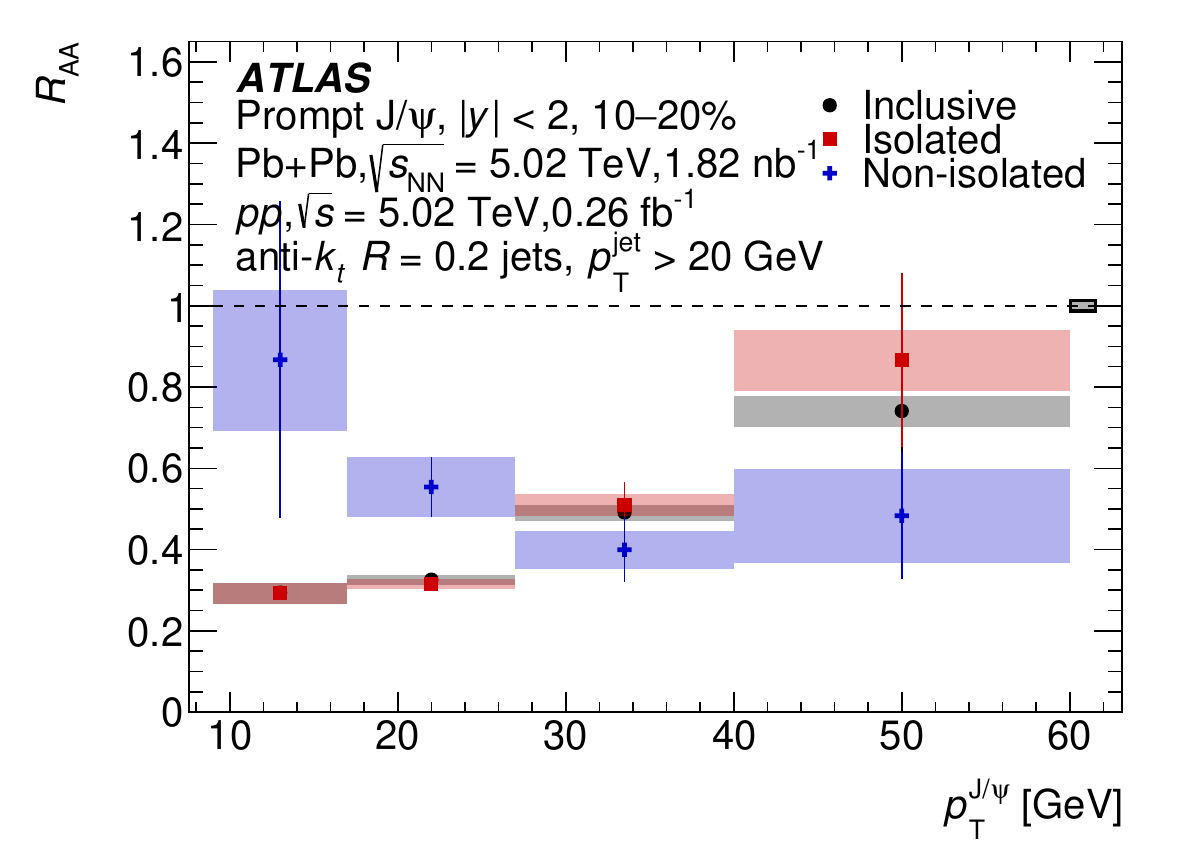}
}

\vspace{0.7em}

\subfloat[]{
\includegraphics[width=0.48\textwidth]{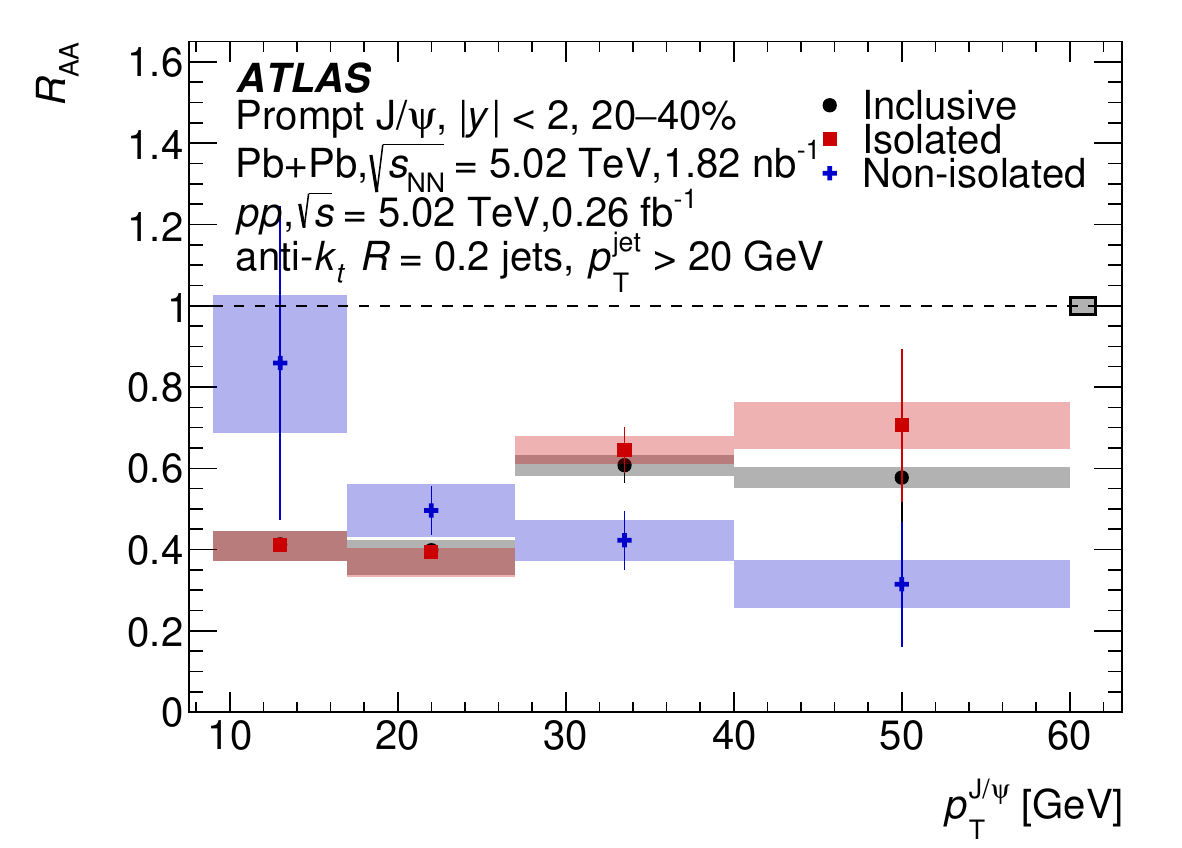}
}
\hfill
\subfloat[]{
\includegraphics[width=0.48\textwidth]{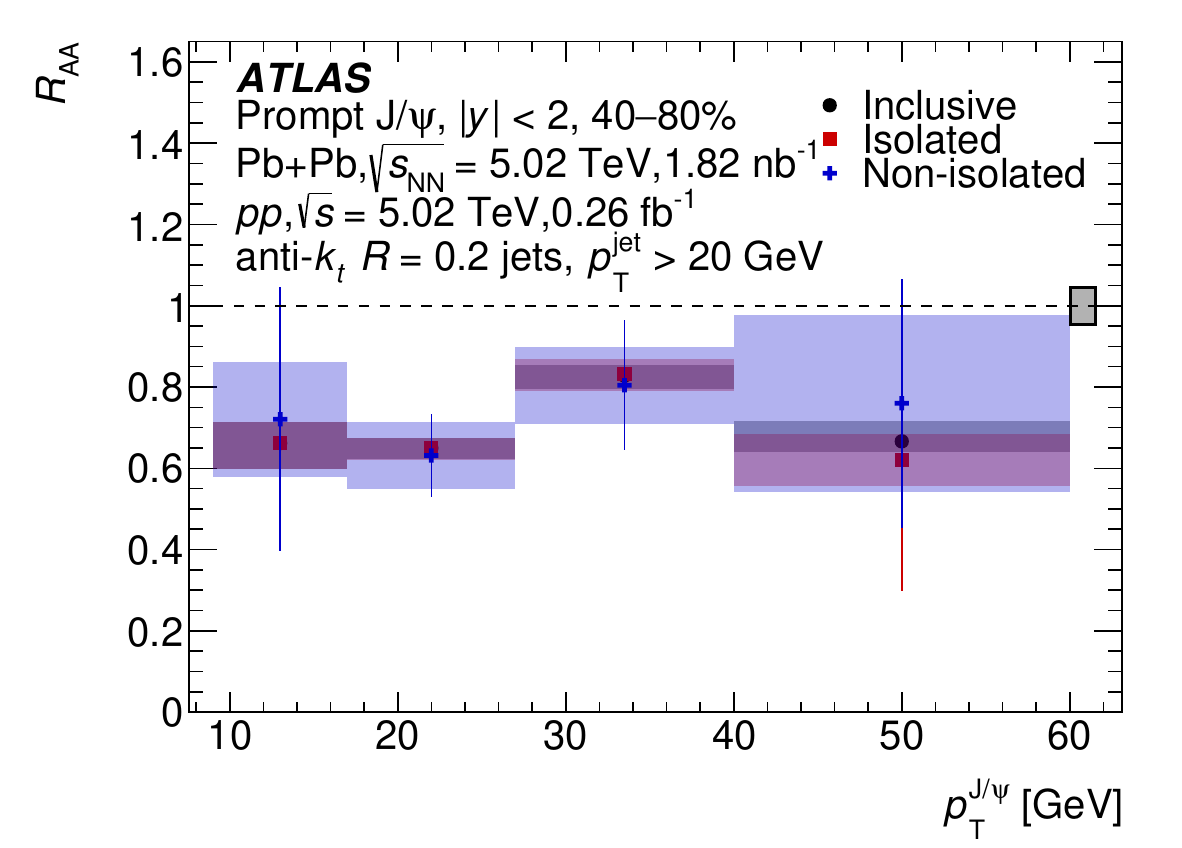}
}

\caption{The nuclear modification factor as a function of \pT for Inclusive, Isolated and Non-isolated prompt $J/\psi$ evaluated for four centrality intervals, namely (a) 0--10\%, (b) 10--20\%, (c) 20--40\%, and (d) 40--80\%. The error bars denote the statistical uncertainties, and the boxes represent the systematic uncertainties. Boxes at $R_{\mathrm{AA}} = 1$ represent global scale uncertainties, namely the $pp$ luminosity uncertainty and \avgTAA uncertainty.}
\label{fig:AppPRaa}
\end{figure}

\begin{figure}[htbp]
\centering

\subfloat[]{
\includegraphics[width=0.48\textwidth]{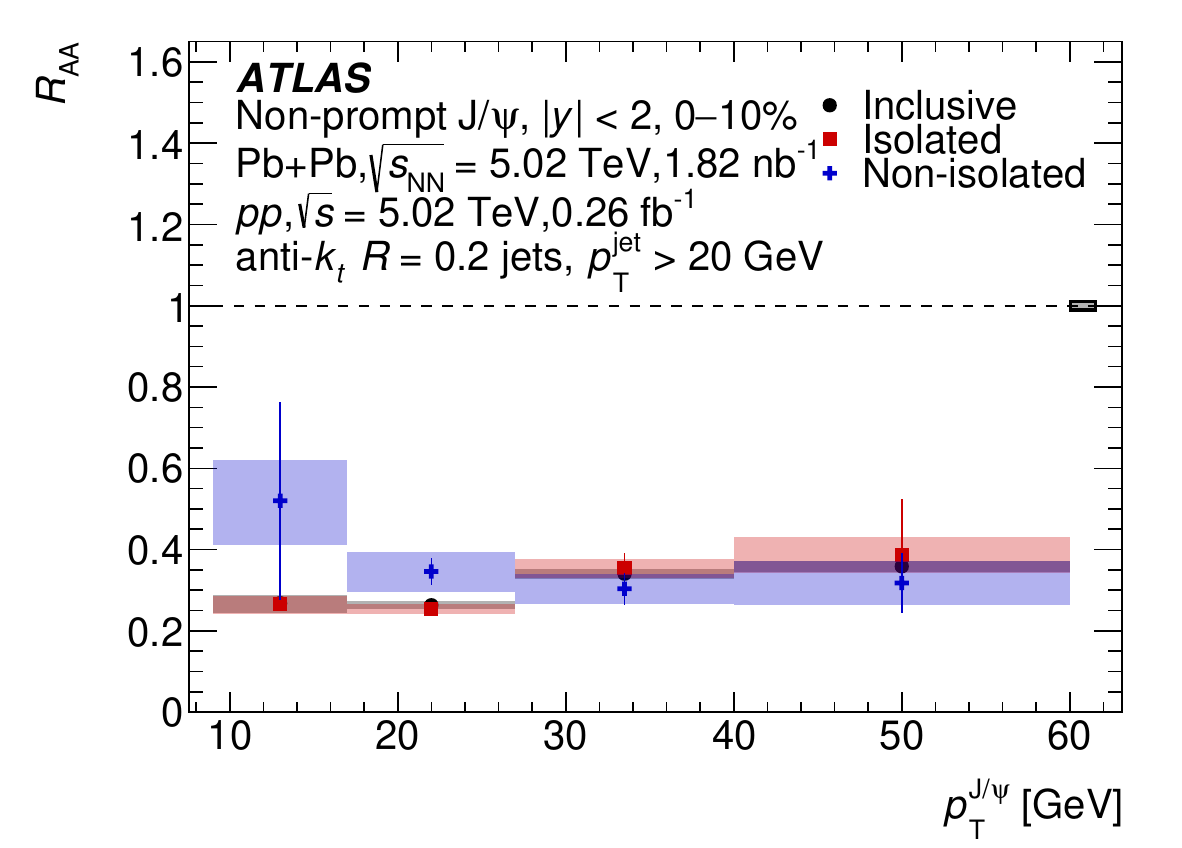}
}
\hfill
\subfloat[]{
\includegraphics[width=0.48\textwidth]{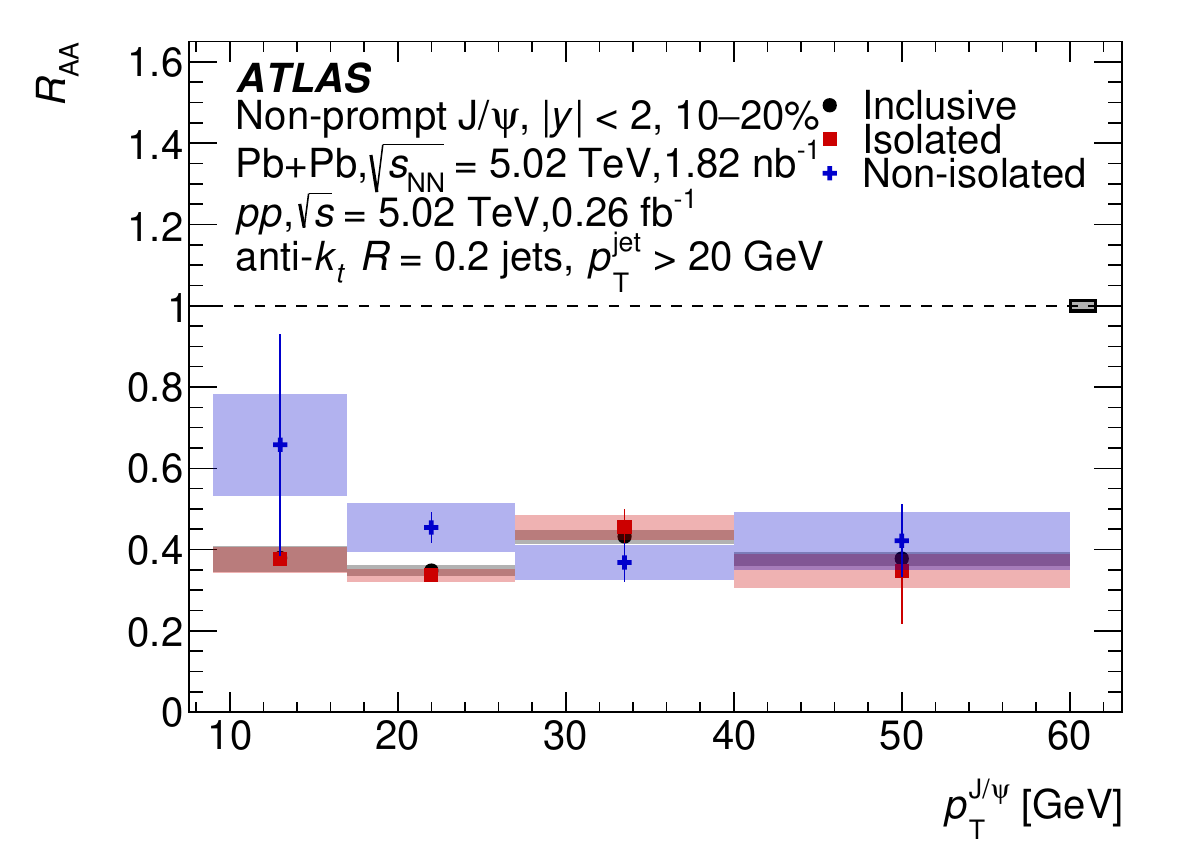}
}

\vspace{0.7em}

\subfloat[]{
\includegraphics[width=0.48\textwidth]{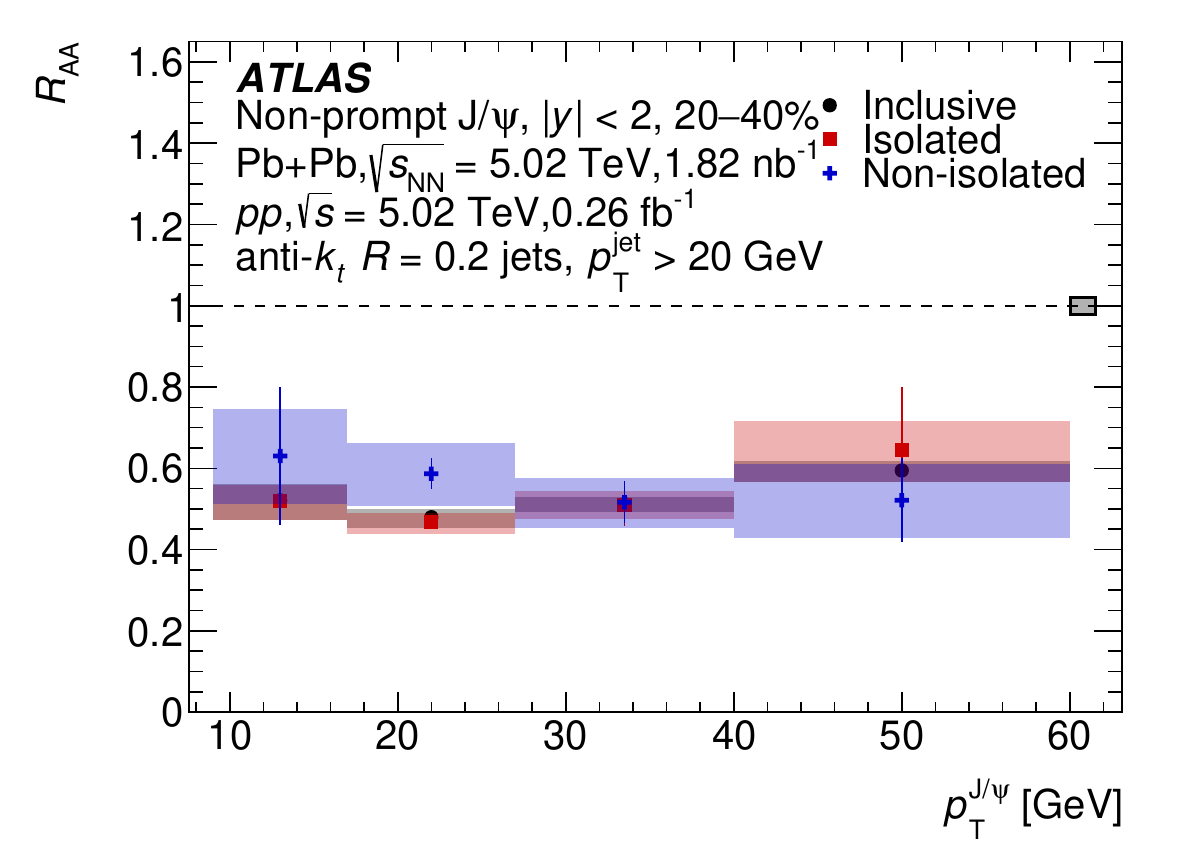}
}
\hfill
\subfloat[]{
\includegraphics[width=0.48\textwidth]{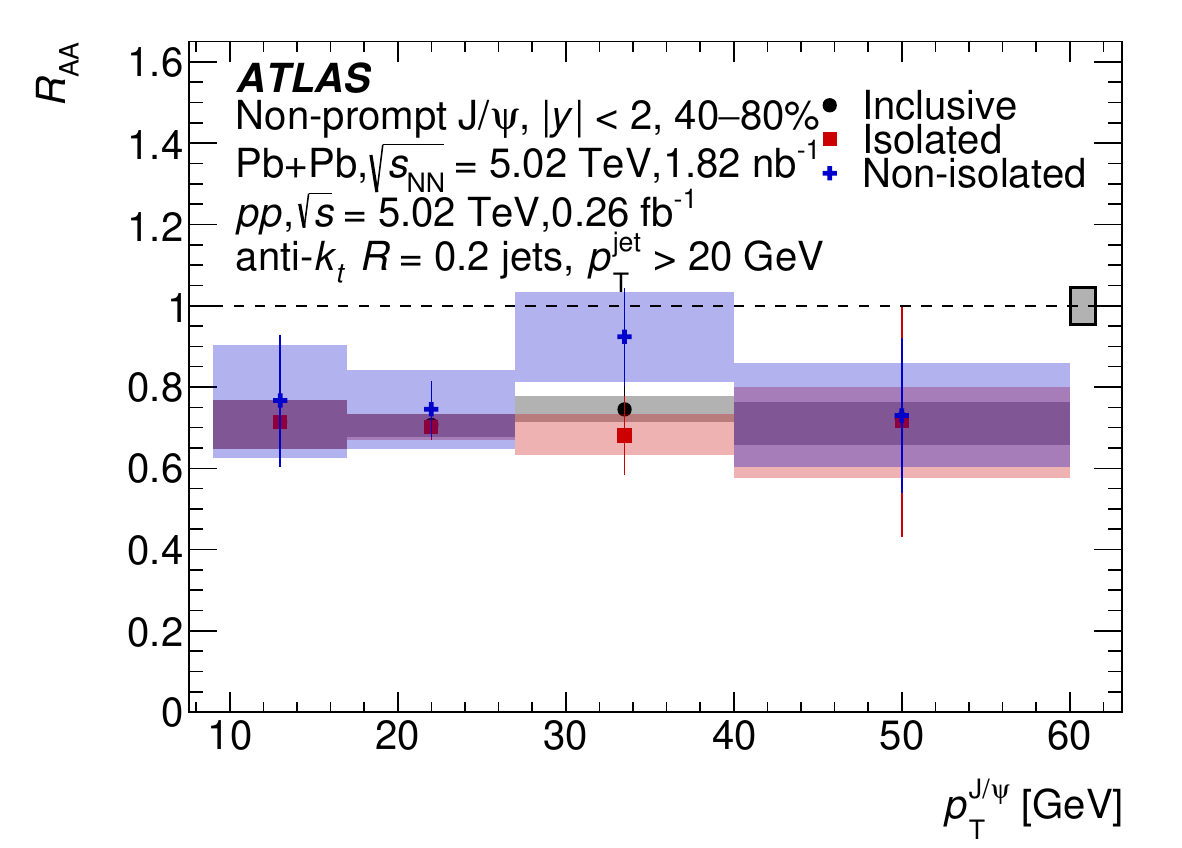}
}

\caption{The nuclear modification factor as a function of \pT for Inclusive, Isolated and Non-isolated non-prompt $J/\psi$ evaluated for four centrality intervals, namely (a) 0--10\%, (b) 10--20\%, (c) 20--40\%, and (d) 40--80\%. The error bars denote the statistical uncertainties, and the boxes represent the systematic uncertainties. Boxes at $R_{\mathrm{AA}} = 1$ represent global scale uncertainties, namely the $pp$ luminosity uncertainty and \avgTAA uncertainty.}
\label{fig:AppNPRaa}
\end{figure}

% The next lines are included from the .//sections/Appendix.tex input file
The $\RAA$ evaluated as a function of \Jpsi $\pt$ for non-isolated, isolated, and inclusive $\jpsi$ in individual centrality intervals is presented for prompt and non-prompt production in Figures~\ref{fig:AppPRaa} and~\ref{fig:AppNPRaa}, respectively. These figures complement the corresponding results for the 0--80\% centrality interval shown in Figure~\ref{fig:RaaPandNP080All}.

\clearpage

\clearpage

% End of text imported from the .//sections/Appendix.tex input file

%

%
%
\printbibliography
\clearpage
\input{atlas_authlist}
%

%
%
%
%

%
%
%
%

\end{document}

%% file: atlas_authlist.tex
% ATLAS Collaboration author list
% Reference date of HION-2018-22 is 2026-05-07
% Author list last updated on date 29-SEP-26
% Data extracted on 29-Sep-2026 for paper reference HION-2018-22
% at 7:21pm
 
\begin{flushleft}
\hypersetup{urlcolor=black}
{\Large The ATLAS Collaboration}

\bigskip

\AtlasOrcid[0000-0002-6665-4934]{G.~Aad}$^\textrm{\scriptsize 102}$,
\AtlasOrcid[0000-0001-7616-1554]{E.~Aakvaag}$^\textrm{\scriptsize 17}$,
\AtlasOrcid[0000-0002-5888-2734]{B.~Abbott}$^\textrm{\scriptsize 121}$,
\AtlasOrcid[0000-0002-0287-5869]{S.~Abdelhameed}$^\textrm{\scriptsize 83b}$,
\AtlasOrcid[0000-0002-1002-1652]{K.~Abeling}$^\textrm{\scriptsize 54}$,
\AtlasOrcid[0000-0001-5763-2760]{N.J.~Abicht}$^\textrm{\scriptsize 48}$,
\AtlasOrcid[0000-0002-8496-9294]{S.H.~Abidi}$^\textrm{\scriptsize 30}$,
\AtlasOrcid[0009-0003-6578-220X]{M.~Aboelela}$^\textrm{\scriptsize 44}$,
\AtlasOrcid[0000-0002-9987-2292]{A.~Aboulhorma}$^\textrm{\scriptsize 36e}$,
\AtlasOrcid[0000-0001-5329-6640]{H.~Abramowicz}$^\textrm{\scriptsize 154}$
% ...
\bigskip
\\

% ...
$^{17}$Department for Physics and Technology, University of Bergen, Bergen; Norway.\\
% ...
$^{30}$Physics Department, Brookhaven National Laboratory, Upton NY; United States of America.\\
% ...
$^{36}$$^{(a)}$Facult\'e des Sciences Ain Chock, Universit\'e Hassan II de Casablanca;$^{(b)}$Facult\'{e} des Sciences, Universit\'{e} Ibn-Tofail, K\'{e}nitra;$^{(c)}$Facult\'e des Sciences Semlalia, Universit\'e Cadi Ayyad, LPHEA-Marrakech;$^{(d)}$LPMR, Facult\'e des Sciences, Universit\'e Mohamed Premier, Oujda;$^{(e)}$Facult\'e des sciences, Universit\'e Mohammed V, Rabat;$^{(f)}$Institute of Applied Physics, Mohammed VI Polytechnic University, Ben Guerir; Morocco.\\
% ...
$^{44}$Physics Department, Southern Methodist University, Dallas TX; United States of America.\\
% ...
$^{48}$Fakult\"{a}t Physik, Technische Universit{\"a}t Dortmund, Dortmund; Germany.\\
% ...
$^{54}$II. Physikalisches Institut, Georg-August-Universit\"{a}t G\"ottingen, G\"ottingen; Germany.\\
% ...
$^{83}$$^{(a)}$Khalifa University of Science and Technology, Abu Dhabi;$^{(b)}$New York University Abu Dhabi, Abu Dhabi;$^{(c)}$United Arab Emirates University, Al Ain;$^{(d)}$University of Sharjah, Sharjah; United Arab Emirates.\\
% ...
$^{102}$CPPM, Aix-Marseille Universit\'e, CNRS/IN2P3, Marseille; France.\\
% ...
$^{121}$Homer L. Dodge Department of Physics and Astronomy, University of Oklahoma, Norman OK; United States of America.\\
% ...
$^{154}$Raymond and Beverly Sackler School of Physics and Astronomy, Tel Aviv University, Tel Aviv; Israel.\\
% ...
$^{a}$ Also at Affiliated with an institute formerly covered by a cooperation agreement with CERN.\\
$^{b}$ Also at An-Najah National University, Nablus; Palestine.\\
$^{c}$ Also at Borough of Manhattan Community College, City University of New York, New York NY; United States of America.\\
$^{d}$ Also at Center for Interdisciplinary Research and Innovation (CIRI-AUTH), Thessaloniki; Greece.\\
$^{e}$ Also at Centre of Physics of the Universities of Minho and Porto (CF-UM-UP); Portugal.\\
$^{f}$ Also at CERN, Geneva; Switzerland.\\
% ...

\end{flushleft}

% Created with Glance <Atlas.Glance@cern.ch>

%% file: ANA-HION-2018-22-PAPER.bib
@article{Bityukov:2008zz,
    author = "Bityukov, Sergey and Krasnikov, Nikolai and Nikitenko, Alexander and Smirnova, Vera",
    editor = "Speer, Thomas and Carminati, Federico and Werlen, Monique",
    title = "{Two approaches to combining significances}",
    doi = "10.22323/1.070.0118",
    journal = "PoS",
    volume = "ACAT08",
    pages = "118",
    year = "2008"
}

@article{Avoni:2025lmu,
    author = "Avoni, Giulio and others",
    title = "{Upgrades of the ATLAS zero degree calorimeter system for Run 3 at the Large Hadron Collider}",
    eprint = "2509.05948",
    archivePrefix = "arXiv",
    primaryClass = "physics.ins-det",
    doi = "10.1088/1748-0221/20/11/P11021",
    journal = "JINST",
    volume = "20",
    number = "11",
    pages = "P11021",
    year = "2025"
}

@article{LHCb:2013izl,
    author = "{LHCb Collaboration}",
    %author = "Aaij, R and others",
    collaboration = "LHCb",
    title = "{Measurement of $J/\psi$ polarization in $pp$ collisions at $\sqrt{s}=7$ TeV}",
    eprint = "1307.6379",
    archivePrefix = "arXiv",
    primaryClass = "hep-ex",
    reportNumber = "LHCB-PAPER-2013-008, CERN-PH-EP-2013-130",
    doi = "10.1140/epjc/s10052-013-2631-3",
    journal = "Eur. Phys. J. C",
    volume = "73",
    number = "11",
    pages = "2631",
    year = "2013"
}

@article{ALICE:2011gej,
    author = "{ALICE Collaboration}",
    %author = "Abelev, Betty and others",
    collaboration = "ALICE",
    title = "{$J/\psi$ polarization in $pp$ collisions at $\sqrt{s}=7$ TeV}",
    eprint = "1111.1630",
    archivePrefix = "arXiv",
    primaryClass = "hep-ex",
    reportNumber = "CERN-PH-EP-2011-182",
    doi = "10.1103/PhysRevLett.108.082001",
    journal = "Phys. Rev. Lett.",
    volume = "108",
    pages = "082001",
    year = "2012"
}

@article{Loizides:2017ack,
    author = "Loizides, Constantin and Kamin, Jason and d'Enterria, David",
    title = "{Improved Monte Carlo Glauber predictions at present and future nuclear colliders}",
    eprint = "1710.07098",
    archivePrefix = "arXiv",
    primaryClass = "nucl-ex",
    doi = "10.1103/PhysRevC.97.054910",
    journal = "Phys. Rev. C",
    volume = "97",
    number = "5",
    pages = "054910",
    year = "2018",
    note = "[Erratum: Phys.Rev.C 99, 019901 (2019)]"
}

@article{Barberio:1990ms,
    author = "Barberio, Elisabetta and van Eijk, Bob and Was, Zbigniew",
    title = "{PHOTOS: A Universal Monte Carlo for QED radiative corrections in decays}",
    reportNumber = "CERN-TH-5857-90",
    doi = "10.1016/0010-4655(91)90012-A",
    journal = "Comput. Phys. Commun.",
    volume = "66",
    pages = "115--128",
    year = "1991"
}

@article{LHCb:2017llq,
    author = "{{LHCb Collaboration}}",
    collaboration = "LHCb",
    title = "{Study of J/{\ensuremath{\psi}} Production in Jets}",
    eprint = "1701.05116",
    archivePrefix = "arXiv",
    primaryClass = "hep-ex",
    reportNumber = "LHCB-PAPER-2016-064, CERN-EP-2017-006",
    doi = "10.1103/PhysRevLett.118.192001",
    journal = "Phys. Rev. Lett.",
    volume = "118",
    number = "19",
    pages = "192001",
    year = "2017"
}

@article{Lansberg:2019adr,
    author = "Lansberg, Jean-Philippe",
    title = "{New Observables in Inclusive Production of Quarkonia}",
    eprint = "1903.09185",
    archivePrefix = "arXiv",
    primaryClass = "hep-ph",
    doi = "10.1016/j.physrep.2020.08.007",
    journal = "Phys. Rept.",
    volume = "889",
    pages = "1--106",
    year = "2020"
}

@article{Brambilla:2010cs,
    author = "Brambilla, N. and others",
    title = "{Heavy Quarkonium: Progress, Puzzles, and Opportunities}",
    eprint = "1010.5827",
    archivePrefix = "arXiv",
    primaryClass = "hep-ph",
    reportNumber = "SLAC-R-996, TUM-EFT-11-10, CLNS-10-2066, ANL-HEP-PR-10-44, ALBERTA-THY-11-10, CP3-10-37, FZJ-IKP-TH-2010-24, INT-PUB-10-059, JLAB-THY-11-1308, FERMILAB-PUB-10-737-T",
    doi = "10.1140/epjc/s10052-010-1534-9",
    journal = "Eur. Phys. J. C",
    volume = "71",
    pages = "1534",
    year = "2011"
}

@article{Bodwin:1994jh,
    author = "Bodwin, Geoffrey T. and Braaten, Eric and Lepage, G. Peter",
    title = "{Rigorous QCD analysis of inclusive annihilation and production of heavy quarkonium}",
    eprint = "hep-ph/9407339",
    archivePrefix = "arXiv",
    reportNumber = "ANL-HEP-PR-94-24, FERMILAB-PUB-94-073-T, NUHEP-TH-94-5",
    doi = "10.1103/PhysRevD.55.5853",
    journal = "Phys. Rev. D",
    volume = "51",
    pages = "1125--1171",
    year = "1995",
    note = "[Erratum: Phys.Rev.D 55, 5853 (1997)]"
}

@article{Cho:1995vh,
    author = "Cho, Peter L. and Leibovich, Adam K.",
    title = "{Color octet quarkonia production}",
    eprint = "hep-ph/9505329",
    archivePrefix = "arXiv",
    reportNumber = "CALT-68-1988",
    doi = "10.1103/PhysRevD.53.150",
    journal = "Phys. Rev. D",
    volume = "53",
    pages = "150--162",
    year = "1996"
}

@article{Cho:1995ce,
    author = "Cho, Peter L. and Leibovich, Adam K.",
    title = "{Color octet quarkonia production. 2.}",
    eprint = "hep-ph/9511315",
    archivePrefix = "arXiv",
    reportNumber = "CALT-68-2026",
    doi = "10.1103/PhysRevD.53.6203",
    journal = "Phys. Rev. D",
    volume = "53",
    pages = "6203--6217",
    year = "1996"
}

@article{Baumgart:2014upa,
    author = "Baumgart, Matthew and Leibovich, Adam K. and Mehen, Thomas and Rothstein, Ira Z.",
    title = "{Probing Quarkonium Production Mechanisms with Jet Substructure}",
    eprint = "1406.2295",
    archivePrefix = "arXiv",
    primaryClass = "hep-ph",
    doi = "10.1007/JHEP11(2014)003",
    journal = "JHEP",
    volume = "11",
    pages = "003",
    year = "2014"
}

@article{Bain:2017wvk,
    author = "Bain, Reggie and Dai, Lin and Leibovich, Adam and Makris, Yiannis and Mehen, Thomas",
    title = "{NRQCD Confronts LHCb Data on Quarkonium Production within Jets}",
    eprint = "1702.05525",
    archivePrefix = "arXiv",
    primaryClass = "hep-ph",
    doi = "10.1103/PhysRevLett.119.032002",
    journal = "Phys. Rev. Lett.",
    volume = "119",
    number = "3",
    pages = "032002",
    year = "2017"
}

@article{Arleo:2017ntr,
    author = "Arleo, Fran{\c{c}}ois",
    title = "{Quenching of Hadron Spectra in Heavy Ion Collisions at the LHC}",
    eprint = "1703.10852",
    archivePrefix = "arXiv",
    primaryClass = "hep-ph",
    doi = "10.1103/PhysRevLett.119.062302",
    journal = "Phys. Rev. Lett.",
    volume = "119",
    number = "6",
    pages = "062302",
    year = "2017"
}

@article{Spousta:2016agr,
    author = "Spousta, Martin",
    title = "{On similarity of jet quenching and charmonia suppression}",
    eprint = "1606.00903",
    archivePrefix = "arXiv",
    primaryClass = "hep-ph",
    doi = "10.1016/j.physletb.2017.01.041",
    journal = "Phys. Lett. B",
    volume = "767",
    pages = "10--15",
    year = "2017"
}

@article{Makris:2019ttx,
    author = "Makris, Yiannis and Vitev, Ivan",
    title = "{An Effective Theory of Quarkonia in QCD Matter}",
    eprint = "1906.04186",
    archivePrefix = "arXiv",
    primaryClass = "hep-ph",
    doi = "10.1007/JHEP10(2019)111",
    journal = "JHEP",
    volume = "10",
    pages = "111",
    year = "2019"
}

@article{ALICE:2019lga,
    author = "{ALICE Collaboration}",
    collaboration = "ALICE",
    title = "{Studies of J/$\psi$ production at forward rapidity in Pb-Pb collisions at $\sqrt{s_{\textrm{NN}}}$ = 5.02 TeV}",
    eprint = "1909.03158",
    archivePrefix = "arXiv",
    primaryClass = "nucl-ex",
    reportNumber = "CERN-EP-2019-187",
    doi = "10.1007/JHEP02(2020)041",
    journal = "JHEP",
    volume = "02",
    pages = "041",
    year = "2020"
}

@article{ALICE:2019nrq,
    author = "{ALICE Collaboration}",
    collaboration = "ALICE",
    title = "{Centrality and transverse momentum dependence of inclusive J/{\ensuremath{\psi}} production at midrapidity in Pb{\textendash}Pb collisions at $\sqrt{s_{\textrm{NN}}}$ = 5.02 TeV}",
    eprint = "1910.14404",
    archivePrefix = "arXiv",
    primaryClass = "nucl-ex",
    reportNumber = "CERN-EP-2019-234",
    doi = "10.1016/j.physletb.2020.135434",
    journal = "Phys. Lett. B",
    volume = "805",
    pages = "135434",
    year = "2020"
}

@article{ALICE:2023gco,
    firstauthor = "Acharya, Shreyasi and others",
    author = "{ALICE Collaboration}",
    collaboration = "ALICE",
    title = "{Measurements of inclusive J/{\ensuremath{\psi}} production at midrapidity and forward rapidity in Pb{\textendash}Pb collisions at $\sqrt{s_\mathrm{NN}} = 5.02$ TeV}",
    eprint = "2303.13361",
    archivePrefix = "arXiv",
    primaryClass = "nucl-ex",
    reportNumber = "CERN-EP-2023-054",
    doi = "10.1016/j.physletb.2024.138451",
    journal = "Phys. Lett. B",
    volume = "849",
    pages = "138451",
    year = "2024"
}

@article{ALICE:2022jeh,
    firstauthor = "Acharya, Shreyasi and others",
    author = "{ALICE Collaboration}",
    collaboration = "ALICE",
    title = "{{\ensuremath{\psi}}(2S) Suppression in Pb-Pb Collisions at the LHC}",
    eprint = "2210.08893",
    archivePrefix = "arXiv",
    primaryClass = "nucl-ex",
    reportNumber = "CERN-EP-2022-210",
    doi = "10.1103/PhysRevLett.132.042301",
    journal = "Phys. Rev. Lett.",
    volume = "132",
    number = "4",
    pages = "042301",
    year = "2024"
}

@article{ALICE:2020wwx,
    firstauthor = "Acharya, Shreyasi and others",
    author = "{ALICE Collaboration}",
    collaboration = "ALICE",
    title = "{{\ensuremath{\Upsilon}} production and nuclear modification at forward rapidity in Pb{\textendash}Pb collisions at $\sqrt{s_\mathrm{NN}} = 5.02$ TeV}",
    eprint = "2011.05758",
    archivePrefix = "arXiv",
    primaryClass = "nucl-ex",
    reportNumber = "CERN-EP-2020-207",
    doi = "10.1016/j.physletb.2021.136579",
    journal = "Phys. Lett. B",
    volume = "822",
    pages = "136579",
    year = "2021"
}

@article{Shuryak:2014zxa,
      author         = "Shuryak, Edward",
      title          = "{Strongly coupled quark-gluon plasma in heavy ion
                        collisions}",
      journal        = "Rev. Mod. Phys.",
      volume         = "89",
      year           = "2017",
      pages          = "035001",
      doi            = "10.1103/RevModPhys.89.035001",
      eprint         = "1412.8393",
      archivePrefix  = "arXiv",
      primaryClass   = "hep-ph",
      SLACcitation   = "%%CITATION = ARXIV:1412.8393;%%"
}

@article{Busza:2018rrf,
    author = "Busza, Wit and Rajagopal, Krishna and van der Schee, Wilke",
    title = "{Heavy Ion Collisions: The Big Picture, and the Big Questions}",
    eprint = "1802.04801",
    archivePrefix = "arXiv",
    primaryClass = "hep-ph",
    reportNumber = "MIT-CTP-4892, MIT-CTP/4892",
    doi = "10.1146/annurev-nucl-101917-020852",
    journal = "Ann. Rev. Nucl. Part. Sci.",
    volume = "68",
    pages = "339--376",
    year = "2018"
}

@article{Connors:2017ptx,
    author = "Connors, Megan and Nattrass, Christine and Reed, Rosi and Salur, Sevil",
    title = "{Jet measurements in heavy ion physics}",
    eprint = "1705.01974",
    archivePrefix = "arXiv",
    primaryClass = "nucl-ex",
    doi = "10.1103/RevModPhys.90.025005",
    journal = "Rev. Mod. Phys.",
    volume = "90",
    pages = "025005",
    year = "2018"
}

@article{Cunqueiro:2021wls,
    author = "Cunqueiro, Leticia and Sickles, Anne M.",
    title = "{Studying the QGP with Jets at the LHC and RHIC}",
    eprint = "2110.14490",
    archivePrefix = "arXiv",
    primaryClass = "nucl-ex",
    doi = "10.1016/j.ppnp.2022.103940",
    journal = "Prog. Part. Nucl. Phys.",
    volume = "124",
    pages = "103940",
    year = "2022"
}

@article{Apolinario:2024equ,
    author = "Apolin\'ario, Lili\'ana and Chien, Yang-Ting and Cunqueiro Mendez, Leticia",
    title = "{Jet substructure}",
    doi = "10.1142/S0218301324300030",
    journal = "Int. J. Mod. Phys. E",
    volume = "33",
    number = "07",
    pages = "2430003",
    year = "2024"
}

@article{Miller:2007ri,
    author = "Miller, Michael L. and Reygers, Klaus and Sanders, Stephen J. and Steinberg, Peter",
    title = "{Glauber modeling in high energy nuclear collisions}",
    eprint = "nucl-ex/0701025",
    archivePrefix = "arXiv",
    doi = "10.1146/annurev.nucl.57.090506.123020",
    journal = "Ann. Rev. Nucl. Part. Sci.",
    volume = "57",
    pages = "205--243",
    year = "2007"
}

@mastersthesis{Oreglia:1980cs,
    author = "Oreglia, M.",
    title = "{A Study of the Reactions $\psi^\prime \to \gamma \gamma \psi$}",
    reportNumber = "SLAC-0236, SLAC-236, UMI-81-08973, SLAC-R-0236, SLAC-R-236",
    type = "Ph.D. Thesis, SLAC-236",
    school = "Stanford University",
    month = "12",
    year = "1980"
}

@article{PDG2024,
  author = {Navas, S. and others (Particle Data Group)},
  title = {Review of Particle Physics},
  journal = {Phys. Rev. D},
  volume = {110},
  number = {3},
  pages = {030001},
  year = {2024},
  doi = {10.1103/PhysRevD.110.030001},
  url = {https://link.aps.org/doi/10.1103/PhysRevD.110.030001}
}

@article{Hahn:2003sc,
    author = "Hahn, H. and others",
    title = "{The RHIC design overview}",
    doi = "10.1016/S0168-9002(02)01938-1",
    journal = "Nucl. Instrum. Meth. A",
    volume = "499",
    pages = "245--263",
    year = "2003"
}

@article{Haas_2017,
doi = {10.1088/1742-6596/898/4/042004},
%url = {https://doi.org/10.1088/1742-6596/898/4/042004},
year = {2017},
month = {oct},
publisher = {IOP Publishing},
volume = {898},
number = {4},
pages = {042004},
author = {Haas, Andrew and on behalf of the ATLAS Collaboration},
title = {ATLAS Simulation using Real Data: Embedding and Overlay},
journal = {Journal of Physics: Conference Series},
}

@article{stouffer1949,
  title={The american soldier: Adjustment during army life.(studies in social psychology in world war ii), vol. 1},
  author={Stouffer, Samuel A and Suchman, Edward A and DeVinney, Leland C and Star, Shirley A and Williams Jr, Robin M},
  year={1949},
  publisher={Princeton Univ. Press}
}


%% file: ATLAS-useful.bib
@Article{Evans:2008zzb,
      author         = "Evans, Lyndon and Bryant, Philip",
      title          = "{LHC Machine}",
      journal        = "JINST",
      volume         = "3",
      pages          = "S08001",
      doi            = "10.1088/1748-0221/3/08/S08001",
      year           = "2008",
      SLACcitation   = "%%CITATION = JINST,3,S08001;%%",
}

@Article{Cacciari:2008gp,
     author    = "Cacciari, Matteo and Salam, Gavin P. and Soyez, Gregory",
     title     = "{The anti-\(k_{t}\) jet clustering algorithm}",
     journal   = "JHEP",
     volume    = "04",
     year      = "2008",
     pages     = "063",
     eprint    = "0802.1189",
     archivePrefix = "arXiv",
     primaryClass  =  "hep-ph",
     doi       = "10.1088/1126-6708/2008/04/063",
     SLACcitation  = "%%CITATION = 0802.1189;%%"
}

@Article{Fastjet,
      author         = "Cacciari, Matteo and Salam, Gavin P. and Soyez, Gregory",
      title          = "{FastJet user manual}",
      journal        = "Eur. Phys. J. C",
      volume         = "72",
      year           = "2012",
      pages          = "1896",
      doi            = "10.1140/epjc/s10052-012-1896-2",
      eprint         = "1111.6097",
      archivePrefix  = "arXiv",
      primaryClass   = "hep-ph",
      reportNumber   = "CERN-PH-TH-2011-297",
      SLACcitation   = "%%CITATION = ARXIV:1111.6097;%%"
}

@Article{Pumplin:2002vw,
     author    = "Pumplin, J. and others",
     title     = "{New Generation of Parton Distributions with Uncertainties from Global QCD Analysis}",
     journal   = "JHEP",
     volume    = "07",
     year      = "2002",
     pages     = "012",
     doi       = "10.1088/1126-6708/2002/07/012",
     eprint    = "hep-ph/0201195",
     archivePrefix = "arXiv",
     SLACcitation  = "%%CITATION = HEP-PH/0201195;%%"
}

@Article{Sjostrand:2007gs,
	author    = "Sj{\"o}strand, T. and Mrenna, S. and Skands, P.",
	title     = "{A brief introduction to PYTHIA 8.1}",
	journal   = "Comput. Phys. Commun.",
	volume    = "178",
	year      = "2008",
	pages     = "852-867",
	eprint    = "0710.3820",
	archivePrefix = "arXiv",
	primaryClass  =  "hep-ph",
	doi       = "10.1016/j.cpc.2008.01.036",
	SLACcitation  = "%%CITATION = 0710.3820;%%"
}

@article{Sjostrand:2014zea,
      author         = "Sj{\"o}strand, Torbj{\"o}rn and Ask, Stefan and Christiansen,
                        Jesper R. and Corke, Richard and Desai, Nishita and Ilten,
                        Philip and Mrenna, Stephen and Prestel, Stefan and
                        Rasmussen, Christine O. and Skands, Peter Z.",
      title          = "{An introduction to PYTHIA 8.2}",
      journal        = "Comput. Phys. Commun.",
      volume         = "191",
      year           = "2015",
      pages          = "159",
      doi            = "10.1016/j.cpc.2015.01.024",
      eprint         = "1410.3012",
      archivePrefix  = "arXiv",
      primaryClass   = "hep-ph",
      reportNumber   = "LU-TP-14-36, MCNET-14-22, CERN-PH-TH-2014-190,
                        FERMILAB-PUB-14-316-CD, DESY-14-178, SLAC-PUB-16122,
                        --FERMILAB-PUB-14-316-CD",
      SLACcitation   = "%%CITATION = ARXIV:1410.3012;%%"
}

@article{Agostinelli:2002hh,
      author         = "Agostinelli, S. and others",
      title          = "{\textsc{Geant4} -- a simulation toolkit}",
      journal        = "Nucl. Instrum. Meth. A",
      volume         = "506",
      year           = "2003",
      pages          = "250",
      doi            = "10.1016/S0168-9002(03)01368-8",
      reportNumber   = "SLAC-PUB-9350, FERMILAB-PUB-03-339",
      SLACcitation   = "%%CITATION = NUIMA,A506,250;%%"
}

@Booklet{RooFit,
      author         = "Verkerke, Wouter and Kirkby, David",
      title          = "{The RooFit toolkit for data modeling}",
      year           = "2003",
      eprint         = "physics/0306116",
      archivePrefix  = "arXiv",
      primaryClass   = "physics.data-an",
}

@article{Barlow:1993dm,
    author = "Barlow, Roger and Beeston, Christine",
    title = "{Fitting using finite Monte Carlo samples}",
    reportNumber = "MAN-HEP-93-1",
    doi = "10.1016/0010-4655(93)90005-W",
    journal = "Comput. Phys. Commun.",
    volume = "77",
    pages = "219",
    year = "1993"
}


%% file: ATLAS.bib
@Article{PERF-2007-01,
    author         = "{ATLAS Collaboration}",
    title          = "{The ATLAS Experiment at the CERN Large Hadron Collider}",
    journal        = "JINST",
    volume         = "3",
    year           = "2008",
    pages          = "S08003",
    doi            = "10.1088/1748-0221/3/08/S08003",
    primaryClass   = "hep-ex",
}

@Article{SOFT-2010-01,
    author         = "{ATLAS Collaboration}",
    title          = "{The ATLAS Simulation Infrastructure}",
    journal        = "Eur. Phys. J. C",
    volume         = "70",
    year           = "2010",
    pages          = "823",
    doi            = "10.1140/epjc/s10052-010-1429-9",
    eprint         = "1005.4568",
    archivePrefix  = "arXiv",
    primaryClass   = "physics.ins-det",
}

@Article{HION-2011-02,
    author         = "{ATLAS Collaboration}",
    title          = "{Measurement of the jet radius and transverse momentum dependence of inclusive jet suppression in lead--lead collisions at \(\sqrt{s_{\text{NN}}} = 2.76\,\text{TeV}\) with the ATLAS detector}",
    journal        = "Phys. Lett. B",
    volume         = "719",
    year           = "2013",
    pages          = "220",
    doi            = "10.1016/j.physletb.2013.01.024",
    reportNumber   = "CERN-PH-EP-2012-134",
    eprint         = "1208.1967",
    archivePrefix  = "arXiv",
    primaryClass   = "hep-ex",
}

@Article{BPHY-2012-02,
    author         = "{ATLAS Collaboration}",
    title          = "{Measurement of the differential cross-sections of prompt and non-prompt production of \(J/\psi\) and \(\psi(2\text{S})\) in \(pp\) collisions at \(\sqrt{s} = 7\) and \(8\,\text{TeV}\) with the ATLAS detector}",
    journal        = "Eur. Phys. J. C",
    volume         = "76",
    year           = "2016",
    pages          = "283",
    doi            = "10.1140/epjc/s10052-016-4050-8",
    reportNumber   = "CERN-PH-EP-2015-292",
    eprint         = "1512.03657",
    archivePrefix  = "arXiv",
    primaryClass   = "hep-ex",
}

@Article{HION-2012-08,
    author         = "{ATLAS Collaboration}",
    title          = "{Measurement of inclusive jet charged-particle fragmentation functions in Pb+Pb collisions at \(\sqrt{s_{\text{NN}}} = 2.76\,\text{TeV}\) with the ATLAS detector}",
    journal        = "Phys. Lett. B",
    volume         = "739",
    year           = "2014",
    pages          = "320",
    doi            = "10.1016/j.physletb.2014.10.065",
    reportNumber   = "CERN-PH-EP-2014-101",
    eprint         = "1406.2979",
    archivePrefix  = "arXiv",
    primaryClass   = "hep-ex",
}

@Article{PERF-2014-02,
    author         = "{ATLAS Collaboration}",
    title          = "{Determination of jet calibration and energy resolution in proton--proton collisions at \(\sqrt{s} = 8\,\text{TeV}\) using the ATLAS detector}",
    journal        = "Eur. Phys. J. C",
    volume         = "80",
    year           = "2020",
    pages          = "1104",
    doi            = "10.1140/epjc/s10052-020-08477-8",
    reportNumber   = "CERN-EP-2019-057",
    eprint         = "1910.04482",
    archivePrefix  = "arXiv",
    primaryClass   = "hep-ex",
}

@Article{PERF-2015-10,
    author         = "{ATLAS Collaboration}",
    title          = "{Muon reconstruction performance of the ATLAS detector in proton--proton collision data at \(\sqrt{s} = 13\,\text{TeV}\)}",
    journal        = "Eur. Phys. J. C",
    volume         = "76",
    year           = "2016",
    pages          = "292",
    doi            = "10.1140/epjc/s10052-016-4120-y",
    reportNumber   = "CERN-EP-2016-033",
    eprint         = "1603.05598",
    archivePrefix  = "arXiv",
    primaryClass   = "hep-ex",
}

@Article{HION-2016-07,
    author         = "{ATLAS Collaboration}",
    title          = "{Prompt and non-prompt \(J/\psi\) and \(\psi(2\text{S})\) suppression at high transverse momentum in \(5.02\,\text{TeV}\) Pb+Pb collisions with the ATLAS experiment}",
    journal        = "Eur. Phys. J. C",
    volume         = "78",
    year           = "2018",
    pages          = "762",
    doi            = "10.1140/epjc/s10052-018-6219-9",
    reportNumber   = "CERN-EP-2018-049",
    eprint         = "1805.04077",
    archivePrefix  = "arXiv",
    primaryClass   = "nucl-ex",
}

@Article{STDM-2016-03,
    author         = "{ATLAS Collaboration}",
    title          = "{Measurement of inclusive jet and dijet cross-sections in proton--proton collisions at \(\sqrt{s} = 13\,\text{TeV}\) with the ATLAS detector}",
    journal        = "JHEP",
    volume         = "05",
    year           = "2018",
    pages          = "195",
    doi            = "10.1007/JHEP05(2018)195",
    reportNumber   = "CERN-EP-2017-157",
    eprint         = "1711.02692",
    archivePrefix  = "arXiv",
    primaryClass   = "hep-ex",
}

@Article{TRIG-2016-01,
    author         = "{ATLAS Collaboration}",
    title          = "{Performance of the ATLAS trigger system in 2015}",
    journal        = "Eur. Phys. J. C",
    volume         = "77",
    year           = "2017",
    pages          = "317",
    doi            = "10.1140/epjc/s10052-017-4852-3",
    reportNumber   = "CERN-EP-2016-241",
    eprint         = "1611.09661",
    archivePrefix  = "arXiv",
    primaryClass   = "hep-ex",
}

@Article{HION-2017-10,
    author         = "{ATLAS Collaboration}",
    title          = "{Measurement of the nuclear modification factor for inclusive jets in Pb+Pb collisions at \(\sqrt{s_{\text{NN}}} = 5.02\,\text{TeV}\) with the ATLAS detector}",
    journal        = "Phys. Lett. B",
    volume         = "790",
    year           = "2019",
    pages          = "108",
    doi            = "10.1016/j.physletb.2018.10.076",
    reportNumber   = "CERN-EP-2018-105",
    eprint         = "1805.05635",
    archivePrefix  = "arXiv",
    primaryClass   = "nucl-ex",
}

@Article{HION-2018-22,
    author         = "{ATLAS Collaboration}",
    title          = "{Measurement of \(J/\psi\)-jet correlations in \(pp\) and Pb+Pb collisions at \(\sqrt{s_{\text{NN}}} = 5.02\,\text{TeV}\) with the ATLAS detector}",
    year           = "2026",
    reportNumber   = "CERN-EP-2026-167",
    eprint         = "2606.31375",
    archivePrefix  = "arXiv",
    primaryClass   = "hep-ex",
}

@Article{JETM-2018-05,
    author         = "{ATLAS Collaboration}",
    title          = "{Jet energy scale and resolution measured in proton--proton collisions at \(\sqrt{s} = 13\,\text{TeV}\) with the ATLAS detector}",
    journal        = "Eur. Phys. J. C",
    volume         = "81",
    year           = "2021",
    pages          = "689",
    doi            = "10.1140/epjc/s10052-021-09402-3",
    reportNumber   = "CERN-EP-2020-083",
    eprint         = "2007.02645",
    archivePrefix  = "arXiv",
    primaryClass   = "hep-ex",
}

@Article{MUON-2018-03,
    author         = "{ATLAS Collaboration}",
    title          = "{Muon reconstruction and identification efficiency in ATLAS using the full Run~2 \(pp\) collision data set at \(\sqrt{s} = 13\,\text{TeV}\)}",
    journal        = "Eur. Phys. J. C",
    volume         = "81",
    year           = "2021",
    pages          = "578",
    doi            = "10.1140/epjc/s10052-021-09233-2",
    reportNumber   = "CERN-EP-2020-199",
    eprint         = "2012.00578",
    archivePrefix  = "arXiv",
    primaryClass   = "hep-ex",
}

@Article{PIX-2018-001,
    author         = "Abbott, B. and others",
    title          = "{Production and integration of the ATLAS Insertable B-Layer}",
    journal        = "JINST",
    volume         = "13",
    year           = "2018",
    pages          = "T05008",
    doi            = "10.1088/1748-0221/13/05/T05008",
    eprint         = "1803.00844",
    archivePrefix  = "arXiv",
    primaryClass   = "physics.ins-det",
}

@Article{TRIG-2018-01,
    author         = "{ATLAS Collaboration}",
    title          = "{Performance of the ATLAS muon triggers in Run~2}",
    journal        = "JINST",
    volume         = "15",
    year           = "2020",
    pages          = "P09015",
    doi            = "10.1088/1748-0221/15/09/p09015",
    reportNumber   = "CERN-EP-2020-031",
    eprint         = "2004.13447",
    archivePrefix  = "arXiv",
    primaryClass   = "physics.ins-det",
}

@Article{HION-2019-58,
    author         = "{ATLAS Collaboration}",
    title          = "{Measurement of the nuclear modification factor for muons from charm and bottom hadrons in Pb+Pb collisions at \(5.02\,\text{TeV}\) with the ATLAS detector}",
    journal        = "Phys. Lett. B",
    volume         = "829",
    year           = "2022",
    pages          = "137077",
    doi            = "10.1016/j.physletb.2022.137077",
    reportNumber   = "CERN-EP-2021-153",
    eprint         = "2109.00411",
    archivePrefix  = "arXiv",
    primaryClass   = "nucl-ex",
}

@Article{HION-2020-06,
    author         = "{ATLAS Collaboration}",
    title          = "{Measurement of substructure-dependent suppression of large-radius jets with charged particles in Pb+Pb collisions with ATLAS}",
    journal        = "Phys. Lett. B",
    volume         = "871",
    year           = "2025",
    pages          = "139929",
    doi            = "10.1016/j.physletb.2025.139929",
    reportNumber   = "CERN-EP-2025-078",
    eprint         = "2504.04805",
    archivePrefix  = "arXiv",
    primaryClass   = "nucl-ex",
}

@Article{HION-2020-10,
    author         = "{ATLAS Collaboration}",
    title          = "{Measurement of muon pairs produced via \(\gamma\gamma\) scattering in nonultraperipheral Pb+Pb collisions at \(\sqrt{s_{\text{NN}}} = 5.02\,\text{TeV}\) with the ATLAS detector}",
    journal        = "Phys. Rev. C",
    volume         = "107",
    year           = "2023",
    pages          = "054907",
    doi            = "10.1103/PhysRevC.107.054907",
    reportNumber   = "CERN-EP-2022-047",
    eprint         = "2206.12594",
    archivePrefix  = "arXiv",
    primaryClass   = "nucl-ex",
}

@Article{DAPR-2021-01,
    author         = "{ATLAS Collaboration}",
    title          = "{Luminosity determination in \(pp\) collisions at \(\sqrt{s} = 13\,\text{TeV}\) using the ATLAS detector at the LHC}",
    journal        = "Eur. Phys. J. C",
    volume         = "83",
    year           = "2023",
    pages          = "982",
    doi            = "10.1140/epjc/s10052-023-11747-w",
    reportNumber   = "CERN-EP-2022-281",
    eprint         = "2212.09379",
    archivePrefix  = "arXiv",
    primaryClass   = "hep-ex",
}

@Article{HION-2021-12,
    author         = "{ATLAS Collaboration}",
    title          = "{Production of \(\varUpsilon(\text{nS})\) mesons in Pb+Pb and \(pp\) collisions at \(5.02\,\text{TeV}\)}",
    journal        = "Phys. Rev. C",
    volume         = "107",
    year           = "2023",
    pages          = "054912",
    doi            = "10.1103/PhysRevC.107.054912",
    reportNumber   = "CERN-EP-2022-045",
    eprint         = "2205.03042",
    archivePrefix  = "arXiv",
    primaryClass   = "nucl-ex",
}

@Article{SOFT-2022-02,
    author         = "{ATLAS Collaboration}",
    title          = "{Software and computing for Run~3 of the ATLAS experiment at the LHC}",
    journal        = "Eur. Phys. J. C",
    volume         = "85",
    year           = "2025",
    pages          = "234",
    doi            = "10.1140/epjc/s10052-024-13701-w",
    reportNumber   = "CERN-EP-2024-100",
    eprint         = "2404.06335",
    archivePrefix  = "arXiv",
    primaryClass   = "hep-ex",
    related        = "SOFT-2022-02-err",
    relatedstring  = "Erratum:",
}

@Booklet{ATL-SOFT-PUB-2026-001,
    author         = "{ATLAS Collaboration}",
    title          = "{ATLAS Computing Acknowledgements}",
    howpublished   = "{ATL-SOFT-PUB-2026-001}",
    url            = "https://cds.cern.ch/record/2952666",
    year           = "2026",
}

@Report{ATLAS-TDR-19,
    author         = "{ATLAS Collaboration}",
    title          = "{ATLAS Insertable B-Layer: Technical Design Report}",
    type           = "ATLAS-TDR-19; CERN-LHCC-2010-013",
    year           = "2010",
    url            = "https://cds.cern.ch/record/1291633",
    related        = "ATLAS-TDR-19-addm",
    relatedstring  = "Addendum:",
}

@Article{LUCID2,
    author         = "Avoni, G. and others",
    title          = "{The new LUCID-2 detector for luminosity measurement and monitoring in ATLAS}",
    journal        = "JINST",
    volume         = "13",
    year           = "2018",
    pages          = "P07017",
    doi            = "10.1088/1748-0221/13/07/P07017",
}


%% file: CMS.bib
@Article{CMS-BPH-13-003,
    author         = "{CMS Collaboration}",
    title          = "{Measurement of the prompt \(J/\psi\) and \(\psi(2\text{S})\) polarizations in \(pp\) collisions at \(\sqrt{s} = 7\,\text{TeV}\)}",
    journal        = "Phys. Lett. B",
    volume         = "727",
    year           = "2013",
    pages          = "381",
    doi            = "10.1016/j.physletb.2013.10.055",
    reportNumber   = "CERN-PH-EP-2013-140",
    eprint         = "1307.6070",
    archivePrefix  = "arXiv",
    primaryClass   = "hep-ex",
}

@Article{CMS-BPH-15-003,
    author         = "{CMS Collaboration}",
    title          = "{Study of \(J/\psi\) meson production inside jets in \(pp\) collisions at \(\sqrt{s} = 8\,\text{TeV}\)}",
    journal        = "Phys. Lett. B",
    volume         = "804",
    year           = "2020",
    pages          = "135409",
    doi            = "10.1016/j.physletb.2020.135409",
    reportNumber   = "CERN-EP-2019-186",
    eprint         = "1910.01686",
    archivePrefix  = "arXiv",
    primaryClass   = "hep-ex",
}

@Article{CMS-SMP-15-007,
    author         = "{CMS Collaboration}",
    title          = "{Measurement of the double-differential inclusive jet cross section in proton--proton collisions at \(\sqrt{s} = 13\,\text{TeV}\)}",
    journal        = "Eur. Phys. J. C",
    volume         = "76",
    year           = "2016",
    pages          = "451",
    doi            = "10.1140/epjc/s10052-016-4286-3",
    reportNumber   = "CERN-EP-2016-104",
    eprint         = "1605.04436",
    archivePrefix  = "arXiv",
    primaryClass   = "hep-ex",
}

@Article{CMS-HIN-16-023,
    author         = "{CMS Collaboration}",
    title          = "{Measurement of nuclear modification factors of \(\Upsilon(1\text{S})\), \(\Upsilon(2\text{S})\), and \(\Upsilon(3\text{S})\) mesons in PbPb collisions at \(\sqrt{s_{\text{NN}}} = 5.02\,\text{TeV}\)}",
    journal        = "Phys. Lett. B",
    volume         = "790",
    year           = "2019",
    pages          = "270",
    doi            = "10.1016/j.physletb.2019.01.006",
    reportNumber   = "CERN-EP-2018-110",
    eprint         = "1805.09215",
    archivePrefix  = "arXiv",
    primaryClass   = "hep-ex",
}

@Article{CMS-HIN-16-025,
    author         = "{CMS Collaboration}",
    title          = "{Measurement of prompt and nonprompt charmonium suppression in PbPb collisions at \(5.02\,\text{TeV}\)}",
    journal        = "Eur. Phys. J. C",
    volume         = "78",
    year           = "2018",
    pages          = "509",
    doi            = "10.1140/epjc/s10052-018-5950-6",
    reportNumber   = "CERN-EP-2017-308",
    eprint         = "1712.08959",
    archivePrefix  = "arXiv",
    primaryClass   = "hep-ex",
}

@Article{CMS-HIN-19-007,
    author         = "{CMS Collaboration}",
    title          = "{Fragmentation of jets containing a prompt \(J/\psi\) meson in PbPb and \(pp\) collisions at \(\sqrt{s_{\text{NN}}} = 5.02\,\text{TeV}\)}",
    journal        = "Phys. Lett. B",
    volume         = "825",
    year           = "2022",
    pages          = "136842",
    doi            = "10.1016/j.physletb.2021.136842",
    reportNumber   = "CERN-EP-2021-104",
    eprint         = "2106.13235",
    archivePrefix  = "arXiv",
    primaryClass   = "hep-ex",
}

@Article{CMS-HIN-21-007,
    author         = "{CMS Collaboration}",
    title          = "{Observation of the \(\Upsilon(3\text{S})\) Meson and Suppression of \(\Upsilon\) States in PbPb Collisions at \(\sqrt{s_{\text{NN}}} = 5.02\,\text{TeV}\)}",
    journal        = "Phys. Rev. Lett.",
    volume         = "133",
    year           = "2024",
    pages          = "022302",
    doi            = "10.1103/PhysRevLett.133.022302",
    reportNumber   = "CERN-EP-2023-011",
    eprint         = "2303.17026",
    archivePrefix  = "arXiv",
    primaryClass   = "hep-ex",
}


%% file: ConfNotes.bib
@Booklet{ATLAS-CONF-2015-016,
    author         = "{ATLAS Collaboration}",
    title          = "{Jet energy scale and its uncertainty for jets reconstructed using the ATLAS heavy ion jet algorithm}",
    howpublished   = "{ATLAS-CONF-2015-016}",
    url            = "https://cds.cern.ch/record/2008677",
    year           = "2015",
}


%% file: PubNotes.bib
@Booklet{ATL-PHYS-PUB-2014-021,
    author         = "{ATLAS Collaboration}",
    title          = "{ATLAS Pythia~8 tunes to \(7\,\text{TeV}\) data}",
    howpublished   = "{ATL-PHYS-PUB-2014-021}",
    url            = "https://cds.cern.ch/record/1966419",
    year           = "2014",
}

@Booklet{ATL-PHYS-PUB-2021-011,
    author         = "{ATLAS Collaboration}",
    title          = "{Evaluating statistical uncertainties and correlations using the bootstrap method}",
    howpublished   = "{ATL-PHYS-PUB-2021-011}",
    url            = "https://cds.cern.ch/record/2759945",
    year           = "2021",
}
